\documentstyle{article}
\begin{document}
\vskip.5in
\centerline{\Large {\bf Similarity Renormalization Group Approach to}}
\vskip.1in
\centerline{\Large {\bf Boost Invariant Hamiltonian Dynamics}}
\vskip.1in
\centerline{December, 1997}

\vskip .3in
\centerline{Stanis{\l}aw D. G{\l}azek}
\vskip .1in
\centerline{Institute of Theoretical Physics, Warsaw University}
\centerline{ul. Ho{\.z}a 69, 00-681 Warsaw}

\vskip.3in
\centerline{\bf Abstract}
\vskip.1in

We outline a method of deriving boost invariant hamiltonians for
effective particles in quantum field theory.  The method is based on the
similarity renormalization group transformation for hamiltonians in the
canonical light-front quantization scheme.  The hamiltonians are defined
and calculated using creation and annihilation operators.  The
renormalization group equations are written for a sequence of unitary
transformations which gradually transform the bare canonical creation
and annihilation operators of a local theory to the creation and
annihilation operators of effective particles in an effective theory
with the same dynamical content but a finite range of energy transfers
due to form factors in the interaction vertices.  The form factors
result from the similarity renormalization group flow of effective
hamiltonians.  The regularized initial hamiltonian and the renormalized
effective hamiltonians possess seven kinematical Poincar\'e symmetries
specific to the light-front quantization scheme.  Thus, the effective
interactions can be used to describe the constituent dynamics in
relativistically moving systems including the rest and the infinite
momentum frame.  Solutions to the general equations for the effective
hamiltonians are illustrated in perturbation theory by second-order
calculations of self-energy and two-particle interaction terms in Yukawa
theory, QED and QCD.  In Yukawa theory, one obtains the generalized
Yukawa potential including its full off-energy-shell extension and form
factors in the vertices.  In QED, the effective hamiltonian eigenvalue
problem converges for small coupling constants to the Schr\"odinger
equation but the typical relativistic ultraviolet singularities at short
distances between constituents are regularized by the similarity form
factors.  In the second-order QCD effective hamiltonian one obtains a
boost invariant logarithmically confining quark-anti-quark interaction
term which may remain uncanceled in the non-abelian dynamics of
effective quarks and gluons.


\newpage

{\bf 1. INTRODUCTION}
\vskip.1in

This paper describes a theory of boost invariant effective hamiltonians
in quantum field theory.  Physical states are assumed to be describable
by solutions to the Schr\"odinger equation with these hamiltonians.

The hamiltonians are derived by integrating first-order differential
equations of the similarity renormalization group.  The initial
condition is provided by the regularized canonical light-front
hamiltonian of a local quantum field theory with counterterms.  The
hamiltonian acts in the Fock space which is constructed by applying the
canonical bare creation operators to a vacuum state.  The vacuum is
annihilated by the corresponding annihilation operators.  Due to the
light-front boost symmetry, this representation of states is useful in a
relativistic theory and the dynamics in any frame resembles dynamics in
the infinite momentum frame.

The similarity renormalization group is defined in terms of running
creation and annihilation operators.  The running operators interpolate
between the bare ones in a local hamiltonian and effective ones in the
effective hamiltonian.  The effective operators are used for
construction of the effective basis states in the Fock space.  The
effective hamiltonians are calculable term by term using methods of
successive approximations and perturbation theory.  One has to study the
role of different terms in the effective Schr\"odinger equation.
Solutions include bound states.

One reason for the renormalization group to play an important role in
the hamiltonian approach is ultraviolet divergences; the initial
expressions for bare hamiltonians contain divergences of local field
theory and the divergences require renormalization.  The most prominent
example is the canonical hamiltonian of QED.  Old-fashioned tree
diagrams of hamiltonian perturbation theory are finite and closely
reproduce experimental data.  Renormalization problems appear when one
sums over intermediate states and the sum diverges.  The divergences
correspond to diverging loop integrals in Feynman diagrams.  However,
the hamiltonian approach greatly differs from the lagrangian
diagrammatic approaches.

One apparent difference is that the sums over intermediate states
involve integrals over a three-dimensional momentum space while the
integrals in the lagrangian calculus are four-dimensional.  Although a
connection exists for finite integrals which are not sensitive to
cutoffs, the connection is broken when the integrals diverge.  In the
hamiltonian approach, we introduce the effective Fock space basis and we
construct hamiltonians using three-dimensional regularization and
renormalization procedures.  In the lagrangian approach, one directly
calculates Green functions using four-dimensional regularization and
renormalization techniques.  Equivalence of the two approaches in
diverging cases remains to be shown, especially when the bound states
are taken into account.

Besides removing divergences, the renormalization group is useful in the
hamiltonian approach because it introduces a hierarchy of scales.
Phenomena of different scales are dealt with in a certain order.  This
enables us to solve problems involving many scales.  Particle theories
contain many, possibly even infinitely many different scales.  Using the
renormalization group approach, we can start from the hamiltonian of a
basic theory that couples all degrees of freedom of all scales and we
can reduce the initial hamiltonian to an effective renormalized
hamiltonian in which the couplings between degrees of freedom of vastly
different scales vanish.  Then, the couplings between different scales
are further reduced in the renormalization group flow to obtain the
effective hamiltonian matrix which is sufficiently narrow in scale so
that its spectrum of eigenstates can be found in practice.  The
similarity transformation is designed to eliminate all large changes of
scale by the effective interactions.

The different scales in the hamiltonian renormalization group approach
are defined by different scales of momentum.  The momentum scales are
defined using the relative momenta of interacting particles.  The
definitions will be given in Section 2. Here we need to mention that the
effective hamiltonian of a small width contains interactions which
couple particles of similar energies only.  The energy changes induced
by the hamiltonian are limited by the running similarity renormalization
group cutoff, denoted by $\lambda$.  The smaller is $\lambda$ the
smaller is the energy width of the effective hamiltonian.  Exact results
for physical quantities are independent of $\lambda$.

A number of model subspaces of the Fock space need to be considered when
one is solving for the spectrum of a field-theoretic effective
hamiltonian because the full space of states is too large for
computations.  Different physical problems require different model
subspaces.  Working within a subspace of interest, one should secure
that the results for physical quantities are independent of the running
cutoff $\lambda$.  The cutoff independence can appear only in a certain
range of cutoffs that corresponds to the physical problem and model
subspace under consideration.  However, once the cutoff independence in
the finite range is achieved, one expects to have solved the theory in
this range.

The following diagram will illustrate the situation.
\vskip.2in
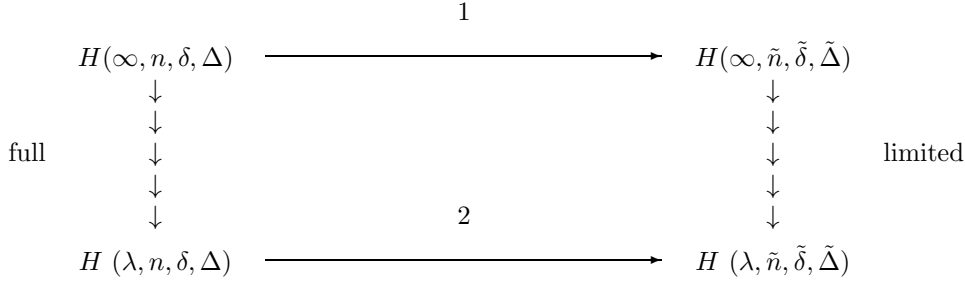
\begin{tabular}{rcccl}
&                 &            1           &                                &\\
& $H(\infty,n,\delta,\Delta)$ & $^{\vector(3,0){150}}$ & $H(\infty,\tilde n,\tilde \delta,\tilde \Delta)$ &\\
& $\downarrow$    &                        & $\downarrow$                   &\\
& $\downarrow$    &                        & $\downarrow$                   &\\
full & $\downarrow$&                       & $\downarrow$ & limited          \\
& $\downarrow$    &                        & $\downarrow$                   &\\
& $\downarrow$    &            2           & $\downarrow$                   &\\
&$H~(\lambda,n,\delta,\Delta)$& $^{\vector(3,0){150}}$ & $H~(\lambda,\tilde n,\tilde \delta, \tilde \Delta)$&
\end{tabular}
\vskip.3in

In this diagram, the vertical arrows indicate evolution in the
renormalization group parameter $\lambda$ which limits the relative
energy transfers in the interaction terms.  It will be explained in
detail in next Sections how the limits are imposed.  $\lambda$ ranges
from infinity in the initial hamiltonian to a finite value in an
effective hamiltonian.  The hamiltonians depend on additional parameters
$n$, $\delta$ and $\Delta$.

$n$ stands for the cutoff on the {\it change} of the particle number.
It defines the limits on the numbers of creation and annihilation
operators that can appear in a single hamiltonian term.  For example,
the canonical expressions for light-front hamiltonians in local field
theories of physical interest have the number of creation and
annihilation operators in a single term limited to 4, and the particle
number cannot change by more than $n = 2$.

$\delta$ stands for the infrared cutoff.  For example, it may be the
lower bound on the longitudinal momentum carried by a particle that
appears or disappears in a single interaction.

$\Delta$ stands for the ultraviolet cutoff which defines the upper limit
on the relative transverse momentum of particles which can appear or
disappear in a single interaction.

The left branch of the diagram is marked ``full'' because it represents
the renormalization group flow calculated using the effective creation
and annihilation operators with no restriction imposed on the space of
states.

The right branch of the figure is marked ``limited'' because it
describes the renormalization group flow in the bare model space which
is limited by parameters $\tilde n$, $\tilde \delta$ and $\tilde
\Delta$.  Imposing the limits is denoted by the arrow marked 1. For
example, $\tilde n$ can limit the number of bare particles, $\tilde
\delta$ can limit the bare particle momenta from below and $\tilde
\Delta$ can limit from above free energies of the states which are taken
into account.

The initial $H(\infty,n,\delta,\Delta)$ contains counterterms which are
constructed using the condition that physical results have well defined
limits when the cutoffs $n$, $\delta$ and $\Delta$ are relaxed.  The
construction of counterterms in perturbation theory will be discussed in
detail in the next Sections.  Once the counterterms remove the
regularization dependence from the effective dynamics the arguments $n$,
$\delta$ and $\Delta$ in $H(\lambda,n,\delta,\Delta)$ in the lower left
corner of the diagram are equivalent to their limiting values, $n =
\infty$, $\delta = 0$ and $\Delta = \infty$.  Thus,
$H(\lambda,n,\delta,\Delta) \equiv H(\lambda)$.  One should stress that
the infrared regulator $\delta$ may still appear in the effective
hamiltonian if there are massless particles in the theory.  This is
important in gauge theories.

Accuracy of the step denoted by the arrow 1 has to be checked by
relaxing the model cutoff parameters $\tilde n$, $\tilde \delta$ and
$\tilde \Delta$ and measuring the resulting changes in the spectrum of
$H(\lambda,\tilde n, \tilde \delta,\tilde \Delta)$.  Naturally, these
cutoffs may have to be varied in a big range because they are introduced
along the arrow 1 for bare particles.

The effective hamiltonians at the bottom of the diagram, namely,
$H(\lambda,n,\delta,\Delta)$ and $H(\lambda,\tilde n,\tilde \delta,
\tilde \Delta)$ with energy transfers limited by finite $\lambda$, are
connected by the arrow marked 2. This arrow denotes the procedure of
introducing the small space cutoffs $\tilde n$, $\tilde \delta$ and
$\tilde \Delta$ which enable us to approximately solve for the spectrum
of the effective hamiltonian $H(\lambda,n,\delta,\Delta)$.  This time,
however, the final computation cutoffs are introduced at the level of
the effective particles, not at the level of the initial bare particles.

The arrow 2 denotes the replacement of the whole effective hamiltonian
matrix by a limited matrix.  The procedure of obtaining the small matrix
will be discussed below.  The spectrum of the small matrix may be very
close to the corresponding part of the spectrum of the full matrix
because $\lambda$ is small (this will become clear later).  The accuracy
of the calculation must be verified by relaxing the cutoffs $\tilde n$,
$\tilde \delta$ and $\tilde \Delta$ and observing convergence of results
as in the case of the arrow 1 and branch ``limited''.  But now, it is
natural to expect that the cutoffs $\tilde n$, $\tilde \delta$ and
$\tilde \Delta$ may have to be varied only in a small range which
corresponds to $\lambda$.  Thus, a finite dynamical problem to solve is
defined.

The ``full'' renormalization group evolution is calculable using the
method described in this paper.  The ``limited'' evolution can be
calculated using the matrix elements techniques introduced earlier by
G{\l}azek and Wilson in Refs.  \cite{GW1} and \cite{GW2} who drew on the
work of Wilson \cite{W1} \cite{W2}.  The matrix elements techniques were
introduced for application to QCD.  \cite{W3} Alternatively, one can
adopt Wegner's flow equations for hamiltonian matrix elements in cases
soluble with the energy-independent width $\lambda$.  \cite{WEG}
\cite{GW4} The present approach can be viewed as a special case of the
general similarity renormalization group for hamiltonians because the
hamiltonians we consider transform by the same unitary transformations
as our creation and annihilation operators.  However, by having defined
the renormalization group transformation for the effective creation and
annihilation operators, we remove the need to consider the model space
dependence of the renormalization group transformation.

The transformation we describe in this paper is partly similar to the
transformation discussed by Melosh.  \cite{Melosh} The important
difference is that we provide a dynamical theory of the transformation
in a form applicable to particles of different kinds.  If one restricts
attention to QCD, the boost invariant calculus is expected to help in
establishing a connection between the constituent quark model, Feynman
parton model, and perturbative quantum chromodynamics.

Both ways in the diagram which start from the initial hamiltonian
$H(\infty,n,\delta,\Delta)$ and go through the arrow 1 and the arrows
``limited'' (called branch 1l) and the arrows ``full'' and the arrow 2
(called branch f2), lead to a finite $H(\lambda,\tilde n,\tilde \delta,
\tilde \Delta)$.  When calculations of some selected matrix elements of
the effective hamiltonian are done in perturbation theory, both ways of
going through the diagram are equivalent.  For only a finite range of
particle numbers and momenta can be reached in a limited number of steps
of size $\lambda$ starting from the finite values selected by the
external states of the matrix elements in question.

Differences arise when one solves for the spectrum of an effective
hamiltonian and when one attempts to vary the model space parameters
$\tilde n$, $\tilde \delta$ and $\tilde \Delta$.  In the ``full''
calculation, one obtains a single effective hamiltonian which one can
solve in successively enlargeable model spaces.  In the ``limited''
calculation, the model space restrictions are imposed at the beginning
and they lead to an effective hamiltonian whose action cannot be
considered in a larger model space without repeating the renormalization
group calculation in the larger space.

An explicit example of a difference between the two branches 1l and f2
in the diagram above is provided in Ref.  \cite{TD} which discusses a
Tamm-Dancoff (TD) approach analogous to the branch 1l (cf.  Ref.
\cite{HO}).  In the TD approach, there are restrictions on the particle
number which naturally lead to the sector-dependent counterterms as
described in Ref.  \cite{TD}, for example, for masses.  On the other
hand, in the procedure of the branch f2 no such sector dependent
counterterms arise.  The present paper describes examples of
sector-independent mass counterterms.

Proportionality to different powers of the coupling constant helps in
estimates of how important are different effective interaction terms and
how to choose the model space.  Finding the basis which can span a good
approximation to the full solution requires trial and error studies.
This general feature can be illustrated by the following $2 \times 2$
matrix.

\begin{center}
$$ \left[ \begin{array}{ccc}
a + b g^2 & g v       \\
g v       & c + d g^2
\end{array} \right] $$
\end{center}

\noindent This matrix is a model of the entire effective hamiltonian
matrix calculated to second order in $g$ including all couplings between
all effective Fock sectors as given by the ``full'' calculation.  Thus,
we have the hamiltonian terms order 1, order $g$ and order $g^2$.  In a
perturbative calculation using matrix elements, which is focused on the
upper sectors, one would calculate only the terms $a$ and $b g^2$.

Assume that $a$ and $c$ are of the same order, and $b$ and $d$ are of
the same order, and calculate eigenvalues of the model matrix neglecting
terms order $g^4$ and higher powers of $g$.  For arbitrarily small $g$,
the eigenvalues are given by $a$ and terms quadratic in $g$.  No terms
linear in $g$ arise.  The quadratic corrections include contributions
due to the term $g v$ which couples different sectors.  The role of this
coupling needs to be estimated.  The presence of $d g^2 $ seems to be
irrelevant because it couples to the upper sector through the
off-diagonal terms order $g$.  Hence, it seems to contribute only in
order $g^4$ to the eigenvalues.

It is well known that the above analysis is wrong in the case with
degenerate diagonal matrix elements no matter how the degeneracy arises.
For example, consider the case of a finite $g$ such that $ a + b g^2 = c
+ d g^2 $. The eigenvalues are equal $ a + b g^2 \pm gv$.  They are
linear in $g$ instead of being quadratic, for arbitrary $v$.  The lowest
eigenvalue eigenstate is a superposition of the upper and the lower
sector instead of being dominated by the upper one.  In this example,
the degeneracy is not visible until the term $d g^2$ is included in the
calculation.  As a second example consider the case with degenerate
matrix elements $ c $ and $ a + b g^2 $ and $d$ not included.  The
simple non-degenerate perturbative expansion is again not applicable.
But the addition of the term $d g^2$ can lift the degeneracy and make
the simple perturbative expansion work.

Corrections due to interaction terms such as $g v$ may be additionally
suppressed for very small $\lambda$ since the range of $v$ in momentum
variables is given by $\lambda$.  If $\lambda$ is reduced in the
renormalization group flow down to a number on the order of some
positive power of $g$ then the resulting interaction can contribute to
the eigenvalues in the order implied by $g$ and $\lambda$ together which
is higher than $g^2$.  In addition, the effective interaction $v$ may
contain small factors.  For example, in the effective $e^+e^-$-sector of
positronium, the emission of photons is proportional to the velocity of
electrons which is order $\alpha$, on average.  The interaction term $g
v$ which couples states with an additional photon, plays no role in the
eigenvalue in order $\alpha^2$ if the width $\lambda$ restricts energy
changes to order $\alpha^2 \, m_{electron}$ (cf.  \cite{JPG}).

Terms such as $d g^2$ have been originally discussed in the light-front
approach to QCD by Perry.  \cite{Perry} Heavy quarkonia are dominated by
the effective $Q\bar Q$ sector.  Terms such as $d g^2$ in other Fock
sectors may lift up energies of effective gluons due to the non-abelian
interactions to a sufficiently high value so that the model hamiltonian
$a + b g^2 $ in the $Q\bar Q$ sector alone may have eigenstates which
approximate the full solution for heavy mesons.  The important
observation made by Perry \cite{Perry} in a frame dependent matrix
elements approach using coupling coherence is that the terms $b g^2$
contain a logarithmically confining potential.  An analogous boost
invariant logarithmic interaction term in the Fock space in our approach
will be discussed in Section 3.

The above diagram and the $2 \times 2$ matrix model illustrate the
structure of our similarity renormalization group approach to the
light-front hamiltonian dynamics in quantum field theory.  We summarize
the steps here.

The first step is the calculation of the effective hamiltonian,

$$   H(\lambda) \quad = \quad S^\dagger_{\lambda,n,\delta,\Delta}
     \,\, H(\infty,n,\delta,\Delta) \,\, S_{\lambda,n,\delta,\Delta}
     \,\, . \eqno(1.1)$$

\noindent $S$ denotes the similarity transformation.  Eq.  (1.1)
corresponds to the arrows marked full in the diagram.

The second step is to solve the effective Schr\"odinger equation

$$   H(\lambda) \,\, |\psi\rangle \quad = \quad E \,\, |\psi\rangle
     \,\, . \eqno(1.2) $$

\noindent $H(\lambda)$ has the same dynamical content and eigenvalues as
$H(\infty,n,\delta,\Delta)$.  The eigenvalue $E$ is independent of the
width $\lambda$.

Equation (1.2) greatly differs from the eigenvalue equation for
$H(\infty,n,\delta,\Delta)$.  The major difference is that the dynamics
of $H(\lambda)$ has a limited range on the energy scale and the
hamiltonian does not contain ultraviolet divergences.  Therefore, one
can attempt to solve the eigenvalue problem scale by scale.  Scattering
processes are described by the same hamiltonian.  Next Sections will
give examples of two fermions scattering in different theories.

Solutions to Eq.  (1.2) provide renormalization conditions for the
finite parts of counterterms.  A general method is necessary for
reducing the full eigenvalue problem to a manageable one.  This step is
marked by the arrow 2 in the diagram.  In the case of the $2 \times 2$
matrix model, this step corresponds to the calculation of the model
space hamiltonian in the upper-left corner of the matrix.  The
similarity renormalization scheme guarantees that this step is free from
ultraviolet divergences because the width $\lambda$ is finite.

In the general case, one can apply the well known Bloch \cite{Bloch}
technique of calculating model space hamiltonians.  Suppose we want to
evaluate a model two-body hamiltonian knowing $H(\lambda)$ with $\lambda
< m$, where $m$ is the effective one-body mass.  We can introduce the
projection operator $P$ on the effective two-particle sector with a
limited center-of-mass energy.  We also introduce the operator $R$ which
generates the multi-particle and high energy components of the
eigenstates from their limited mass two-body part.  By assumption, $R$
satisfies the conditions $(1-P)R = RP = R$ and $PR = R(1-P) = 0$ and the
equation $( P + R - 1 ) \, H(\lambda) \, (P + R) = 0$.  Then, the model
two-body dynamics is described by the hamiltonian \cite{W2}

$$  H_2 \quad = \quad (P + R^\dagger R)^{-1/2}
\,\, (P+R^\dagger) \,\, H(\lambda) \,\, (P + R) \,\,
     (P + R^\dagger R)^{-1/2} \,\, . \eqno(1.3) $$

\noindent The same approach can be used for larger model spaces.  The
model space is characterized by the parameters $\tilde n$, $\tilde
\delta$ and $\tilde \Delta$ in the diagram.  So, the operation $R$
depends on these parameters.  But the resulting spectrum in the range of
interest should not depend on the model space boundary when the width
$\lambda$ is small and the model space contains the dynamically dominant
basis states in the selected range of scales.  The heuristic Eq.  (1.3)
can be applied in perturbation theory in the effective interaction even
for sizable coupling constants since the effective interaction strength
is considerably reduced by the similarity factors.

The scheme outlined above is still prone to the infrared regularization
dependence for massless particles.  This is particularly important in
gauge theories.  However, the effective hamiltonian dynamics is expected
to lead to infrared convergent results for gauge invariant quantities.
There is also a possibility that new effective interactions are
generated from the infrared region and they bring in effects normally
associated with a nontrivial vacuum state.  \cite{W3} \cite{SUS}
\cite{GSR} We shall make comments on the issue of long distance
phenomena in the present approach in Section 2.b where we describe the
range of scales involved in the theory.  The reader should refer to
\cite{SUS}, \cite{GSR}, \cite{W3} and \cite{BPP} for discussions of the
ground state, spontaneous symmetry breaking and zero-modes problems.

The remaining part of this paper is organized as follows.  Section 2
describes the hamiltonian formalism in three subsections.  Namely,
Section 2.a introduces the similarity renormalization group equations
and describes methods of solution, Section 2.b describes regularization
factors, 2.c deals with renormalization conditions.  Section 3 contains
examples of lowest order applications of the formalism.  Our derivation
of the generalized Yukawa potential is given in Section 3.a.  Section
3.b describes the Schr\"odinger equation for positronium in QED.
Section 3.c discusses a confining term for constituent quarks in QCD.
Section 4 concludes the paper.  The list of references is focused on the
similarity renormalization group approach to hamiltonian dynamics in the
light-front Fock space.  The reader should be aware of this limitation.
Examples of other approaches to quantum field theory in the light-front
form of dynamics can be found in Ref.  \cite{BPP}.

\vskip.3in
{\bf 2. EFFECTIVE HAMILTONIANS}
\vskip.1in

This Section is divided into three parts.  The first part describes our
method of calculating effective hamiltonians in the Fock space.  The
second part presents our regularization scheme for initial hamiltonians.
The last part discusses renormalization conditions and the effective
eigenvalue problem.

\vskip.3in
{\bf 2.a Similarity transformation}
\vskip.1in

We construct a family of effective hamiltonians in the light-front Fock
space.  The family is parameterized by a scale parameter $\lambda$ which
ranges from infinity to a finite value.  $\lambda$ limits energy
transfers in the interaction terms.

The hamiltonians are built of sums of ordered products of creation and
annihilation operators.  The hamiltonian labeled by $\lambda$ is
expressed in terms of creation and annihilation operators which
correspond to $\lambda$.  We commonly denote these operators by
$q_\lambda$.  In addition, the creation and annihilation operators carry
labels of quantum numbers such as momentum, spin, flavor or color.  We
will not indicate those numbers in the initial presentation, unless it
is necessary.

All hamiltonians in the family are assumed to be equal. Thus,

$$        H_{\lambda_1} (q_{\lambda_1}) = H_{\lambda_2} (q_{\lambda_2}).
\eqno (2.1) $$

For $\lambda = \infty$, the hamiltonians $H_\infty$ are expressed in
terms of operators creating and annihilating bare particles, $q_\infty$.
Hamiltonians $H_\infty$ can be constructed from the canonical
field theoretic expressions for the energy-momentum density tensors.

Unfortunately, expressions for $H_\infty$ in local field theories are
divergent.  They need to be regularized by introducing a bare
ultraviolet cutoff which we shall denote by $\epsilon$.  The ultraviolet
cutoff $\Delta$ from the previous Section corresponds to
$\Lambda^2/\epsilon$ where $\Lambda$ is an arbitrary finite constant
which carries the necessary dimension of a mass.  The limit of removing
the bare ultraviolet cutoff will correspond to $\epsilon \rightarrow 0$.

$H_{\lambda= \infty} = H_\epsilon$ for all values of $\epsilon$.  For
the limit $\epsilon \rightarrow 0$ to exist the hamiltonians $H_\infty$
must include a number of additional terms (called counterterms) whose
structure will be determined later.

$H_\infty$ may include an infrared regulator, generically denoted by
$\delta$.  For example, this is required in QED with massless photons
and in QCD with massless gluons.  $\delta \rightarrow 0$ when the
infrared regularization is removed.  The parameter $\delta$ is indicated
explicitly if needed.

Our key assumption is that the particle degrees of freedom for all
different scales $\lambda$ are unitarily equivalent to the bare particle
degrees of freedom:

$$   q_{\lambda} = U_\lambda q_\infty U^\dagger_\lambda \, .\eqno (2.2) $$

\noindent This assumption says that the quantum numbers of bare and
effective particles are the same for all values of $\lambda$.  The
following examples explain the origin of this assumption.  (1)
Constituent quarks have the same quantum numbers as current quarks.  (2)
We use the same quantum numbers for photons and electrons independently
of the kind of processes we consider in QED or in related effective
theories such as the nonrelativistic Schr\"odinger equation with Coulomb
potentials between charges.  (3) Pions and nucleons in nuclear physics
have the same quantum numbers quite independently of what kind of
interactions, pion-nucleon vertex form factors or other dynamical
assumptions one uses.

It follows from Eq.  (2.2) that creation and annihilation operators for
$\lambda_1$ and $\lambda_2$ are unitarily equivalent and connected by
transformations of the form $U_{\lambda_1} U^\dagger_{\lambda_2}$.  The
transformations $U_{\lambda_1}$ or $ U_{\lambda_2}$ will depend on the
bare cutoffs but the transformation $U_{\lambda_1}
U^\dagger_{\lambda_2}$ for finite $\lambda_1$ and $\lambda_2$ will have
to be finite in the limit $\epsilon \rightarrow 0$.

The transformation $U_\lambda$ is defined indirectly through a
differential equation of the type used in the similarity renormalization
scheme for hamiltonians of G{\l}azek and Wilson \cite{GW1} \cite{GW2}.
That scheme was originally developed for application to QCD \cite{W3}.
Hamiltonians with labels $\lambda_1$ and $\lambda_2$ are connected by
integration of the differential equation from $\lambda_1$ to
$\lambda_2$.  Our guiding principle in writing the differential equation
for effective hamiltonians is that {\it the resulting interactions
between effective particles with considerably different scales of
relative momenta are suppressed}.

This principle has its origin in the following examples.  (1) Emission
and absorption of short wavelength photons are not essential in the
formation of atoms.  (2) Emission and absorption of hard pions by
nucleons is not important in nonrelativistic nuclear physics.  (3)
Constituent quarks have moderate momenta and their effective dynamics
seem to be independent of the very hard gluon emissions.  (4) High
momentum transfer phenomena are independent of the small momentum
transfer effects such as binding.  A standard way of achieving this kind
of picture in theoretical models is to include form factors in the
interaction vertices.  The form factors quickly tend to zero when
momenta change by more then the size of a specific cutoff parameter.

The cutoff parameter in the form factors sets the scale for allowed
changes of momenta.  It determines the range or width of the interaction
in momentum space.  That width is the origin of our scale $\lambda$
which labels renormalized effective hamiltonians.  Our similarity
factors are analogous to the vertex form factors which are commonly used
in nonlocal models (see also Ref.  \cite {BGP}).  The large momentum
transfer dynamics is integrated out through the similarity
renormalization group equation.

Boost invariance requires that the individual momenta of effective
particles are not restricted because boosts change those unlimitedly.
The hamiltonian width restricts only relative momenta of effective
particles.  Also, the larger is a relative momentum the larger change is
generated by a boost.  Therefore, when the free energy of interacting
particles in their center-of-mass frame (i.e. the free light-front
invariant mass) is much larger than $\lambda$ the immediate change of
energy due to interaction is limited by the large energy itself instead
of $\lambda$.  At the same time, this condition takes care of the
property of wave mechanics that strong interference occurs between waves
of similar wavelengths within a range of wavelengths on the order of the
wavelengths themselves.  \cite{W1}

In our construction, strong dynamical interference effects for states of
similar free energies are made not to contribute in the derivation of
effective hamiltonians.  For example, the similarity transformation is
constructed in such a way that only large energy denominators can appear
in the perturbative calculations of effective hamiltonians and small
denominators are excluded.  \cite{GW1} \cite{GW2} Namely, only large
free energy changes are integrated out.  In this approach, calculations
of strong coherence effects for nearly degenerate states are relegated
to a later step of solving for eigenstates of the effective hamiltonian.
That step may be non-perturbative.  For example, the Coulomb potential
of QED is formally of the first order in $\alpha$ and leads to a variety
of bound atomic structures beyond perturbation theory.

Our differential equations require a separation of the changes in
creation and annihilation operators from changes in coefficients in
front of products of the operators.  In order to define this separation
we assume that terms with a large number of the operators in a product
do not dominate or mediate the effective dynamics of interest.  If the
latter assumption turns out to be invalid our formalism merely provides
a way to approach the resulting problems.  The comment due here is that
if the dynamics leads to spontaneous symmetry breaking, or condensates,
we will have a well defined renormalized hamiltonian theory to study
those phenomena in the desired detail, cf.  Refs.  \cite{SUS},
\cite{GSR} and \cite{W3}.

The unitary equivalence of creation and annihilation operators at the scale
$\lambda$ and at the infinite scale, i.e. those appearing in $H_\infty =
H_\epsilon$, together with the equality of hamiltonians at all scales
imply that

$$ H_\lambda (q_\lambda) = U_\lambda H_\lambda(q_\infty) U^\dagger_\lambda
   = H_\infty (q_\infty). \eqno (2.3) $$

\noindent We denote $H_\lambda(q_\infty) = {\cal H} _\lambda$
and obtain

$$  {\cal H} _\lambda = U^\dagger_\lambda H_\infty U_\lambda  . \eqno (2.4) $$

\noindent Thus, the effective hamiltonian $H_\lambda$ is obtained from
the hamiltonian ${\cal H} _\lambda$ by replacing creation and
annihilation operators for bare particles by creation and annihilation
operators for effective particles with the same quantum numbers.  The
bare creation and annihilation operators are independent of $\lambda$.
One calculates $\lambda$-dependent coefficients in front of the products
of $q_\infty$ in ${\cal H}_\lambda$.

The differential equation for ${\cal H}_\lambda$ is \cite {GW2}

$$ {d \over d\lambda} {\cal H}_\lambda = [ {\cal H}_\lambda, {\cal
T}_\lambda] \, ,\eqno (2.5) $$

\noindent where

$$ {\cal T}_\lambda = U^\dagger_\lambda {d \over d\lambda} U_\lambda .
\eqno (2.6) $$

\noindent ${\cal H}_\lambda$ has the following structure

$$ {\cal H}_\lambda = F_\lambda[{\cal G}_\lambda].  \eqno (2.7) $$

\noindent $F_\lambda[ {\cal G}_\lambda ]$ denotes the similarity form
factors in $ {\cal H}_\lambda$ to be described below.  Using the unitary
equivalence, we also have

$$ H_\lambda(q_\lambda)= F_\lambda[G_\lambda(q_\lambda)], \eqno (2.8) $$

\noindent where

$$ G_\lambda(q_\lambda)= U_\lambda {\cal G}_\lambda U^\dagger_\lambda .
\eqno (2.9) $$

\noindent A similar relation holds for $T_\lambda(q_\lambda)$ and ${\cal
T}_\lambda$ since the latter is expressed in terms of $q_\infty$.

The operation $F_\lambda$ acts on the operator ${\cal G}_\lambda$ equal
to a superposition of terms each of which is an ordered product of
creation and annihilation operators.  The ordering is arbitrary but
needs to be determined.  We adopt the order from left to right of
creators of fermions, creators of bosons, creators of anti-fermions,
annihilators of anti-fermions, annihilators of bosons, annihilators of
fermions.  At least two operators must appear in a product and at least
one creation and one annihilation operator must appear.  No product
contains only creation or only annihilation operators.  This is a
special property of light-front hamiltonians.  Hamiltonians in other
forms of dynamics do not have this property and lead to the necessity of
solving the ground state formation problem before other states can be
considered because the pure creation or annihilation terms produce
disconnected vacuum dynamics.

The operator ${\cal G}_\lambda$ is divided into two parts, ${\cal
G}_{1\lambda}$ and ${\cal G}_{2\lambda}$.  ${\cal G}_{1\lambda}$ is a
superposition of all terms of the form $a^\dagger a$ for $a$ equal
$q_\infty$ of any kind.  In principle, one could also include in ${\cal
G}_{1\lambda}$ some chosen terms with a larger number of creation and
annihilation operators, e.g. terms containing two creation and two
annihilation operators.  However, plane-wave Fock space basis states are
not eigenstates of relevant operators of such type and we limit ${\cal
G}_{1\lambda}$ to terms $a^\dagger a$ to avoid the difficulty in present
calculations.

${\cal G}_{1\lambda}$ becomes the effective free part of $G_\lambda$,
denoted $G_{1\lambda}$, after $q_\infty$ is replaced by $q_\lambda$.
The effective free hamiltonian part $H_{1\lambda}$ is equal to
$G_{1\lambda}$ because $G_{1\lambda}$ is not changed by the operation
$F$.  Eigenvalues of $G_{1\lambda}$ are called {\it free energies}.

The remaining part ${\cal G}_{2\lambda} = {\cal G}_\lambda - {\cal
G}_{1\lambda}$ gives the interaction part of the effective hamiltonian
$H_\lambda$.  One replaces $q_\infty$ by $q_\lambda$ and obtains
$G_{2\lambda}$.  Then, one applies the operation $F_\lambda$ which
inserts the vertex form factors defined as follows.

Let the momentum labels of all creation operators in a single product in
an interaction term be $k_1, k_2, ..., k_I$ and the momentum labels of
all annihilation operators be $k'_1, k'_2, ..., k'_J$.  Each momentum
has three components, $k^+ $ ranging from 0 to $\infty$ and two
transverse components $k^\perp = (k^1,k^2)$, both ranging from $ -
\infty$ to $+ \infty$.  The $z$-axis is distinguished by our choice of
the light-front.  The sum of momentum labels of creation operators,
$\sum_{i=1}^I k_i$, equals the sum of momentum labels of annihilation
operators, $\sum_{j=1}^J k'_j$.  We denote these sums by $P = (P^+,
P^\perp)$, ($P^+$ is positive).  Each $k^+$ or ${k'}^+$ is a positive
fraction of $P^+$; $x_i= k^+_i /P^+ $, $1 > x_i > 0$ and $x'_j =
k'^+_j/P^+ $, $1 > x'_j > 0$.  We have $\sum_{i=1}^I x_i = \sum_{j=1}^J
x'_j = 1$.  We also define

$$ \kappa^\perp_n = k^\perp_n - x_n P^\perp \eqno (2.10) $$

\noindent for all momenta in the hamiltonian term.  $\sum_{i=1}^I
\kappa^\perp_i = \sum_{j=1}^J {\kappa'}^\perp_j = 0$.

These momentum variables appear standard but the way they are used here
is not.  Namely, $P$ is usually not equal to a total momentum of a
physical state.  It characterizes the interaction term whose action
redistributes $P$ from the set of momenta of the annihilated particles
to the set of momenta of the created particles.

Thus, each term in the hamiltonian is characterized by $P$ and two sets
of variables, $X_I=\{(x_i, \kappa^\perp_i)\}^{i=I}_{i=1}$ for creation
operators and $X'_J=\{(x'_j,{\kappa'}^\perp_j)\}^{j=J}_{j=1}$ for
annihilation operators.  For example, in a product of two creation
operators and one annihilation operator we have $x_1 = x$, $x_2 = 1-x$
and $x'_1 = 1$.  Also, $\kappa^\perp_1 = - \kappa^\perp_2 =
\kappa^\perp$ and ${\kappa'}^\perp_1 = 0$.  $P$ can be arbitrary and the
term in question replaces one particle of momentum $P$ by two particles
of momenta $x P + \kappa$ and $(1-x) P - \kappa$ for $+$ and $\perp$
components, $\kappa^+=0$.  It is convenient to speak of $P$ as a {\it
parent} momentum and about the individual particle momenta as {\it
daughter} momenta.  The parent momentum in a hamiltonian term equals one
half of the sum of momenta labeling all creation and annihilation
operators in the term.  Each daughter particle carries a fraction of the
parent momentum.  The parent momentum may be carried by one or more
particles.

The operation $F_\lambda$ acting on a product of creation and
annihilation operators produces

$$  F_\lambda \left[
\prod_{i=1}^I a^\dagger_{k_i} \prod_{j=1}^J a_{k'_j} \right] =
    f_\lambda(X_I,X'_J)
\prod_{i=1}^I a^\dagger_{k_i} \prod_{j=1}^J a_{k'_j}
      . \eqno (2.11) $$

The function $f_\lambda (X_I,X'_J)$ is a suitable function which
represents our physical intuition about form factors.  The arguments of
$f_\lambda$ are invariant with respect to seven kinematical Poincar\'e
transformations of the light-front frame.  This feature results in the
boost symmetry of our theory.  We impose three conditions on the
function $f_\lambda$.

The first condition is that $f_\lambda$ is expressible through the
eigenvalues of $G_{1\lambda}$ corresponding to the sets $X_I$ and $X'_J$
so that $f_\lambda$ equals 1 for small differences between the
eigenvalues and quickly goes to zero when the differences become large.
This is the basic condition of the similarity renormalization scheme for
hamiltonians.\cite{GW1} \cite{GW2} The width of $f_\lambda$ is set by
$\lambda$.  One can consider functions $f_\lambda$ which depend on $X_I$
and $X'_J$ in a more general way than through the eigenvalues of the
free hamiltonian but that option will not be investigated here.

The first condition defines the effective nature of the hamiltonian
labeled by $\lambda$.  Namely, the effective particle states which are
separated by the free energy gap which is much larger than $\lambda$
are not directly coupled by the interactions.  In other words, $\lambda$
limits the free energy changes induced by the effective interaction.
Moreover, as a consequence of $f_\lambda \sim 1$ for similar energies,
$1-f_\lambda$ is close to zero for the similar energies and it vanishes
proportionally to a power of the energy difference.  The higher is the
power the smaller is the role of states of similar energies in the
calculation of the effective hamiltonian.  This will be explained later.
Consequently, the higher is the power the smaller is the role of
non-perturbative phenomena due to energy changes below the scale
$\lambda$ in the calculation of the effective hamiltonian.

Thus, there is a chance for the full hamiltonian diagonalization process
to be divided into two parts:  a perturbative calculation of the
effective renormalized hamiltonian and a non-perturbative
diagonalization of that effective hamiltonian.  This is our
factorization hypothesis in the hamiltonian approach.

The second condition is that both, $1-f_\lambda$ and $d f_\lambda / d
\lambda $, must vanish faster than linearly in the free energy
difference.  This condition is required to exclude the small energy
denominators in perturbation theory and will be explained below.  The
second condition implies that $1-f_\lambda$ vanishes as at least second
power of the energy difference near zero.

The third condition is defined by saying that multi-particle
interactions (especially interactions that change the number of
effective particles by many) should not be important in the effective
hamiltonian dynamics which is characterized by changes of energies below
the scale $\lambda$.  This may be possible if {\it $f_\lambda$ as a
function of the daughter variables approximates the shape of one
particle irreducible vertices which is characteristic to the theory
under consideration}.  Structure of $G_\lambda$ depends on the choice of
the function $f_\lambda$.  Some choices will lead to more complicated
effective interactions than others.  The best choices for the most
efficient description of physical phenomena at some scale $\lambda$ are
such that the effective particles interact in a way that is most easy to
understand and which can be parametrized with the least possible number
of parameters over the range of scales of physical interest.  One can
conceive variational estimates for the best choices of $f_\lambda$ that
minimize complexity of the effective hamiltonians.  For example, it is
clear that creation of effective particles will be suppressed when the
width $\lambda$ becomes comparable to the effective masses of those
particles.

To satisfy the first condition above in a boost invariant way we define
a boost invariant gap between free energy eigenvalues for effective
particles which is to be compared with the running cutoff parameter
$\lambda$.  The light-front quantization scheme does not explicitly
preserve rotational symmetry.  Nevertheless, it has been shown that if
counterterms provide enough freedom through their finite parts and
multi-particle effects are suppressed one can obtain rotationally
invariant results.  \cite{GP} \cite{JUNKO} \cite{JPG}

The free energy eigenvalues relevant to a particular hamiltonian
term with daughter variables $X_I$ and $X'_J$ are

$$ \sum_{i=1}^I { k^{\perp 2}_i + m^2_i(\lambda) \over k^+_i}
= { P^{\perp 2} + {\cal M}^2_I  \over P^+ } \eqno (2.12) $$

\noindent and

$$ \sum_{j=1}^J { {k'}^{\perp 2}_j + m^2_j(\lambda) \over {k'}^+_j}
= { P^{\perp 2} + {\cal M}^2_J  \over P^+ }  , \eqno (2.13) $$

\noindent where

$$ {\cal M}^2_I = \sum_{i=1}^I { \kappa^{\perp 2}_i + m^2_i(\lambda)
\over
   x_i } \eqno (2.14) $$

\noindent and

$$ {{\cal M}'}^2_J = \sum_{j=1}^J { {\kappa'}^{\perp 2}_j +
m^2_j(\lambda)
    \over x'_j } . \eqno (2.15) $$

\noindent
The individual effective particle masses are allowed to
depend on the effective hamiltonian width parameter $\lambda$.
We define {\it the mass difference} for a hamiltonian term to be

$$  \Delta {\cal M}^2 = {{\cal M}'}^2_J - {\cal M}^2_I , \eqno (2.16) $$

\noindent and {\it the mass sum} to be

$$  \Sigma {\cal M}^2 = {{\cal M}'}^2_J + {\cal M}^2_I. \eqno (2.17)  $$

To be specific, we define details of the function $f_\lambda(X_I,X'_J)$
introducing a parameter $z_\lambda$.  Following the similarity
renormalization scheme \cite{GW1} \cite{GW2}, $z_\lambda$ can be chosen
in such a way that $z_\lambda$ is close to zero for $\Delta {\cal M}^2$
small in comparison to $\lambda^2$ or $\Sigma {\cal M}^2$ and
$|z_\lambda |$ is close to 1 for $\Delta {\cal M}^2$ large in comparison
to $\lambda^2$ or comparable to $\Sigma {\cal M}^2$.  For example,

$$ z_\lambda = {\Delta {\cal M}^2 \over \Sigma {\cal M}^2 + \lambda^2 }.
\eqno (2.18) $$

\noindent The definition includes $\Sigma {\cal M}^2$) to ease estimates
in high order perturbation theory, especially in the analysis of
overlapping divergences.  \cite{GW1} The new feature here is that the
introduction of $\Sigma {\cal M}^2$ does not violate the light-front
boost invariance and basic cluster decomposition properties.
$f_\lambda(X_I,X'_J)$ is defined for the purpose of this article to be a
function of $z_\lambda^{2^n}$, $ n \geq 1$, which is analytic in the
vicinity of the interval $[0,1]$ on the real axis, equals 1 for
$z_\lambda=0$ and quickly approaches 0 for $z_\lambda \sim 1$;

$$ f_\lambda(X_I,X'_J) = f(z_\lambda^{2^n}) \, . \eqno (2.19) $$

\noindent For example,

$$ f(u) = \left[ 1 + \left({u (1-u_0) \over u_0 (1-u)}\right)^{2^m}
          \right]^{-1}, \eqno (2.20) $$

\noindent where $1 > u_0 > 0$ and $m \geq 1$.  The larger the exponent
$m$ the closer $f(u)$ approaches $\theta(u_0 - u)$ for $0 \leq u \leq
1$.  Eq.  (2.20) concludes our definition of the operation $F_\lambda$.

The smallest possible value of $\Sigma {\cal M}^2$ in Eq.  (2.18) is
$\left[ \sum_{i=1}^I m_i(\lambda) \right]^2 + \left[ \sum_{j=1}^J
m_j(\lambda) \right]^2 $. Thus, $z_\lambda$ is small for small positive
$\lambda^2$ when $\Delta {\cal M}^2$ is small in comparison to particle
masses.  Therefore, $u_0$ must be much smaller than 1 to force $\Delta
{\cal M}^2$ to be small in comparison to $\Sigma {\cal M}^2$ when
$\lambda^2$ is small.  One can also force $\Delta {\cal M}^2$ to be
small in comparison to the particle masses by making $\lambda^2$
negative so that it subtracts from $\Sigma {\cal M}^2$ its minimal
value.  Then, the mass difference is compared to the sum of kinetic
energies due to the relative motion only.  It is also useful to limit
the small mass differences by choosing an infinitesimally small $u_0$
and introducing $\lambda^2 = u_0^{-1/2^n} \tilde \lambda^2$.  Then,
$|\Delta {\cal M}^2| \leq \tilde \lambda^2$ in the $\theta$-function
limit.  In this case, the band-diagonal hamiltonian width becomes
independent of the mass sum for as long as the latter is small in
comparison to $\lambda^2$.

The infinitesimal transformation ${\cal T}_\lambda$ in Eq.  (2.5) is
defined as follows.  Eq.  (2.5) is rewritten using Eqs.  (2.8) and
(2.9), into the form

$$  {\cal H}' = f'{\cal G} + f {\cal G}' = [f{\cal G}_1, {\cal T}]
+ [f{\cal G}_2, {\cal T}] \, . \eqno (2.21) $$

\noindent The prime denotes differentiation with respect to $\lambda$.
We have simplified the notation of $F_\lambda[{\cal G}_\lambda]$ to
$f{\cal G}$.  Three universal relations $f{\cal G}_1 = {\cal G}_1$ and
$(1-f){\cal G}_1 = f'{\cal G}_1 = 0$ are then used without saying.

Equation (2.21) involves two unknowns, $\cal G$ and $\cal T$.
Additional condition is required to define $\cal T$.  One recalls that
if the interaction is absent, i.e. when ${\cal G}_2 = 0$, then no
evolution with $\lambda$ may appear.  Therefore, in the limit of
negligible interactions, both ${\cal G}' = 0$ and ${\cal T} =0$ . ${\cal
G}'$ should differ from zero if and only if the interactions are
important.  The first term on the right-hand side is order ${\cal T}$
since ${\cal G}_1$ contains terms independent of interactions.  The
second term on the right-hand side is at least of second order in
interactions.  The first term can be used as a seed for defining ${\cal
T}$ through a series of powers of the interaction.

We associate the derivative of $\cal G$ with the second term on the
right-hand side.  The first term on the right-hand side and a part of
the second term which is left after the derivative of $\cal G$ is
defined, together determine $\cal T$. $\cal T$ is defined through the
commutator $[{\cal G}_1, {\cal T}]$ using a curly bracket notation.  We
write

$$ A = \{ B \}_{{\cal G}_1} \eqno (2.22) $$

\noindent when

$$ [A, {\cal G}_1] = B   . \eqno (2.23) $$

\noindent Subscripts of such curly brackets are often omitted in later
discussion.  Suppose B contains a term which involves a product

$$ \prod_{i=1}^I a^\dagger_{k_i} \prod_{j=1}^J a_{k'_j} . \eqno
(2.24) $$

\noindent Then, $\{B\}_{{\cal G}_{1\lambda}}$ contains the same product
(as a part of the same expression) with an additional factor equal

$$ \left[ \sum_{j=1}^J { {k'}^{\perp 2}_j + m^2_j(\lambda) \over
{k'}^+_j} -
\sum_{i=1}^I { k^{\perp 2}_i + m^2_i(\lambda) \over k^+_i}
\right]^{-1}    .  \eqno (2.25) $$

\noindent The sums of individual energies satisfy Eqs.  (2.12) and
(2.13), respectively, and the factor (2.25) equals

$$ \left[ \Delta {\cal M}^2 \over P^+ \right]^{-1} ,  \eqno (2.26) $$

\noindent where $P^+$ is the parent momentum for the product under
consideration and the mass difference is defined in Eq.  (2.16).
All terms in the operator $B$ are multiplied by the corresponding
factors.

The factor (2.25) explodes to infinity when the denominator approaches
zero.  Hence, for the operator $A$ to be well defined, the coefficients
of products of the form (2.24) in the operator $B$ must vanish at least
as fast as the energy denominator itself when the denominator approaches
zero.  Therefore, our definition of $[{\cal G}_1, {\cal T}]$ is given in
terms of an operator which has such property.  Eq.  (2.21) is split into
two equations as follows.

$$  f{\cal G}' = f[f{\cal G}_2, {\cal T}] \,  , \eqno (2.27) $$

$$ [{\cal T}, {\cal G}_1] = (1-f) [f{\cal G}_2, {\cal T}] - f' {\cal G}
 \,  .  \eqno (2.28) $$

\noindent Our second condition introduced below Eq.  (2.11) on the
functions $1 - f_\lambda(X_I, X'_J)$ and $f'_\lambda(X_I, X'_J)$
guarantees that ${\cal T}$ is well defined and tends to zero in the
region of vanishing energy denominators because the right-hand side of
Eq.  (2.28) vanishes at least as fast as the first power of the energy
differences.

Equation (2.27) is a first order differential equation.  One has to
provide an initial condition to define a theory.  The initial conditions
are set in this paper by canonical light-front hamiltonians plus
counterterms.  The latter are determined from the condition that the
effective hamiltonians have well defined limits when the bare cutoff is
removed.  In other words, one has to find the class of initial
conditions at $\lambda = \infty$ which imply $\epsilon$-independent
$H_\lambda$'s for all finite $\lambda$'s when $\epsilon \rightarrow 0$.

A general iterative procedure for calculating the effective hamiltonians
is analogous to the one from Refs.  \cite{GW1} and \cite{GW2}.  However,
instead of iterating two coupled equations for $H_\lambda$ and
$T_\lambda$ we base iteration on an equivalent single equation for
$H_\lambda$ with an explicit solution for $T_\lambda$ already built in.
Simple algebra and substitution of Eq.  (2.28) into Eq.  (2.27), lead to

$$     {d \over d\lambda} {\cal G}_\lambda =
\left[ f_\lambda {\cal G}_{2\lambda}, \left\{ {d \over d\lambda}
(1-f_\lambda) {\cal G}_{2\lambda} \right\}_{{\cal G}_{1\lambda}} \right] .
\eqno (2.29) $$

\noindent Equation (2.29) drives the renormalization group formalism in
this paper.  Note that the right-hand side is given in terms of a
commutator.  Therefore, {\it the effective renormalized hamiltonians
contain only connected interactions}.  This is essential for cluster
decomposition properties of the effective hamiltonians.  \cite{WE1}

Equation (2.29) is of the form

$$     {d \over d\lambda} {\cal G}_\lambda =
   T_\lambda[{\cal G}_\lambda] . \eqno (2.30) $$

\noindent The right hand side contains terms which are bilinear in the
effective interaction strength.  The initial condition for Eq.  (2.29),
or (2.30), is given at $\lambda = \infty$:  ${\cal G}_{\lambda=\infty} =
{\cal G}_\epsilon$.

With accuracy to the first order in powers of the interaction strength,
${\cal G}_\lambda$ is independent of $\lambda$ and ${\cal G}_\epsilon$
is equal to the initial regularized hamiltonian expression one intends
to study, denoted by $H^{(0)}_\epsilon$.  In this initial approximation,
${\cal H}^{(0)}_\lambda = f^{(0)}_\lambda {\cal G}^{(0)}_{\lambda}$,
where ${\cal G}^{(0)}_\lambda = H^{(0)}_\epsilon$ and $f^{(0)}_\lambda$
is the similarity factor calculated using eigenvalues of ${\cal
G}^{(0)}_{1 \lambda}$.  ${\cal H}^{(0)}_\lambda$ forms our first
approximation to the similarity renormalization group trajectory of
operators ${\cal G}_\lambda$ parametrized by $\lambda$.

Eq.  (2.30) can then be written in the iterative form for successive
approximations to the trajectory ${\cal G}_\lambda$.  Namely,

$$ {d \over d\lambda} {\cal G}^{(n+1)}_\lambda =
   T^{(n)}_\lambda[{\cal G}^{(n)}_\lambda]
 .  \eqno (2.31)  $$

\noindent This is an abbreviated notation for

$$ {d \over d\lambda} {\cal G}^{(n+1)}_\lambda = \left[ f^{(n)}_\lambda
{\cal G}^{(n)} _{2\lambda}, \left\{ {d \over d\lambda} (1-f^{(n)}_\lambda)
{\cal G}^{(n)} _{2\lambda} \right\}_{{\cal G}^{(n)}_{1\lambda}} \right]
. \eqno (2.32) $$

\noindent $f^{(n)}_\lambda$ denotes a function of $z^{(n)}_\lambda$
expressed through eigenvalues of ${\cal G}^{(n)}_{1\lambda}$, such as in
Eqs.  (2.19) and (2.20).  The initial condition is set by ${\cal
G}^{(n+1)}_\infty = {\cal G}^{(n+1)}_\epsilon$.  Thus, the solution is

$$ {\cal G}^{(n+1)}_\lambda = {\cal G}^{(n+1)}_\epsilon -
\int_\lambda^\infty T^{(n)}_s[{\cal G}^{(n)}_s] . \eqno (2.33) $$

${\cal G}_\infty$ contains the $\epsilon$-regulated canonical
hamiltonian terms and counterterms.  The counterterms remove the part of
the integral in Eq.  (2.33) which diverges for finite $\lambda$ when
$\epsilon \rightarrow 0$.  Matrix elements of the hamiltonian of the
effective theory are required to have a limit when $\epsilon$ is made
very small.  The condition that the necessary ${\cal G}_\infty$ exists
is the hamiltonian version of renormalizability.  It does not require
the number of counterterms to be finite, although a finite number has
the clear advantage of simplicity.

The part of the integrand in Eq.  (2.33) which leads to the divergence
is denoted by $\left[ T^{(n)}_s[{\cal G}^{(n)}_s] \right]_{div}$, and
the remaining part by $\left[ T^{(n)}_s[{\cal G}^{(n)}_s]
\right]_{conv}$.  ${\cal G} ^{(n+1)}_\epsilon$ contains the initial
regulated hamiltonian terms and counterterms.  The counterterms in
${\cal G}^{(n+1)}_\epsilon$ are discovered from inspection of
$F^{(n+1)}_\lambda [G^{(n+1)}_{\lambda}] $ dependence on $\epsilon$ when
$\epsilon \rightarrow 0$ in the absence of counterterms.

Note that $F^{(n+1)}_\lambda [G^{(n+1)}_{1 \lambda}] = G^{(n+1)}_{1
\lambda} $ and it is not necessary to know $F^{(n+1)}_\lambda$ to
calculate $G^{(n+1)}_{1 \lambda}$.  One calculates $F^{(n+1)}_\lambda$
after $G^{(n+1)}_{1 \lambda}$ is made independent of $\epsilon$ when
$\epsilon \rightarrow 0$.

The diverging dependence on $\epsilon$ when $\epsilon \rightarrow 0$, is
typically of the form $\epsilon^{-1}$ or $\log{\epsilon}$ times operator
coefficients.  The operator coefficients can be found by integrating the
diverging part of the integrand from some arbitrary finite value of
$\lambda$, say $\lambda_0$, to infinity.  The divergence originates from
the upper limit of the integration and it is independent of $\lambda_0$.
The remaining finite part of the integral is sensitive to the lower
limit of integration and depends on $\lambda_0$.  The counterterm does
not depend on $\lambda_0$ but it contains an arbitrary finite part which
emerges in the following way.

The counterterm subtracts the diverging part of the integral.  But
subtracting terms with diverging functions of $\epsilon$ times known
operators does not tell us what finite parts times the same operators to
keep.  Thus, one needs to add arbitrary finite parts to the numbers
$1/\epsilon$ and $\log{\epsilon}$ in the counterterms.  These finite
parts are unknown theoretically and have to be fitted to data.  In
particular, observed symmetries may impose powerful constraints on the
finite parts.

The diverging part of the integrand is such that the lower limit of its
integration produces the same operator structure as the upper limit but
the diverging numbers such as $\epsilon^{-1}$ or $\log{\epsilon}$ from
the upper limit are replaced by finite numbers at the lower limit.
Those finite numbers depend on $\lambda_0$ but, once they are replaced
by the required unknown finite parts, one obtains a valid expression for
the counterterm.  The replacement is achieved by adding to the integral
the same operators multiplied by the numbers which are equal to the
differences between the unknown numbers and the numbers resulting from
the lower limit of the integration.  Thus, the unknown numbers we need
to add to the integral of the diverging part of the integrand from
$\lambda_0$ to infinity depend on $\lambda_0$.  One can write the
$\lambda_0$-independent ${\cal G}^{(n+1)}_{\epsilon}$ as ${\cal
G}^{(n+1)}_{\epsilon\,{finite}}(\lambda_0) + \int_{\lambda_0}^\infty
\left[ T^{(n)}_s[{\cal G}^{(n)}_s] \right]_{div}$.  The free finite
parts of the counterterms are contained in ${\cal
G}^{(n+1)}_{\epsilon\,{finite}}(\lambda_0)$ and one can fit them to data
using predictions obtained from effective hamiltonians at some
convenient scales $\lambda$.

More than one scale $\lambda$ may become necessary for accurate
determination of the free parameters when their values have to be of
considerably different orders of magnitude and require knowledge of
physical phenomena at different scales.  In the present work a single
scale $\lambda = \lambda_0$ is sufficient for practical calculations.
The renormalization conditions are set using $H_{\lambda_0}$.  One may
also consider renormalization conditions for parameters in
$H_{\lambda_0}$ which are set using another effective hamiltonian at
some nearby scale $\lambda_1 \neq \lambda_0$.  This will be illustrated
in the next Section.

The complete recursion including construction of counterterms in Eq.
(2.33) is given by

$$ {\cal G}^{(n+1)}_\lambda = {\cal G}^{(n+1)}_{\epsilon\,
finite}(\lambda_0) + \int_{\lambda_0}^\lambda ds
\left[ T^{(n)}_s[{\cal G}^{(n)}_s] \right]_{div}
- \int_\lambda^\infty ds
\left[ T^{(n)}_s[{\cal G}^{(n)}_s] \right]_{conv}
 . \eqno (2.34) $$

\noindent In the limit $n \rightarrow \infty$, if the limit exists, one
obtains

$$ {\cal G}_\lambda = {\cal G}_{\epsilon\,
finite}(\lambda_0) + \int_{\lambda_0}^\lambda ds
\left[ T_s[{\cal G}_s] \right]_{div}
- \int_\lambda^\infty ds
\left[ T_s[{\cal G}_s] \right]_{conv}
   . \eqno (2.35) $$

\noindent $H_\lambda$ is obtained from Eq.  (2.35) through the
replacement of $q_\infty$ by $q_\lambda$ (to obtain $G_\lambda$) and
action of $F_\lambda$ on $ G_\lambda$.

Perturbative calculations of renormalized effective hamiltonians are
based on the observation that the rate of change of ${\cal G}_\lambda$
with $\lambda$ can be expanded in a power series in the effective
interaction ${\cal G}_{2\lambda}$ at the same running scale $\lambda$.
This is obtained by repeated application of Eq.  (2.29).  One rewrites
Eq.  (2.29) as

$$ {d {\cal G}_\lambda \over d\lambda} = \left[ f_\lambda {\cal
G}_{2\lambda}, \left\{ -f'_\lambda {\cal G}_{2\lambda} \right\}_{{\cal
G}_{1\lambda}} \right] + \left[ f_\lambda {\cal G}_{2\lambda}, \left\{
(1-f_\lambda) {d {\cal G}_{\lambda} \over d\lambda}  \right\}_{{\cal
G}_{1\lambda}} \right] . \eqno (2.36) $$

\noindent Then, one replaces ${\cal G}'_\lambda $ in the last term on
the right-hand side of Eq.  (2.36) by the preceding terms.  Two
successive substitutions produce an expression for $ {\cal G}'_\lambda $
with four explicit powers of the effective interactions and the
remaining terms are of higher order [note that $ (1 - f_\lambda) {\cal
G}'_{1 \lambda} = 0 $].

$$     {d {\cal G}_\lambda \over d\lambda} =
\left[ f{\cal G}, \left\{ -f'{\cal G}\right\} \right]
+ \left[ f{\cal G}, \left\{ (1-f)
\left[ f{\cal G}, \left\{ -f'{\cal G}\right\} \right]
\right\} \right] $$
$$+ \left[ f{\cal G}, \left\{ (1-f)  \left[ f{\cal G}, \left\{ (1-f)
\left[ f{\cal G}, \left\{ -f'{\cal G}\right\} \right] \right\}
\right] \right\} \right] + o({\cal G}^5)  . \eqno (2.37) $$

\noindent We have omitted subscripts 2, $\lambda$ and ${\cal
G}_{1\lambda}$ on the right-hand side.  All the subscripts appear in the
same pattern as in Eq.  (2.36).  Correspondingly, the infinite chain of
substitutions produces an expression ordered by explicit powers of the
effective interactions, to infinity.

$$ {d \over d\lambda} {\cal G}_\lambda =
   \sum_{n=0}^\infty \left[ f{\cal G},
   ( \left\{(1-f)\left[ f{\cal G}, )^{(n)}
   \{-f'{\cal G}\} ( \right] \right\} )^{(n)} \right] .
   \eqno (2.38) $$

\noindent The round bracket raised to the $n$-th power means $n$
consecutive repetitions of the symbols from within the round bracket.
The subscripts are omitted for clarity as in Eq.  (2.37).

The above expansion in powers of the effective interactions provides a
systematic order by order algorithm for building an expression for the
effective hamiltonian.  The energy denominators and functions
$f_\lambda$ are calculated using eigenvalues of ${\cal G}_{1\lambda}$.
Therefore, in Eqs.  (2.37) and (2.38), the derivatives of the function
$f_\lambda$ contain two kinds of contributions:  those resulting from
differentiating the explicit $\lambda$ dependence in the arguments
$z_\lambda$ (for example, $\lambda^2$ in Eq.  (2.18) ), and those
resulting from differentiating the free energy eigenvalues (for example,
$\Delta {\cal M}^2$ in Eq.  (2.18)).  Since the free energy terms
include effective masses which depend on the width $\lambda$, the
derivatives of the effective masses appear in the equations on the
right-hand side.  Moving them to the left-hand side leads to coupled
nonlinear differential equations for the effective hamiltonians.

The general iterative approach in Eq.  (2.34) or the expansion in Eq.
(2.38), can be analysed using expansion in the running coupling
constants.  One can divide ${\cal G}_{1\lambda}$ into two parts:  one
which is independent of the coupling constants and another one which
vanishes when the coupling constants are put equal to zero.  The parts
depending on the coupling constants are moved over to ${\cal
G}_{2\lambda}$ and treated as an interaction.  After ${\cal
G}_{1\lambda}$ is reduced to the part which is independent of the
interactions, the derivatives of $f_\lambda$ in Eqs.  (2.36) to (2.38)
do not introduce additional powers of the interaction strength and the
series is strictly ordered in powers of the interactions according to
their explicit appearance in the formula (2.38).  This series then
provides the perturbative expansion in terms of the running coupling
constants.

The simplest case of the perturbative expansion involves a single
coupling constant at a single scale.  Firstly, one expands the
renormalization group equations into a series of terms ordered by powers
of the bare coupling $g_0$.  Secondly, one evaluates the effective
coupling $g_1$ at the chosen scale $\lambda_1$ as a power series in the
bare coupling.  Thirdly, the latter series is inverted and the bare
coupling is expressed as a series in the effective coupling $g_1$.
Then, one can pursue perturbative calculations in terms of the effective
coupling.  In particular, one can reduce the hamiltonian width to
$\lambda_2 < \lambda_1$ and calculate $g_2$ as a series in $g_1$.  Such
steps can be repeated.  For example, one can reduce the width in each
step by a factor 2. \cite{W1} \cite{W2} $N$ steps will reduce the width
by the factor $2^{-N}$.  This way one can build the renormalization
group flow indicated by the chains of small arrows in the diagram
discussed in Section 1. If many coupling constants appear but they can
be reduced to functions of a finite set of independent running coupling
constants the finite set determines the theory and one speaks of
coupling coherence.  \cite{PWcc}

\vskip.3in
{\bf 2.b Regularization}
\vskip.1in

A canonical bare hamiltonian obtained from a local field theory is
divergent.  This Section describes how the ultraviolet singularities in
the canonical hamiltonian are regularized with the bare cutoff
$\epsilon$.  We also introduce infrared regularization.  Our
presentation is ordered as follows.  First, we briefly explain
connection between the ultraviolet and infrared regularizations in
light-front dynamics.  Then, we proceed with definitions of the
canonical hamiltonian terms.  For that purpose, we have to discuss the
fundamental set of scales in the hamiltonian approach and explain the
role of lagrangian densities for classical fields in the construction.
Then, we describe details of the ultraviolet and, subsequently, infrared
regularizations.

The ultraviolet and infrared regularizations are connected through
masses.  The infrared structure is influenced by the masses in the
initial hamiltonian $H^{(0)}_\epsilon$.  $H^{(0)}_\epsilon$ carries the
superscript 0 to indicate that it is the initial hamiltonian which does
not yet include counterterms.  If the ultraviolet counterterms change
the masses the infrared behavior is changed too.

An initial mass value is generically denoted by $m^{(0)}_\epsilon$.  A
light-front energy, $p^-_m$, of a free particle with a four-momentum
$p_m = \left(p^+, p^\perp, p^-_m = (p^{\perp 2} + m^2 )/p^+ \right)$,
tends to infinity when $m^2 > 0$ and $p^+$ tends to zero.  But $p^-_m$
may be finite or even approach zero in this limit if $m^2 = 0$ and
$p^\perp$ approaches zero too.  The limit of small momentum $p^+$ is
always a high-energy limit when $m^2 > 0$.  But it ceases to be the
high-energy limit for the small transverse momenta if $m^2 \rightarrow
0$.  Thus, $m^{(0)}_\epsilon$ in the initial hamiltonian is capable of
switching from the high-energy regime in the longitudinal direction to
the low-energy one when we take the limit $m^{(0)}_\epsilon \rightarrow
0$.  Conversely, introducing masses turns the infrared low-energy regime
into the high-energy regime.

We begin the construction of $H^{(0)}_\epsilon$ with an enumeration of
momentum scales.  We distinguish scales related to the boundary
conditions for fields at spatial infinity, small momentum cutoffs,
phenomenological parameters and large momentum cutoffs.

The bare hamiltonian $H^{(0)}_\epsilon$ is defined in terms of the
operators $q_\infty$.  The bare quantum fields are built from these
operators with plane-wave coefficients.  \cite{WE1} The initial basis in
the Fock space is built from the vacuum state $|0\rangle$ using
$q^\dagger_\infty$.  Fermion, anti-fermion and boson creation and
annihilation operators are denoted by $b^\dagger$, $d^\dagger$,
$a^\dagger$, $b$, $d$ and $a$, respectively.  For example, $|k \sigma> =
b^\dagger_{k\sigma} |0\rangle$ denotes a state of one bare fermion of
momentum $k=(k^+, k^\perp)$.  Spin $z$-axis projection, flavor, color or
other quantum numbers, are denoted by a common symbol $\sigma$.  The
momentum variable in the subscript is distinguished in order to describe
the scales involved in the definition of the hamiltonian.  The order of
scales in momentum space is reverse to the order of scales in position
space.

The largest scale in the position space is the quantization volume.  In
other words, the momenta can be thought of as discrete when necessary.
But we insure by our choices of scales that the granulation in momentum
space is never noticeable and the quantization volume is effectively
infinite for all our purposes.  Thus, we universally adopt continuous
notation for momentum variables.

Potentially related to the boundary conditions, hypotheses about
zero-modes and spontaneously broken symmetry in light-front quantum
field theories were recently discussed in Ref.  \cite{W3} which quotes
important earlier literature on the subject.  Basically, one may expect
that new terms emerge in the effective hamiltonians and the new terms
account for the large scale dynamical effects.  Susskind and co-workers
have proposed a way to think about the wee parton dynamics in a model.
\cite{SUS} Ref.  \cite{GSR} describes QCD sum rules technique using the
notion of vacuum condensates in the light-front scheme.  However, the
original quantum dynamics of the vacuum formation and spontaneous
symmetry breaking are not yet understood and cannot be discussed further
here.

The next smaller size in position space is the inverse of the infrared
regulator.  Two types of the infrared regulator appear.  The first is a
lower bound, denoted by $\delta$, on the parent $+$-momentum fraction
that can be carried by an operator in an interaction term.  The second
is a mass parameter $\mu_\delta$.  $\mu_\delta$ appears as the mass
parameter $m^{(0)}_\epsilon$ in the initial hamiltonian
$H^{(0)}_\epsilon$.  It is introduced for massless bare particles.
$\mu_\delta$ cuts off the small longitudinal momentum region at a small
scale order $\mu^2_\delta/ \Delta^2$, where $\Delta^2$ is the invariant
mass upper bound.  When the infrared regularization is being removed,
$\delta$ or $\mu_\delta$ are sent to zero but their inverse is always
kept negligible in comparison with the quantization volume scale.

The next smaller scale in position space is set by the size of the
volume used for preparation of incoming and detection of outgoing
particles (including bound states) and the corresponding time scale.
Physics is contained within this scale and observables are allowed to
depend on this scale since the preparation and detection of states is a
part of a physical process.

The order of magnitude of momenta larger than the experimental wave
packet widths are characterized in terms of three different scales, (1)
masses of particles, (2) the width of the effective hamiltonian (i.e.
$\lambda$), and (3) the bare cutoff scale $\epsilon^{-1}$.  When solving
for the hamiltonian spectrum, a new scale may emerge dynamically,
determined by the effective coupling constants, masses and width of the
effective hamiltonian.  Scale invariance at large momenta may be
violated through a dimensional transmutation even if all mass scales are
negligible in comparison to the momenta and $\lambda$.

The width $\lambda$ ranges from $\infty$ to convenient finite values.
Description of physical phenomena involving energy-momentum transfers of
the order of $k$ require $\lambda$ to be larger than $k$.  It is also
useful to use $\lambda$ not too large in comparison to $k$ in order to
avoid too much detail in the dynamics.  For example, useful values of
$\lambda$ in nonrelativistic systems are smaller than effective masses.
In QED, the convenient $\lambda$ in the hydrogen description is much
larger than the binding energy and much smaller than the electron mass.

The bare cutoff scale $\epsilon^{-1}$ is the largest momentum scale in
the theory.  The formal limit $\lambda \rightarrow \infty$ is used only
to remove $\lambda$ dependence from the hamiltonian regulated by
$\epsilon$.  In other words, no $\lambda$ dependence appears in the
hamiltonians with $\lambda$ larger than the scale implied by
$\epsilon^{-1}$.  No physical quantity depends on $\epsilon$ when
$\epsilon \rightarrow 0$.  The similarity renormalization scheme for
hamiltonians is built to achieve this goal to all orders in perturbation
theory (cf. \cite {GW1} \cite{GW2}).

We proceed to the explicit construction of simplest terms in the
hamiltonian $H^{(0)}_\epsilon$.  Details of counterterms are not known
from the outset.  Light-front power counting rules are helpful \cite{W3}
in determining possible structures of the counterterms but more is
required in practice.  The similarity renormalization group provides the
required details.

All starting hamiltonians in quantum field theories contain a free part
which we denote by ${\cal G}^{(0)}_1$.  The free part for fermions and
bosons has the form

$$ {\cal G}^{(0)}_1 = \sum_\sigma \int [k] \left[ {k^{\perp 2} +
m^{(0)\, 2}_\epsilon \over k^+} ( b^\dagger_{k\sigma} b_{k\sigma} +
d^\dagger_{k\sigma} d_{k\sigma}) + {k^{\perp 2} + \mu^{(0)\,2}_\epsilon
\over k^+} a^\dagger_{k\sigma} a_{k\sigma} \right] . \eqno (2.39) $$

\noindent We adopt the following conventions.  Summation over $\sigma$
denotes a sum over all quantum numbers except the momentum.

$$ \int [k] = {1 \over 16 \pi^3} \int_0^\infty {dk^+ \over k^+}
\int d^2 k^\perp . \eqno (2.40) $$

\noindent The creation and annihilation operators in Eq.  (2.39) are the
bare ones denoted in Section 2.a by $q_\infty$.  They satisfy standard
commutation or anti-commutation relations

$$ \left[ a_{k\sigma}, a^\dagger_{k'\sigma'} \right] =
   \left\{ b_{k\sigma}, b^\dagger_{k'\sigma'} \right\} =
   \left\{ d_{k\sigma}, d^\dagger_{k'\sigma'} \right\} =
   16 \pi^3 k^+ \delta^3 (k-k') \delta_{\sigma \sigma'}   \eqno (2.41)$$

\noindent with all other commutators or anti-commutators equal zero as
dictated by the spin and statistics assignments of Yukawa theory, QED or
QCD.

The initial mass parameters $m^{(0)}_\epsilon$ and $\mu^{(0)}_\epsilon$
do not include effects of any interactions and are independent of the
interaction strength.  We may have to consider limits where the mass
parameters are close to zero, in comparison to all other quantities of
relevance to physics.  For example, $\mu^{(0)}_\epsilon$ may be the
infrared regulator mass denoted by $\mu_\delta$.  Recall that the
subscript $\epsilon$ indicates that the mass parameters stand in the
hamiltonian with $\lambda = \infty$.

The initial hamiltonian contains an interaction part, ${\cal G}^{(0)}_2
= H^{(0)}_\epsilon - {\cal G}^{(0)}_1$.  For example, electrons may emit
photons.  One writes the corresponding interaction term in QED as

$$ \sum_{\sigma_1 \sigma_2 \sigma'_1} \int [k_1][k_2][k'_1] 16\pi^3
\delta^3(k_1+k_2-k'_1) {\bar u}_{m^{(0)}_\epsilon k_1 \sigma_1} e
\not\!\varepsilon^*_{k_2 \sigma_2} u_{m^{(0)}_\epsilon k'_1 \sigma'_1}
b^\dagger_{k_1 \sigma_1} a^\dagger_{k_2 \sigma_2} b_{k'_1 \sigma'_1} .
\eqno (2.42) $$

\noindent We use conventions to be specified shortly.  The hamiltonian
term (2.42) is contained in the expression

$$ h = \int dx^- \,d^2x^\perp \,\, \left[e \bar \psi_{m ^{(0)}
_\epsilon}(x) \not\!\!  A (x) \psi_{m^{(0)}_\epsilon}(x) \right]_{x^+=0}
\eqno (2.43) $$

\noindent where the fields $\psi_{m ^{(0)} _\epsilon}(x)$ and $A^\nu
(x)$ for $x^+=0$ are defined by writing

$$ \psi_{m ^{(0)} _\epsilon}(x) = \sum_{\sigma} \int [k] \left[
u_{m^{(0)}_\epsilon k \sigma} b_{k \sigma} e^{-ikx} + v_{m ^{(0)}
_\epsilon k \sigma} d^\dagger_{k \sigma} e^{ikx} \right] \eqno (2.44) $$

\noindent and

$$ A^\nu (x) = \sum_{\sigma} \int [k] \left[ \varepsilon^\nu_{k \sigma}
a_{k \sigma} e^{-ikx} + \varepsilon^{\nu *}_{ k \sigma} a^\dagger_{k
\sigma} e^{ikx} \right] . \eqno (2.45) $$

\noindent Spinors $u_{m k \sigma}$ and $v_{m k \sigma}$ are defined by
boosting spinors for fermions at rest, $u_{m \sigma}$ and $v_{m
\sigma}$, to the momentum $k$, as if the fermion mass were $m$.  This is
done using the light-front kinematical boost representation for fermions

$$ S(m,k) = (mk^+)^{-1/2} [\Lambda_+ k^+ + \Lambda_- (m + \alpha^\perp
k^\perp)]. \eqno (2.46) $$

\noindent Namely, $u_{m k \sigma} = S(m,k) u_{m \sigma}$ and $v_{m k
\sigma} = S(m,k) v_{m \sigma}$.  Solving constraint equations for the
free fermion fields in canonical field theory amounts to using these
spinors.  The same boost operation defines the polarization vectors for
photons which are independent of the photon mass.  We have
$\varepsilon_{k \sigma} = \left( \varepsilon^+_{k \sigma} = 0,
\varepsilon^-_{k \sigma} = 2 k^\perp \varepsilon^\perp_\sigma/k^+ ,
\varepsilon^\perp_{k \sigma} = \varepsilon^\perp_\sigma \right)$.  The
spin label $\sigma$ denotes the spin projection on the z-axis.  We adopt
a number of conventions from Ref.  \cite{BL}.  It is useful to work with
the above spinors and polarization vectors because they provide insight
into the physical interpretation of the calculated matrix elements.  For
example, the spinors and polarization vectors help in tracing
cancelations which result from the current conservation (e.g. see Eq.
(3.103) etc. in the next Section).

Equation (2.43) includes 5 terms in addition to (2.42).  The other 5
terms lead to emission of photons by positrons, absorption of photons by
electrons or positrons, or to transitions between electron-positron
pairs and photons.  There is no term leading to creation of an
electron-positron pair and a photon, or to annihilation of such three
particles.  This is the distinguished property of the light-front
hamiltonians:  conservation of momentum $k^+ > 0$ excludes a possibility
that the three momenta sum up to zero.

Strictly speaking, one has to limit each $k^+$ from below by a nonzero
positive lower bound in order to make sure that the three $+$-momentum
components cannot add up to zero.  This lower bound is provided by the
inverse of the quantization volume.  Our cutoffs and scale hierarchy
ensure that this largest of spatial scales in the theory does not need
to be invoked in the description of physical phenomena.  The
regularization procedure cuts off such small momenta long before they
have a chance to become relevant.  If high-order perturbation theory
subsequently leads to effective hamiltonians which describe universal
low momentum components in all physical states the notion of a
nontrivial vacuum has to be taken seriously into account for practical
computational reasons.  A priori, we cannot exclude this will happen.
But we postpone considerations of such a situation until it becomes
necessary in the future work.

The product $\bar \psi \not\!\!  A \psi$ denotes a sum of 6 basic
interactions.  The products of creation and annihilation operators are
ordered as indicated at the beginning of this Section.  However, Eq.
(2.43) requires additional steps before one can assign it a well defined
meaning because operators such as (2.42) can easily produce states of
infinite norm.  One needs to define the individual terms such as (2.42)
in order to provide meaning to the whole combination of similar terms in
Eq.  (2.43)

There are inverse powers of $k^+$ in Eq.  (2.42) and $k^+$ may be
arbitrarily close to 0. For example, when $k_3^+$ and $k_1^+$ in (2.42)
are similar (and they are allowed to be arbitrarily close to each other
no matter what their own size is), the photon momentum $k_2^+=k_3^+ -
k_1^+$ is arbitrarily close to zero.  The problem is that the photon
momentum appears in the photon polarization vector in the denominator:
$\epsilon^-_{k_2 \sigma_2} = 2 k_2^\perp
\epsilon^\perp_{\sigma_2}/k_2^+$.  Unless $k_2^\perp$ is close to zero
the resulting emission strength approaches $\infty$ for $k_2^+
\rightarrow 0$.  Therefore, even for a very small coupling constant $e$,
the interaction can be arbitrarily strong.  This divergence is canceled
in special circumstances.  For example, in the tree diagrams for the
S-matrix elements in QED, the cancelation is a consequence of the
presence of more terms in the hamiltonian and the energy and charge
current conservation in physical processes.  However, for the
off-energy-shell matrix elements of the $T$-matrix, in loop diagrams, or
in bound state equations, such cancelations will not be ensured
automatically and could lead to ill-defined expressions.

In particular, one has to keep in mind that in the perturbative
calculation of the S-matrix it is possible to apply energy and momentum
conservation laws for incoming and outgoing particles on their
mass-shells.  In contrast, in the bound state calculations, the
individual particle momenta cannot simultaneously be on the individual
mass-shells and still sum up to the bound state momentum - the bound
state dynamics is always off-shell and the on-shell perturbative
mechanisms for cancelations cease to be sufficient.

In Eq.  (2.42), the inverse powers of the longitudinal momentum also
appear in the fermion spinors.  These can be a source of divergences
too.  However, the examples we describe in this article do not lead to
problems with infrared fermion divergences and we do not dwell on this
subject here.

The spinor matrix elements depend on the transverse momenta of the
fermions and the boson polarization vector depends on the transverse
momentum of the boson.  The strength of the interaction grows when the
relative transverse momenta grow and leads to divergences.  The
divergence problem manifests itself clearly when one attempts to
evaluate the ratio of norms of the states $h |k\sigma\rangle$ and
$|k\sigma\rangle$.  This ratio is certainly not finite and it is not
well defined.

One might ask if it is useful to consider the ill-defined hamiltonian
term (2.43).  The answer is unambiguous yes because scattering
amplitudes calculated using this term in combination with two other
terms in second order perturbation theory agree very well with
observable scattering of electrons and photons.  No loop integration
appears in these calculations to indicate the divergence problem.

It is well known that the terms one should put into the light-front
hamiltonian are provided by the formal lagrangian density for
electrodynamics ${\cal L} = -{1 \over 4} F^{\mu \nu} F_{\mu \nu} + \bar
\psi (i\not\!\!  D - m )\psi$.  One can rewrite the lagrangian density
into a corresponding light-front hamiltonian density by using an
expression for the energy-momentum tensor density $T^{\mu\nu}$.
Integrating $T^{+-}$ over the light-front hyper-plane gives the
expression one starts from in building the light-front hamiltonian for
QED.

The initial hamiltonian $H^{(0)}_\epsilon$ for QED results from formal
operations on fields $\psi_+$ and $A^\perp$.  \cite{Y} One uses the
gauge $A^+=0$ and solves the constraint equations, substitutes
expansions of the form (2.44) and (2.45) into the formal expression for
$T^{+-}$, integrates the density over the light-front hyper-plane and
normal-orders all terms.  The normal-ordering produces terms that
involve numerically divergent momentum integrals.  The classical field
theory does not tell us what to do with the divergences resulting from
the ordering of operators.

To deal with the divergences one has to regularize the hamiltonian
theory from the outset.  The naive connection between the classical
theory and the quantum theory as given by the quantization rules is
broken by the regularization.  The regularization turns out to force new
terms in the hamiltonian.  To gain control on the regularization effects
one has to construct a renormalization theory for hamiltonians.  The
diverging terms which result from normal ordering can be safely dropped
in the form they appear ill-defined in the canonical approach because
the renormalization procedure introduces other terms of the same
operator structure to replace them.

The regularization for light-front hamiltonians which we apply to
expressions resulting from field theoretic lagrangian densities is first
described for the term (2.42).  In that term, the parent momentum $P$
equals $k'_1$.  The spinors and polarization vectors conveniently group
a number of terms with different momentum dependences into a combination
which is invariant under light-front kinematical symmetry
transformations.  Among those terms there are terms containing masses,
terms which in field theory result from derivatives $i\partial^\perp$ or
$i\partial^+$ or from inverting the operator $i \partial^+$.  All those
derivatives are replaced in the term (2.42) by momenta of particles
created or destroyed by that term.

We first introduce the daughter momentum variables for the created
electron and photon.  We have introduced daughter momenta in a similar
configuration in Section 2.a while defining the similarity functions
$f_\lambda$.  Here, we use the daughter momentum variables for the
purpose of regularization.  The variables are

$$  x_1 = k_1^+/{k'}^+_1 = x , \eqno (2.47.a)  $$

$$  x_2 = k_2^+/{k'}^+_1 = 1-x , \eqno (2.47.b)  $$

$$  x'_1 = {k'}^+_1 / {k'}^+_1 = 1, \eqno (2.47.c) $$

$$ \kappa^\perp_1 = k_1^\perp - x_1 P^\perp = \kappa^\perp, \eqno
(2.47.d) $$

$$ \kappa^\perp_2 = k_2^\perp - x_2 P^\perp = - \kappa^\perp , \eqno
(2.47.e) $$

$$ {\kappa'}^{\perp}_1 = {k'}_1^\perp - x'_1 P^\perp = 0 . \eqno
(2.47.f) $$

\noindent For each creation and annihilation operator in the interaction
term (2.42) we define a {\it daughter energy} variable.  Namely,

$$ e_1 = { \kappa_1^{\perp 2} + m^{(0) \,2}_\epsilon \over x_1} =
         {\kappa^{\perp 2} + m^{(0)\,2}
         _\epsilon \over x}, \eqno (2.48.a) $$

$$ e_2 = { \kappa_2^{\perp 2} + \mu^{(0)\,2}_\epsilon \over x_2} =
         {\kappa^{\perp 2} + \mu^{(0)\,2}
         _\epsilon \over 1-x}, \eqno (2.48.b) $$

$$ e'_1 = { {\kappa'}_1^{\perp 2} + m^{(0)\,2}_\epsilon \over x'_1} =
          m^{(0)\,2}_\epsilon . \eqno (2.48.c) $$

\noindent For each creation and annihilation operator in the interaction
term (2.42) we introduce a factor which is a function, $r(y_i)$, of the
variable $y_i = \epsilon e_i/\Lambda^2$, where the subscript $i$ denotes
the operator in question.  In the no cutoff limit, $\epsilon \rightarrow
0$.  $\Lambda$ is an arbitrary constant with dimension of a mass ($\hbar
= c = 1$).  All masses and momenta are measured in units of $\Lambda$.
In this article, we choose $r(y) = (1+y)^{-1}$.  Thus, the term (2.42)
is regulated by the factor

$$ (1+ \epsilon e_1/\Lambda^2)^{-1} (1+ \epsilon e_2/\Lambda^2)^{-1} (1+
\epsilon e_3/\Lambda^2)^{-1} \eqno (2.49) $$

\noindent under the integral.  The third factor in the above expression
can be replaced by 1, since $m^{(0)}_\epsilon$ is a finite constant and
it cannot compensate the smallness of $\epsilon$.  We shall make such
replacements wherever the parent momentum is carried by a single
creation or annihilation operator.

In the case of terms which contain only 1 creation and 1 annihilation
operator, i.e. in ${\cal G}^{(0)}_1$, no regularization is introduced.
Restrictions on the particle momenta in these terms would violate
kinematical symmetries of the light-front hamiltonian dynamics because
momenta in these terms are equal to the parent momenta and limiting the
parent momenta violates the light-front boost invariance.

In the initial expressions for hamiltonian densities of Yukawa theory,
QED or QCD, only terms with products of up to four fields appear.
Therefore, we have only two more situations to consider in addition to
the cases such as ${\cal G}^{(0)}_1$ and terms of the type (2.42).  In
the first situation we have three creation operators and one
annihilation operator or vice versa, and in the second situation we have
two creation and two annihilation operators. Both cases are regularized
using the same general rule.

Independently of the number of creation and annihilation operators in a
product, the regularization is introduced by multiplying every creation
and annihilation operator in the product by a function $r(y)$ such as in
the factor (2.49), where $y=\epsilon e_d /\Lambda^2$ and $e_d$ is the
corresponding daughter energy variable.  Later, after counterterms are
calculated, the same regularization factors are introduced in the
counterterms.

An additional step is required in the case of hamiltonian terms which
originate from the products of four fields including inverse powers of
$i\partial^+$ acting on a product of two fields.  We introduce two kinds
of a fifth daughter momentum and two corresponding daughter energy
variables, $e_{512}$ and $e_{534}$.  The numbering originates from
assigning numbers to the fields in the product according to the
schematic notation $\phi_1 \phi_2 (i\partial^+)^{-n} \phi_3 \phi_4$.
One of the fifth daughter energy variables is associated with the
operators coming from the fields number 1 and 2, and the other one is
associated with the operators coming from the fields number 3 and 4. The
regularized terms will contain an additional product of functions
$r(y_{512})$ and $r(y_{534})$ with the arguments $y_{512} = \epsilon
e_{512}/\Lambda^2$ and $y_{534} = \epsilon e_{534}/\Lambda^2$.

The auxiliary daughter energy variables $e_{512}$ and $e_{534}$ are
calculated as if they represented daughter energy variables for an
intermediate particle, a boson or a fermion, created and annihilated in
the vertices which contained the products $\phi_1 \phi_2$ and $\phi_3
\phi_4$, respectively.  Those vertices are treated as if each of them
contained three fields instead of two but the field of the intermediate
particle was contracted so that the corresponding creation operator and
the corresponding annihilation operator are absent in the resulting
term.  This particular definition of a gedanken intermediate particle
does not refer to any particular Fock state and remains valid when the
operators $q_\infty$ are replaced with $q_\lambda$ by the unitary
transformation $U_\lambda$.  The definition was inspired by Refs.
\cite{Y} and \cite{BRS} where the correspondence between the
intermediate states with backward moving particles with spin in the
infinite momentum frame and the light-front seagull interaction terms is
extensively described.

Mathematically, the definition of $e_{512}$ and $e_{534}$ is introduced
in the following way.  Every creation and annihilation operator in the
fields $\phi_1$, $\phi_2$, $\phi_3$ and $\phi_4$ is assigned a
corresponding number $s_i$, i=1,2,3,4.  $s_i$ equals $+1$ for a creation
operator and $s_i$ equals $-1$ for an annihilation operator.  We define
$k^+_5 = |s_3 k_3^+ + s_4 k_4^+|$ and $s_5 = (-s_3 k_3^+ - s_4
k_4^+)/k_5^+$.  The gedanken particle is thought to be created in the
product of fields including $\phi_3 \phi_4$ when $s_5 = 1$ and it is
thought to be annihilated in that product when $s_5 = -1$.  We define
the momentum $k_5= (k_5^+, k_5^\perp)$ by the relation $s_5 k_5 = - s_3
k_3 - s_4 k_4 = s_1 k_1 + s_2 k_2$.  We also introduce two auxiliary
parent momenta, $P_{34} = {1 \over 2} (k_5 + k_3 + k_4)$ and $P_{12} =
{1 \over 2} (k_5 + k_1 + k_2)$.  Then, we introduce the daughter
momentum and energy variables

$$ x_{512} = k^+_5 /P^+_{12}   , \eqno (2.50.a) $$

$$ \kappa^\perp_{512} = k^\perp_5 - x_{512} P^\perp_{12},
   \eqno (2.50.b) $$

$$ e_{512} = {\kappa^{\perp 2}_{512} + m^{(0)\,2}_{\epsilon 5} \over
x_{512} }, \eqno (2.50.c) $$

$$ x_{534} = k^+_5 /P^+_{34}    , \eqno (2.50.d) $$

$$ \kappa^\perp_{534} = k^\perp_5 - x_{534} P^\perp_{34} ,
   \eqno (2.50.e) $$

$$ e_{534} = {\kappa^{\perp 2}_{534} + m^{(0)\,2}_{\epsilon 5} \over
x_{534} } , \eqno (2.50.f) $$

\noindent where $m^{(0)}_{\epsilon 5}$ equals $m^{(0)}_\epsilon$ for
regularization of the terms involving $(i\partial^+)^{-1}$ and
$m^{(0)}_{\epsilon 5}$ equals $\mu^{(0)}_\epsilon$ for regularization of
the terms involving $(i\partial^+)^{-2}$.  This step completes our
definition of the ultraviolet regularization of initial hamiltonians.

We proceed to the definition of the infrared regularization.  Inverse
powers of $i\partial^+$ for massive particles are already regulated when
the ultraviolet regularization is imposed.  This was explained above.

For each creation and annihilation operator of an initially massless
particle we introduce a factor which limits the daughter momentum
fraction $x$ for that operator to be greater than $\delta$.  An example
of such a factor is given by $(1 +\delta/x)^{-1}$.  Note that our
definition also implies that the same regularization factor is inserted
for the gedanken particles with $x_{512}$ defined in Eq.  (2.50.a) and
$x_{534}$ defined in Eq.  (2.50.d).

Besides introducing the cutoff $\delta$ on the momentum fractions
carried by massless particles, we can also introduce for each initially
massless particle a finite regularization mass term which is denoted by
$m_\delta$.  In other words, in the case of the initially massless
particles, $m^{(0)}_\epsilon = m_\delta$.  Such finite masses in the
daughter energies lead to additional suppression of the infrared
longitudinal momentum region.  The additional mass terms are introduced
through mass counterterms which contain unknown finite parts.  Since the
finite parts are arbitrary and not known to be zero we introduce the
finite mass terms and investigate their role.

\vskip.3in
{\bf 2.c Renormalization conditions}
\vskip.1in

The free finite parts of counterterms are determined by renormalization
conditions which result from comparison of theoretical predictions with
data.  Calculations of observables require solutions to bound state or
scattering problems using renormalized hamiltonians.  In principle, one
could work with hamiltonians of any width $\lambda$.  In practice, one
is limited to consider some subspaces in the Fock space.  Therefore, the
issue of setting renormalization conditions is subtle.

In theories with small coupling constants and without confinement one
has an option of defining on-mass-shell renormalization conditions for
single particles and scattering states in perturbation theory.  It means
that one can determine free parameters in the effective hamiltonians by
demanding that single particle eigenstates of an effective hamiltonian
and the $S$-matrix calculated using this hamiltonian have the required
properties.  The key examples to be discussed in detail in the next
Section are Yukawa theory (pseudoscalar coupling) and QED.

In theories with confinement one has to choose mass parameters for
confined particles and these are not directly observable.  We suggest in
this case to use similar renormalization conditions in perturbative
calculations of effective hamiltonians as in QED.  Details are described
in Section 3.c.  Besides ultraviolet the perturbative self-interactions
of quarks and gluons diverge also in the infrared region where
intermediate states have similar energies to the outer states and the
effective dynamics is no longer perturbative.  Therefore, the
perturbatively renormalized mass terms introduce large infrared effects
in the effective dynamics where perturbative cancelation mechanisms are
no longer valid.  These large infrared effects are welcome as a source
of confining potentials suggested by Perry.  \cite{Perry}

The key question we have to answer in practice is how many effective
particles have to be taken into account to solve the effective
eigenvalue problem and how many can be included in a doable calculation.
A good example of a theoretical problem one can think of is how the
momentum or spin of a proton is shared by its constituents.  The
phenomenology of deep inelastic scattering of leptons and nucleons
suggests a considerable number of constituents even at moderate momentum
transfers.  If the number of effective constituents has to be large one
may encounter ambiguities in the determination of free parts of
counterterms because observables will be calculable only through
complicated procedures.  On the other hand, the constituent quark model
suggests that the large number of constituents is not needed to explain
main features of the spectrum of hadrons.  Therefore, one can expect
that many states are important in the large width hamiltonian dynamics
but only a few effective particles appear in the small width case.  The
renormalization conditions set through the small width dynamics will use
a small number of constituents but require non-perturbative solutions
for the spectrum.

${\cal G}_\lambda$ in Eq.  (2.35) contains the unknown finite parts of
counterterms in ${\cal G}_{\epsilon\,finite}(\lambda_0)$.  The
hamiltonian $H_{\lambda_0}$ can be used to calculate scattering
amplitudes and bound state properties.  The most familiar example of QED
is largely perturbative as far as renormalization is concerned in order
$\alpha$.  One can calculate the physical electron energy defined as the
lowest eigenvalue of the effective hamiltonian for the eigenstates with
electron quantum numbers.  Thanks to the symmetries of the light-front
dynamics the eigenvalue has the form $(p^{\perp 2} + m^2_e)/p^+$ and
$m_e$ has to be equal to the physical electron mass.  Note also that the
effective mass term for the interacting photons must be different from
zero (and growing with $\lambda$) in order to obtain massless photon
eigenstates.  Examples of the renormalization conditions for QED are
presented in the next Section.  The same procedure in QCD is expected to
lead to strong infrared effects because the non-abelian QCD interactions
prevent the same cancelation of infrared divergences as in QED.

Hamiltonian belongs to the algebra of Poincar\'e generators.  The
Poincar{\'e} algebra commutation relations can be studied order by order
in perturbation theory to find out constraints the algebra imposes on
the counterterms.  The general structure of the similarity
transformation for creation and annihilation operators allows extension
of the hamiltonian renormalization procedure to the whole algebra.  The
renormalization group evolution is given by the same Eq.  (2.5) for all
generators.  Renormalization of the Poincar\'e algebra is not further
analysed in this article.  \cite{Maslowski}

\vskip.3in
{\bf 3. EXAMPLES OF APPLICATION}
\vskip.1in

This Section describes a set of examples of lowest order calculations of
renormalized effective hamiltonians using the scheme from Section 2. We
begin by the description of generic rules for calculating the right-hand
side of the renormalization group equation (2.29).  The rules follow
from the commutator structure.  Then, we discuss examples from Yukawa
theory, QED and QCD.

\nopagebreak
\vskip.3in
{\bf 3.a Evaluation of commutators}
\vskip.1in
\nopagebreak

The right-hand sides of Eqs.  (2.29) and (2.38) are commutators.  This
implies that the interactions which appear in the hamiltonians
$H_\lambda(q_\lambda)$ and in the counterterms in $H_\epsilon$ are
connected.  This Section explains how this result comes about.

The commutators can be evaluated in a number of equivalent ways but some
of the ways are more convenient than others.  Suppose we are to evaluate

$$      {\hat H} = [ {\hat A}, \{ {\hat B} \}_{\hat C} ]   . \eqno (3.1)$$

\noindent
${\hat A} = A(X,Y) \prod_{i=1}^{I_A}
a^\dagger_{x_i}\prod_{j=1}^{J_A} a_{y_j}$,
${\hat B} = B(V, W) \prod_{k=1}^{I_B}
a^\dagger_{v_k}\prod_{l=1}^{J_B} a_{w_l}$
and ${\hat C} = \sum_z E(z) a^\dagger_z a_z$.
The right-hand side of Eq. (3.1) equals

$$ {\hat H} =
    A(X,Y)\prod_{i=1}^{I_A} a^\dagger_{x_i} \prod_{j=1}^{J_A} a_{y_j}
   {B(V,W)\over E_w - E_v}
          \prod_{k=1}^{I_B} a^\dagger_{v_k}\prod_{l=1}^{J_B} a_{w_l}$$
$$-{B(V,W)\over E_w - E_v}
          \prod_{k=1}^{I_B} a^\dagger_{v_k}\prod_{l=1}^{J_B} a_{w_l}
    A(X,Y)\prod_{i=1}^{I_A} a^\dagger_{x_i}\prod_{j=1}^{J_A} a_{y_j}
 , \eqno(3.2) $$

\noindent where $E_w = \sum_{l=1}^{J_B} E(w_l)$ and $E_v =
\sum_{k=1}^{I_B} E(v_k)$.  By commuting $\prod_{j=1}^{J_A} a_{y_j}$ in
the first term through $\prod_{k=1}^{I_B} a^\dagger_{v_k}$ one generates
the contracted terms with a number of contractions ranging from 1 to the
smaller of the numbers $J_A$ and $I_B$, and a term with
$\prod_{j=1}^{J_A} a_{y_j}$ standing to the right of $\prod_{k=1}^{I_B}
a^\dagger_{v_k}$.  Then, by commuting $\prod_{l=1}^{J_B} a_{w_l}$ in the
latter term through $\prod_{i=1}^{I_A} a^\dagger_{x_i}$, one obtains new
contracted terms with the number of contractions ranging from 1 to the
smaller of the numbers $I_A$ and $J_B$, and a term equal to the second
term in Eq.  (3.2) with an opposite sign which thereby is canceled out
leaving only connected terms in the result for $\hat H$.  This result
holds despite anti-commutation relations for fermions because
interactions contain even numbers of fermion operators.

After the second term in Eq.  (3.2) is canceled one is left with a
number of partially contracted terms in which annihilation operators may
still stand to the left of creation operators. A number of ordering
transpositions need to be done before a generic ordering of operators
adopted in the previous Section is achieved.  In fact, the process of
commuting factors in $\hat A$ through factors in $\{ {\hat B} \}_{\hat
C}$ in the first term on the right-hand side of Eq.  (3.2) produced
above a number of terms with creation operators moved to the right of
annihilation operators unnecessarily. These transpositions have to be
undone to recover final answers with the adopted ordering.
Nevertheless, it is visible that disconnected terms cannot appear and
the following rule simplifies the calculations.

The right-hand side of Eq.  (3.2) equals the sum of the contracted terms
which result from ${\hat A} \{ {\hat B} \}_{\hat C}$ by moving
$\prod_{j=1}^{J_A} a_{y_j}$ through $\prod_{k=1}^{I_B} a^\dagger_{v_k}$
and, the contracted terms which result from $-\{ {\hat B} \}_{\hat C}
{\hat A}$ by moving $\prod_{l=1}^{J_B} a_{w_l}$ through
$\prod_{i=1}^{I_A} a^\dagger_{x_i}$.  All other terms cancel out.

\vskip.3in
{\bf 3.b Yukawa theory}
\vskip.1in
\nopagebreak

The standard procedure from Ref.  \cite{Y} leads from the lagrangian
density ${\cal L}_Y= {\bar \psi} (i\not\!\partial - m - g\phi)\psi +
{1\over 2}(\partial^\mu \phi \partial_\mu \phi - \mu^2 \phi^2)$ to the
light-front hamiltonian expression of the form

$$ H_Y = \int dx^- d^2 x^\perp \left[ \bar \psi_m \gamma^+
{-\partial^{\perp 2} +
m^2 \over 2 i\partial^+} \psi_m +
{1\over 2}\phi (-\partial^{\perp 2} + \mu^2 )\phi \right.$$
$$ \left. + g \bar \psi_m \psi_m \phi + g^2\bar \psi_m \phi {\gamma^+
\over 2i\partial^+}\phi \psi_m \right]_{x^+=0}. \eqno(3.3) $$

\noindent We replace fields $\psi_m(x)$ and $\phi(x)$ for $x^+=0$ by the
Fourier superpositions of creation and annihilation operators, order the
operators in all terms and drop the terms containing divergent integrals
which result from the contractions.  Then, we introduce the
regularization factors.

In the course of calculating effective hamiltonians we will also add new
terms to $H_Y$ due to the presence of the regularization, in accord with
the renormalization theory from the previous Section.  For example, we
will add a small term $\delta m^2_\epsilon = m^2_\epsilon - m^2$ to
$m^2$ in the first term and $\delta \mu^2_\epsilon = \mu^2_\epsilon -
\mu^2 $ to $\mu^2$ in the second term.  We will calculate these terms
below using the renormalization theory to order $g^2$.

In order to consider particles with quantum numbers of nucleons and
pions one needs to include the isospin and replace the scalar coupling
by $i\gamma_5$.  \cite{BGP} However, for the purpose of the illustration
of the renormalization procedure to second order in the coupling $g$, we
do not have to introduce these explicitly.  The additional factors
merely lead to somewhat different algebra which can be traced throughout
the whole calculation and final results including isospin and
$i\gamma_5$ can be read from the results in the Yukawa theory.  In this
Section we assume $m > \mu > 0 $.

\vskip.3in
\centerline {\bf Meson mass squared}
\vskip.1in

The simplest example of a second order expression for a term in an
effective hamiltonian in the Yukawa theory is provided by the meson mass
squared.  We first describe steps which produce this expression.  The
number of distinct steps in the procedure is 10:  defining the
regularized initial hamiltonian, calculation of the effective
hamiltonian, analysis of the cutoff dependence of finite matrix elements
of the calculated terms and extraction of the structure of the
divergence, evaluation of the counterterm, isolation of the finite part,
calculation of the effective hamiltonian knowing the structure of the
counterterm, solving a physical problem such as an eigenvalue problem or
a scattering problem using the effective hamiltonian, adjusting the
finite part of the counterterm to match data (including adjustments for
the observed symmetries), and computing the final expression for the
effective hamiltonian with the counterterm finite part determined from
the fit to data.

The simplest example is described in full detail of the 10 steps.  Such
extensive presentation is not provided in later examples where more
complicated expressions would require too much space.  The first example
is discussed in such detail despite the fact that in this case it is
easy to predict the answer.

For example, one might propose the structure of the counterterm using,
as is usually done, some scattering amplitude instead of the matrix
elements of an effective hamiltonian.  Note that one can also impose
renormalization condition using a scattering amplitude which results
from a calculation performed without use of the effective hamiltonian.

However, the systematic approach from Section 2 is the only tool we have
for dealing with more complicated cases of light-front hamiltonians and
their eigenvalue equations.  In other words, the simplest available case
is used to present all the steps in detail because it illustrates the
procedure in a familiar setting.  When we proceed to more complicated
interactions details of the calculation are discussed only where a new
feature appears.

Equation (2.36) implies to second order in ${\cal G}_{2\lambda}$ that

$$ {d \over d\lambda} {\cal G}_{1\lambda} =
\left[  {\cal G}_{12\lambda}
{{d \over d\lambda} f^2(z^2_\lambda) \over  {\cal
G}_{1\lambda}
- E_{1\lambda} } {\cal G}_{21\lambda} \right]_{11}
 + \sum_{p=3}^\infty \left[  {\cal G}_{1p\lambda}
{{d \over d\lambda} f^2(z^2_\lambda) \over  {\cal G}_{1\lambda}
- E_{1\lambda} } {\cal G}_{p1\lambda} \right]_{11} ,   \eqno(3.4) $$

\noindent where the double-digit subscripts refer to the number of
creation and annihilation operators (in that order) and the bracket
subscript denotes the part which contributes to the rate of change of
${\cal G}_{1\lambda}$ with $\lambda$.  $E_{1\lambda}$ is the eigenvalue
of ${\cal G}_{1\lambda}$ which corresponds to the creation and
annihilation operators indicated by the subscript 11.  The reason for
that only one free energy eigenvalue appears in the denominators is that
${\cal G}_{1\lambda}$ of Eq.  (3.4) is a one-body operator and quantum
numbers which label creation and annihilation operators in ${\cal
G}_{1\lambda}$ are the same, including momentum.  Therefore, the free
energy eigenvalues are also the same:  both are equal to $E_{1\lambda}$.
Consequently, all commutators are written on the right-hand side of Eq.
(3.4) in the simplified form.  The numerator similarity factors reduce
to the derivative of $f^2_\lambda$ (we have chosen $n=1$ in Eq.
(2.19)).  Terms with more than two intermediate particles ($p \geq 3$)
are of order $g^4$ or higher.

Assuming that $g$ in Eq.  (3.3) is extremely small, writing ${\cal
G}_{1\lambda}$ as a series in powers of $g$ and keeping only terms order
$g^2$, we obtain the following result from Eq. (3.4) for the meson free
energy term.

$$  {\cal G}_{1\,meson\,\lambda} =  \int[k] { k^{\perp 2} +
\mu^2_\lambda \over k^+ } a^\dagger_k a_k . \eqno (3.5) $$

\noindent A remarkable feature in this result is that no correction
arises to the term $k^{\perp 2} /k^+$ which is protected by the
kinematical symmetries; the total transverse momentum does not appear in
a boost invariant expression.

The width dependence of $\mu_\lambda$ is determined by the equation

$$ {d \mu^2_\lambda \over d\lambda} = g^2 \int[x\kappa]
{d f^2(z^2_\lambda) \over d\lambda}
{ 8(x-{1\over 2})^2{\cal M}^2 \over {\cal M}^2 -\mu^2 }
r_\epsilon(x,\kappa) , \eqno (3.6) $$

\noindent where ${\cal M}^2 = (\kappa^2 + m^2) / x(1-x) $. $m^2$ and
$\mu^2$ are the original bare mass squared parameters from Eq.  (3.3).
They do not include terms order $g^2$ and higher because such terms
would lead to higher order corrections than $g^2$ for the whole
expression.  The terms order $g^2$ and higher are treated as
interactions in the perturbative calculation.

In terms of graphs for the effective hamiltonian calculus, Eq.  (3.6)
represents the contribution of a fermion loop on a meson line.  However,
the graphs are not provided in order to avoid confusion with other
diagrammatic techniques.

$$   \int [x\kappa] = (16\pi^3)^{-1} \int_0^1 {dx \over
x(1-x)} \int d^2\kappa^\perp . \eqno (3.7) $$

\noindent Using Eq.  (2.20) in the limit of a $\theta$-function, $f(u) =
\theta(u_0 - u)$, one obtains

$$ f^2(z^2_\lambda) = \theta\left[ \lambda^2 + {1+\sqrt{u_0} \over
\sqrt{u_0}}\mu^2 - {1-\sqrt{u_0} \over \sqrt{u_0}} {\cal M}^2 \right].
\eqno (3.8.a) $$

\noindent For example, for $u_0 = {1 \over 4}$ one has $f^2(z^2_\lambda)
= \theta[\lambda^2 + 3\mu^2 - {\cal M}^2]$.  Therefore, the derivative
of $f_\lambda$ with respect to $\lambda$ forces the invariant mass of
the fermion-anti-fermion pair, ${\cal M}^2$, to be equal $\lambda^2 +
3\mu^2$.  The derivative selects the range of energies in the integral
where the similarity function changes most rapidly.  The regions where
the function approaches a constant, i.e. 1 near the diagonal and 0
beyond the hamiltonian width, are strongly suppressed.  The region that
contributes is the edge of the diagonal proximum.  \cite{GW1} The
derivative of $f_\lambda$ is large and positive in this region and it
approaches a $\delta$-function in the limit of Eq.  (3.8).

In the limit of an infinitesimally small $u_0$, as discussed below Eq.
(2.20), one would substitute $\lambda^2 = u_0^{-1/2} \tilde \lambda^2$.
Then,

$$ f^2(z^2_\lambda) = \theta\left[\tilde \lambda^2 + \mu^2 - {\cal M}^2
\right]\, . \eqno (3.8.b) $$

The numerator factor in the square bracket in Eq.  (3.6) originates from
spinors of the intermediate fermions, $Tr(\not\!\!p_m + m)(\not\!\!{\bar
p}_m - m)$ with $p_m^2 = {\bar p}_m^{\,2} = m^2$.  The subscript $m$
indicates that the $-$ component is calculated from the mass-shell
condition knowing $+$ and $\perp$ components.  $+$ and $\perp$
components of $p$ and $\bar p$ are constrained by the light-front
spatial momentum conservation law, $p + {\bar p} = k$, where $k$ is the
meson momentum.  The pseudoscalar interaction with $i\gamma_5$ gives the
same result with an additional term $+8m^2$ in the numerator.

According to Eq.  (2.49),

$$ r_\epsilon(x,\kappa) = \left[ 1 + \epsilon{{\cal M}^2 \over
\Lambda^2} + \left(\epsilon{{\cal M}^2 \over \Lambda^2}\right)^2
x(1-x)\right]^{-2} . \eqno (3.9) $$

\noindent No infrared regularization is required in Yukawa theory
with massive particles, $m > 0$ and $\mu > 0$.

If the regularization factors in Eq.  (2.49) contain $e_i$ divided by
$1-x_i$ in place of $e_i$ one obtains here

$$ r_\epsilon(x,\kappa) = \left[1+ \epsilon \left({{\cal M}^2 \over
\Lambda^2}\right) \right]^{-4} \eqno (3.10) $$

\noindent instead of Eq.  (3.9).  The spinor bracket on the right-hand
side of Eq.  (3.6) equals $8 (p^3)^2$ when one changes variables from
$x$ and $\kappa^\perp$ to $\vec p$ with $p^\perp = \kappa^\perp$ and
$p^3$ is determined from the relation ${\cal M} = 2 \sqrt{m^2 + {\vec
p}^{\,2}}$.  Thus, with the modified regularization factor of Eq.
(3.10) the spinor bracket can be replaced by ${1 \over 3} {\vec p}^{\,2}
= {1\over 3}({1\over 4}{\cal M}^2 -m^2)$ and the integrand function of
$x$ and $\kappa^2$ can be reduced to a function of ${\cal M}^2$.  Such
simplifications are helpful in a qualitative analysis of the cutoff
dependence.

In the limit of Eq. (3.8.a) for $u_0={1\over 4}$ one obtains

$$ {d \mu^2_\lambda \over d\lambda^2} = {\alpha \over 16 \pi} \left(1 +
{\mu^2 \over \lambda^2 + 2\mu^2} \right) \theta(z^2_0)
\left({2\over a}\right)^4 \int_0^{z_0} dz{z^2 \over \left[
(1+ 2/a)^2 - z^2\right]^2} , \eqno (3.11) $$

\noindent where $a=\epsilon (\lambda^2 + 3\mu^2)/\Lambda^2$ and $z_0=
\sqrt{1-4m^2/(\lambda^2 + 3\mu^2)}$.  Note that for $\lambda^2 \leq 4m^2
-3\mu^2$ the derivative of the effective meson mass equals zero and the
mass stays at the width independent value $\mu^2_{4m^2 -3\mu^2}$.  If
one uses Eq.  (3.10) instead, the corresponding result is

$$ {d \mu^2_\lambda \over d\lambda^2} = {\alpha \over 48 \pi} \left(1 +
{\mu^2 \over \lambda^2 + 2\mu^2} \right) \theta(z^2_0) z^3_0 (1+a)^{-4}.
\eqno (3.12) $$

\noindent Eqs. (3.11) and (3.12) are the same for $a \ll 1$ which is the
limit of removing the regularization cutoff, $\epsilon \rightarrow 0$,
for a fixed hamiltonian width $\lambda$. In this limit one has

$$ {d \mu^2_\lambda \over d\lambda^2} = {\alpha \over 48 \pi} \left(1 +
{\mu^2 \over \lambda^2 + 2\mu^2} \right) \left(1-{4m^2\over\lambda^2
+3\mu^2}\right)^{3/2} \theta(\lambda^2 + 3\mu^2 -4m^2) \, .\eqno (3.13)
$$

If one assumes that the meson mass squared parameter in the effective
hamiltonian has some finite value, $\mu^2_0=\mu^2_{\lambda_0}$ at some
$\lambda_0$ such that $\lambda_0^2 \geq 4m^2 -3\mu^2$ then, the
integration of Eq.  (3.13) demonstrates that

$$ \mu^2_\lambda = \mu_0^2 + {\alpha \over 48 \pi} (\lambda^2 -
\lambda_0^2) + {\alpha \over 48 \pi} (\mu^2 - 6m^2)\log{\lambda^2 \over
\lambda_0^2} + \mu^2_{conv}(\lambda,\lambda_0) \, . \eqno (3.14) $$

\noindent $\mu^2_{conv}(\lambda,\lambda_0)$ denotes the result of
integrating the convergent part of the integrand,

$$ \mu^2_{conv} (\lambda,\lambda_0) = {\alpha \over 48 \pi}
\int_{\lambda_0^2}^{\lambda^2} ds\left[ \left(1 + {\mu^2 \over s +
\mu^2} \right) \left(1-{4m^2\over s +3\mu^2}\right)^{3/ 2} - 1 -
{\mu^2 -6m^2 \over s} \right] . \eqno (3.14.a)$$

\noindent $\mu^2_{conv} (\lambda,\lambda_0) $ has a finite (i.e.
$\lambda$-independent) limit for large $\lambda$.  It contains the terms
which vanish for large $\lambda$ as inverse powers of $\lambda^2$.  The
dependence of $\mu^2_{conv} (\lambda,\lambda_0)$ on $m$ and $\mu$ is not
indicated explicitly because we will not need it in the discussion of
counterterms.  However, one should keep in mind that the mass parameters
determine the value of $\lambda = \sqrt{4m^2-3\mu^2}$ where the
effective mass stabilizes.  We simplify our notation assuming that the
effective cutoffs are always above the point of stabilization.  Below
the stabilization point, the meson mass has the constant value which is
independent of $\lambda$.  We will show later that the constant value is
equal to the physical meson mass.

The mass squared term in the effective hamiltonian with a non-negligible
coupling $g$ grows linearly with $\lambda^2$.  A logarithmic correction
appears with an opposite sign due to the factor $z_0^3$ in Eq.  (3.13),
as indicated in Eq.  (3.14).  However, one cannot make contact in Eq.
(3.14) with the initial hamiltonian by letting $\lambda$ grow to
infinity because one would obtain an ill-defined result.  The factors
depending on $a$ in Eqs.  (3.11) or (3.12) remove the infinite growth of
$\mu_\lambda$ when $\lambda \rightarrow \infty$.  Eq.  (3.12) is simpler
than Eq.  (3.11) and illustrates the same idea so we start with Eq.
(3.12).

Equation (3.12) can be integrated over $\lambda$ from infinity to any
finite value because the factor $(1+a)^{-4}$ provides convergence for
$\lambda^2 > \Lambda^2/\epsilon$.  Nevertheless, the integral diverges
as a function of $\epsilon$ when $\epsilon \rightarrow 0$.  The
divergence appears as a single number.  Therefore, the counterterm is
also a number.  We add $\mu^2_\epsilon - \mu^2$ to $\mu^2$ in the
initial hamiltonian.  We also write $\mu^2_\epsilon$ as a series in
powers of $g$, $\mu^2_\epsilon = \mu^2 + \delta \mu^2_\epsilon + o(g^4)$
so that $\delta \mu^2_\epsilon \sim g^2$.  Thus,

$$ \mu_\lambda^2 = \mu^2 + \delta\mu^2_\epsilon -
\int_{\lambda^2}^\infty {\alpha \over 48 \pi} \left(1+{\mu^2\over s+
2\mu^2}\right) \left(1-{4m^2\over s +3\mu^2}\right)^{3/ 2}
\left(1+\epsilon {s+ 3\mu^2 \over \Lambda^2}\right)^{-4} +o(g^4).  \eqno
(3.15) $$

\noindent This is an example of Eq.  (2.33) in a perturbative
application to second order in powers of $g$ in Yukawa theory.  The
counterterm $\delta \mu^2_\epsilon$ will be calculated following the
steps described below Eq.  (2.33).

The diverging part of the integrand equals $\alpha/48\pi \,\,
[1+(\mu^2-6m^2)/s]$ and the remaining part is convergent.  The
convergent part of the integrand has the same structure as in
$\mu^2_{conv}(\lambda,\lambda_0)$ but the integral now extends from
$\lambda^2$ to infinity instead of from $\lambda_0^2$ to $\lambda^2$.
In the convergent part, one can replace the regulating factor by 1.
Simplifications occur in the limit $\epsilon \rightarrow 0$ and the
result of integration in Eq.  (3.15) is

$$ \mu_\lambda^2 = \mu^2 +\delta\mu^2_\epsilon + {\alpha \over 48
\pi}\left[ -{\Lambda^2 \over 3}{1 \over \epsilon} + (\lambda^2 +3\mu^2)
+(\mu^2-6m^2)\left( \log{\epsilon{\lambda^2\over\Lambda^2}} +{11\over
6}\right)\right]$$
$$ - \mu^2_{conv}(\infty,\lambda) + o(g^4), \eqno (3.16) $$

\noindent where the square bracket originates from the diverging
part.

Following the procedure described below Eq.  (2.33), we define the
counterterm $\delta \mu^2_\epsilon$ as the negative of the integral of
the diverging integrand for some arbitrarily chosen $\lambda=\lambda_0$
plus an unknown finite piece corresponding to $\lambda_0$ and denoted by
$\delta \mu^2_{\epsilon \, finite}(\lambda_0)$.  Namely,

$$ \delta \mu_\epsilon^2 = {\alpha \over 48 \pi}\left[{\Lambda^2 \over
3}{1 \over \epsilon}+ (\mu^2 - 6m^2) \log{1\over\epsilon} - (\lambda_0^2
+3\mu^2) - (\mu^2 - 6m^2) \left(\log{\lambda_0^2\over\Lambda^2}
+{11\over 6}\right)\right]$$

$$+\delta \mu^2_{\epsilon\,finite}(\lambda_0)+o(g^4),\eqno (3.17)$$

\noindent where

$$\delta \mu^2_{\epsilon\,finite}(\lambda_0) = {\alpha \over 48\pi}
\left[ {\Lambda^2 \over 3} c_1 + (\mu^2 - 6m^2)c_2 + (\lambda^2_0
+3\mu^2) + (\mu^2-6m^2) (\log{\lambda_0^2 \over \Lambda^2} + {11\over
6}) \right] + o(g^4) \eqno (3.17.a) $$

\noindent with the unknown finite numbers $c_1$ and $c_2$.  So, in fact,

$$\delta \mu^2_\epsilon = {\alpha \over 48\pi} \left[ \Lambda^2
({1\over\epsilon} + c_1) + (\mu^2-6m^2) (\log{1\over \epsilon} + c_2 )
\right] . \eqno(3.17.b)$$

Since the whole expression on the right-hand side of Eq.  (3.17) is
merely a number, it is not necessary to find $c_1$ and $c_2$ or any
other part of it separately.  One can easily find the number $\delta
\mu^2_{\epsilon\,finite}(\lambda_0)$ from the knowledge of
$\mu^2_\lambda$ at some value of $\lambda$.  The resulting counterterm
will render well defined finite boson mass squared parameter in the
effective hamiltonians in the limit $\epsilon \rightarrow 0$.  Using
Eqs.  (3.16) and (3.17) one obtains

$$ \mu^2_\lambda = \mu^2 +\delta\mu^2_{\epsilon\,finite}(\lambda_0) +
{\alpha \over 48\pi} \left[\lambda^2 -\lambda_0^2 +(\mu^2-6m^2)
\log{\lambda^2 \over \lambda_0^2}\right] - \mu^2_{conv}(\infty,\lambda)
+ o(g^4).  \eqno (3.18) $$

\noindent Equation (3.18) is an example of Eq. (2.35).

The unknown finite term $\delta \mu^2_{\epsilon \, finite}(\lambda_0)$
has to be found by comparison with data, which might include symmetry
requirements.  We shall discuss an example of a renormalization
condition later in this Section.

Let us assume now that at some arbitrarily chosen value of $\lambda =
\lambda_1$ the effective meson mass squared required in Eq.  (3.5) by a
fit to data equals $\mu^2_1$, i.e.  $\mu^2_{\lambda_1} = \mu^2_1$.  We
can calculate $\delta \mu^2_{\epsilon \, finite}(\lambda_0)$ using Eq.
(3.18) with $\lambda= \lambda_1$ and $\mu^2_{\lambda_1}$ on the
left-hand side replaced by the number $\mu^2_1$ inferred from the
experimental data.  The result is

$$ \delta \mu^2_{\epsilon\,finite} (\lambda_0) = \mu_1^2 - \mu^2 -
{\alpha \over 48\pi} \left[\lambda_1^2 -\lambda_0^2 +(\mu^2-6m^2)
\log{\lambda_1^2 \over \lambda_0^2}\right] +
\mu^2_{conv}(\infty,\lambda_1) + o(g^4) . \eqno (3.19) $$

\noindent Note that  one has to include the contribution of the
convergent terms in the determination of the arbitrary constants
when using the value of $\mu_1^2$.  Knowing $\delta \mu^2_{\epsilon \,
finite}(\lambda_0)$ one can calculate $\mu^2_\lambda$.  Namely,

$$ \mu^2_\lambda = \mu_1^2 + {\alpha \over 48\pi} \left[\lambda^2
-\lambda_1^2 + (\mu^2 -6m^2) \log{\lambda^2 \over \lambda_1^2}\right] +
\mu^2_{conv}(\lambda,\lambda_1) + o(g^4) . \eqno (3.20) $$

\noindent As expected, Eq.  (3.20) is the same as Eq.  (3.14) when
$\lambda_1=\lambda_0$ and $\mu_1 = \mu_0$.  One can also trace the
origin of all the terms from Eq.  (3.14); the diverging and converging
terms and the counterterm in Eq.  (3.15).

We can now analyze Eq.  (3.11) analogously to Eq.  (3.12) without
calculating all integrals explicitly.  Integrating both sides of Eq.
(3.11), we have

$$ \mu_\lambda^2 = \mu^2_\epsilon - \int_{\lambda^2}^\infty ds {d\mu^2_s
\over ds}.  \eqno (3.21) $$

\noindent By demanding that $\mu^2_\epsilon$ removes the diverging (i.e.
$\epsilon$-dependent) part of the integral, and leaving the finite part
of $\mu^2_\epsilon$ free so that at some $\lambda=\lambda_0$ the
effective boson mass squared parameter has a desired value $\mu^2_0$, we
obtain

$$ \mu_\lambda^2 = \mu^2_0 + \int_{\lambda_0^2}^{\lambda^2} ds {d\mu^2_s
\over ds} \, . \eqno (3.22) $$

\noindent The integrand is given by the right-hand side of Eq.  (3.11)
with $\lambda^2=s$.  Since $s$ ranges only from $\lambda^2_0$ to
$\lambda^2$ and both are finite we can take the limit $\epsilon
\rightarrow 0$ under the integral sign and the integrand becomes equal
to the right-hand side of Eq.  (3.13) with $\lambda^2$ replaced with
$s$.  Integration over $s$ produces exactly the same answer as given by
Eq.  (3.14).  Thus, we see that the result of Eq.  (3.14) is independent
of the regularization scheme.  It is determined by the initial
hamiltonian of Yukawa theory as given by Eq.  (3.3).  The only unknown
in Eq.  (3.14) is the value of $\mu^2_0$.  More precisely, we know that
$\mu^2_0 = \mu^2 + \delta \mu_0^2 + o(g^4)$ and the unknown term is
$\delta \mu^2_0 \sim g^2$.

Note that the above calculations can be carried out in a different way
using the following observation.  Equation (3.6) in the lowest order of
perturbation theory has a particularly simple structure.  Namely, the
only dependence on $\lambda$ comes from the similarity function
$f(z^2_\lambda)$ and both sides of the equation are equal to the derivatives
with respect to $\lambda$.  Therefore, one can directly integrate both
sides and obtain a compact integral expression for $\mu^2_\lambda$ for
arbitrary functions $f$.

One should remember that such simplifications do not occur in higher
order calculations or beyond simple perturbative expansion when the
coupling constant and mass parameters depend on $\lambda$ themselves.
Therefore, we stress that the dominant contribution to the rate of
change of $\mu^2_\lambda$ with $\lambda$ comes from the edge of the
diagonal proximum.  This fact remains generally valid and the procedure
applied above represents a generic situation despite simplicity of the
example.  This example does not involve a distinction between the bare
coupling $g$ and a renormalized coupling because to order $g^2$ there is
none.

In order to determine $\delta \mu_0^2$ we need to specify a
renormalization condition.  A natural requirement for $\mu^2_0$ is that
the effective hamiltonian has one boson eigenstates parametrized by
momenta $p^+$ and $p^\perp$ with eigenvalues of the form $p^- =
(p^{\perp 2} + {\tilde \mu}^2)/p^+$ where $\tilde \mu$ is equal to the
physical boson mass.  Our approach preserves kinematical symmetries of
the light-front frame explicitly and the eigenvalue is guaranteed to
appear in that form.  Therefore, one can calculate a whole spectrum of
eigenvalues for eigenstates of different momenta by calculating the
single mass parameter $\tilde \mu$.

In order to write the effective eigenvalue equation and find out
$\mu^2_0$ which leads to the desired value of $\tilde \mu$ (if it is
possible), the following steps need to be taken.

One inserts Eq.  (3.14) into Eq.  (3.5).  Then, one replaces the bare
operators $a^\dagger_k$ and $a_k$ in the whole ${\cal G}_{\lambda}$ by
the effective ones, $a^\dagger_{\lambda k}$ and $a_{\lambda k}$,
obtaining $G_\lambda$.  Next, by applying the operation $F_\lambda$, one
obtains the effective hamiltonian with the form factors in the vertices,
$H_\lambda = F_\lambda[G_\lambda]$.

The effective eigenvalue equation for bosons is an example of Eq.
(1.2).  Here, it is sufficient to consider the eigenvalue equation for
$H_\lambda$ in the expansion to second order in $g$ since we are
interested in $\delta \mu^2_0$ which is proportional to $g^2$.  The only
terms that contribute are the free energy term including the effective
mass squared and the interaction terms which change the particle number
by one.  The latter equal the canonical interactions with the similarity
form factors.

To zeroth order in $g$, a physical meson state equals a single
effective meson state, and ${\tilde \mu}^2 = \mu^2$.

No terms order $g$ arise in ${\tilde \mu}^2$ and the next correction is
order $g^2$.  This correction can be calculated using the operation $R$
and the model hamiltonian defined in a perturbative expansion from Eq.
(1.3).  Using expansion into a series of powers of $g$ to second order,
one can restrict the model space to the single effective boson sector.
The effect of coupling to the fermion-anti-fermion pair states is
included in perturbation theory.

Another method is to arbitrarily limit the number of effective Fock
sectors and diagonalize the effective hamiltonian in that limited space.
Such procedure could be called the effective Tamm-Dancoff approach
(ETD), cf.  Refs.  \cite{TD} and \cite{HO}.  The term of the second
order in $g$ in the eigenvalue will determine $\delta \mu^2_0$.  One can
limit the space of states to one effective boson and effective
fermion-anti-fermion pairs.  Note that interactions in the ETD are
tempered by the similarity factors of width $\lambda$ which is on the
order of particle masses and no ultraviolet divergence exists.  This way
our ETD approach overcomes the old problem of ultraviolet mass
renormalization in the TD approach.

The model calculation using $R$ and the ETD calculation have to agree
with each other for small coupling constants in the presence of a finite
energy gap between states with different numbers of particles, which is
the case here for $0 < \mu < 2m$.  We discuss only second order
corrections in $g$.  Therefore, we can focus on a straight-forward
perturbation theory anyways.  Nevertheless, our simple calculations have
two interesting aspects.

The first one is that no coupling renormalization effects arise to order
$g^2$.  Therefore, the expansion in powers of $g$ up to $g^2$ is
equivalent to the expansion in powers of an effective coupling, no
matter how the latter is defined.  It is important to realize that the
expansion in powers of $g$ is understood here to be the expansion in
powers of the effective coupling which appears in the effective
hamiltonian of width $\lambda$; $g_\lambda = g$ to order $g^2$.  It is
not meant to be the expansion in the initial field theory coupling
constant.

The second aspect is following.  The perturbative expansions applied in
the effective eigenvalue problem are expansions in the interaction which
is suppressed in strength by the similarity vertex form factor of width
$\lambda$.  Therefore, the range in momentum space of the effective
interaction terms is infinitely smaller than the momentum range of the
analogous interaction in the bare hamiltonian.  In other words, the
effective strength of the interactions is greatly reduced and much
smaller than the value of $g$ itself would imply if it stood in the
initial bare hamiltonian.  Our initial expansion in powers of $g$ can
now be understood as a shortcut to the expansion in powers of the
effective coupling.  The latter expansion may have a considerable range
of rapid convergence because the form factors reduce the size of
coefficients in the expansion.  The effective coefficients are much
smaller than in the case of the initial hamiltonian without form
factors.

Thus, the operation $R$ on $H_\lambda$ expanded in powers of the
effective interaction (the coupling constant itself can be sizable),
opens new options for studying the effective eigenvalue problem in the
whole Fock space using the basis built with the effective creation and
annihilation operators corresponding to the width $\lambda$.  One can
estimate contributions of various components by making different choices
of the model spaces and solving model dynamical problems numerically.
Wave functions are expected to fall off sharply for large momenta and
particle numbers if $g$ is not too large.  The effective coupling
constant needs to be set equal to the right value at $\lambda$.

The second order expression in perturbation theory implies

$$ {p^{\perp 2}+ {\tilde \mu}^2 \over p^+} \langle p'|p\rangle =
{p^{\perp 2}+ \mu^2_\lambda \over p^+}\langle p'|p \rangle -
\langle p'|F_{\lambda}[G_{12\lambda}]{1\over G_{1\lambda} -
{(p^{\perp 2}+ \mu^2 ) / p^+}}F_{\lambda}[G_{21\lambda}]|p\rangle \,
. \eqno(3.23) $$

\noindent $|p\rangle$ denotes a single effective meson state with
momentum $p^+$ and $p^\perp$, $\langle p'|p \rangle = 16\pi^3 p^+
\delta^3(p'-p)$.  The term order $g^2$ produces

$$ {\tilde \mu}^2 = \mu^2_\lambda - \int[x\kappa] gf(z^2_\lambda)
{8(x-{1\over 2})^2{\cal M}^2 \over {\cal M}^2 -\mu^2} gf(z^2_\lambda) +
o(g^4) , \eqno(3.24) $$

\noindent where the notation is the same as in Eq.  (3.6).  Using Eq.
(3.8) with $u_0= {1\over 4}$ at $\lambda=\lambda_0$ one obtains

$$ \mu^2_0 = {\tilde \mu}^2 + {\alpha \over 4\pi} \int_0^1dx
\int_0^\infty d\kappa^2 {8(x-{1\over 2})^2{\cal M}^2 \over x(1-x)({\cal
M}^2 -\mu^2)} \theta(\lambda^2_0 + 3\mu^2 -{\cal M}^2) + o(g^4) .
\eqno(3.25) $$

For $\lambda_0 \leq \sqrt{4m^2 - 3\mu^2}$ the effective meson mass
parameter equals the physical meson mass, as promised.  For fermions
with masses order 0.9 GeV, this implies no corrections to a light meson
mass such as $\mu_\pi$ for cutoffs smaller than 1.8 GeV.  But one has to
remember that the correction for the pseudoscalar $\pi N$ interaction is
different from Eq.  (3.25), i.e. the spin factor has to be enlarged by
$8m^2$ (see comments above Eq.  (3.9)).

However, the actual measure of the off-shell effects is not given
directly by $\mu^2_0$ but by the sum of $\mu^2_0$ and the self-energy
resulting from the effective interactions.  According to Eqs.  (3.22)
and (3.24), the sum of both contributions in the physical pion mass
itself is actually equal zero to order $\alpha$.

Using Eqs.  (3.21), (3.22) and (3.25), one can express the meson mass
squared term in the initial renormalized hamiltonian in terms of the
physical meson mass $\tilde \mu$ and the initial mass parameter $\mu$.
Namely, $\mu^2 = {\tilde \mu}^2 + o(g^2)$ and

$$ \mu^2_\epsilon = {\tilde \mu}^2 + g^2 \int[x\kappa] {8(x-{1\over
2})^2{\cal M}^2 \over {\cal M}^2 -\mu^2}r_\epsilon(x,\kappa) + o(g^4)
\, .\eqno(3.26) $$

\vskip.3in
\centerline {\bf Fermion mass squared}
\vskip.1in

In complete analogy to Eqs.  (3.4) to (3.6) one obtains the fermion
energy operator,

$$ {\cal G}_{1\,fermion\,\lambda} = \sum_\sigma \int[k] { k^{\perp 2} +
m^2_\lambda \over k^+ } (b^\dagger_{k\sigma} b_{k\sigma} +
d^\dagger_{k\sigma} d_{k\sigma})  .\eqno(3.27) $$

\noindent Results for fermions and anti-fermions are identical.  We have

$$ {d m^2_\lambda \over d\lambda} = g^2 \int[x\kappa]
{df^2(z^2_\lambda) \over d\lambda}
{ {\bar u}_{m\sigma k}(\not\!p_m + m)u_{m\sigma k}
\over {\cal M}^2 - m^2}
\, r_\epsilon(x,\kappa) .\eqno (3.28) $$

\noindent ${\cal M}^2 = {(m^2 + \kappa^2)/ x} + {(\mu^2 + \kappa^2) /
(1-x)}$.  The regularization factor of Eq.  (2.49) implies

$$ r_\epsilon(x,\kappa) =\left[ 1+{\epsilon\over\Lambda^2} {\cal M}^2 +
\left({\epsilon\over\Lambda^2}\right)^2{\kappa^2 +m^2 \over x}{\kappa^2
+\mu^2\over 1-x}\right]^{-2} . \eqno(3.29) $$

\noindent  The spin factor in Eq. (3.28) can be rewritten as

$$ {\bar u}_{m\sigma k}(\not\!p_m + m)u_{m\sigma k} = {\bar u}_{m\sigma
k}[ x\not\!k_m + m + {1\over 2} \gamma^+ (p^-_m - x k^-_m) ]u_{m\sigma
k}.  \eqno(3.30) $$

\noindent $\not\!k_m$ between spinors is equivalent to $m$.  The term
with $\gamma^+$ is typical in light-front calculations. Its part
proportional to $k^{\perp 2} / k^+$ cancels out.  The term linear in
$k^\perp$ does not contribute because it is odd in $\kappa^\perp$ and
all other factors including the regularization factor depend only on
the modulus of $\kappa^\perp$. The spin factor is thus reduced to

$$ {\bar u}_{m\sigma k}\left[ (x + 1) m + {\gamma^+ \over 2k^+}
{\kappa^2 +(1-x^2)m^2 \over x} \right] u_{m\sigma k}= {\kappa^2+(1+x)^2
m^2 \over x} . \eqno(3.31) $$

\noindent Result for a pseudoscalar interaction with $i\gamma_5$ is the
same except for the opposite sign in front of $x$ in the numerator.
Inclusion of the isospin introduces the number of bosons in the theory
in front of the integral in Eq.  (3.28).

We observe a similar structure in Eq.  (3.28) as in the meson mass
dependence on $\lambda^2$ in Eq.  (3.6).  Namely, there are terms
diverging linearly and logarithmically and there is a series of
convergent terms.  We observe that the divergences amount to a number
which grows when $\epsilon \rightarrow 0$ and integrate both sides of
Eq.  (3.28) to obtain

$$ m^2_\lambda = m^2_\epsilon - g^2 \int[x\kappa] \left[1 -
f^2(z^2_\lambda)\right]
{ \kappa^2 + (1+x)^2 m^2 \over x ({\cal M}^2 - m^2) } \,
r_\epsilon(x,\kappa) ,\eqno (3.32) $$

\noindent where, according to Eq.  (2.18), $z_\lambda = ({\cal M}^2 -
m^2)/({\cal M}^2 + m^2 + \lambda^2)$.  The $\epsilon$-dependent terms
originate from 1 in the bracket which is independent of $\lambda$.  The
counterterm $\delta m^2_\epsilon \sim g^2$ in $m^2_\epsilon = m^2 +
\delta m^2_\epsilon + o(g^4)$ removes the divergence.  The finite part
of the counterterm is left to be determined by data.

$$ m^2_\lambda = m^2 + \delta m^2_{\epsilon \, finite} + g^2
\int[x\kappa] f^2(z^2_\lambda) { \kappa^2+(1+x)^2 m^2 \over x
({\cal M}^2 - m^2) } , \eqno (3.33) $$

\noindent The value of $\delta m^2_{\epsilon \, finite}$ is determined
from the value of $m^2_\lambda$ required in the effective hamiltonian
$H_\lambda$ by some physical condition.  If we had defined the divergent
part by an integral from $\lambda_0$ we would have to take into account
that $m^2_{\epsilon \, finite}$ depends on $\lambda_0$ to compensate for
the $\lambda_0$ dependence of the diverging integral.  When we define
the counterterm to be given by the whole $\lambda$-independent part of
the integral in Eq.  (3.32), plus a finite constant to be determined by
data, then $\delta m^2_{\epsilon \, finite}$ does not depend on
$\lambda_0$.  Nevertheless, it can be expressed in terms of
$m^2_{\lambda_0}$.  For example, if the effective fermion mass squared
at $\lambda = \lambda_0$ should be $m^2_{\lambda_0} = m^2_0$ then

$$ m^2_\lambda = m^2_0 + g^2 \int[x\kappa] \left[ f^2(z^2_\lambda) -
f^2(z^2_{\lambda_0}\right] {\kappa^2+(1+x)^2 m^2
\over x ({\cal M}^2 - m^2) } . \eqno (3.34) $$

\noindent $m^2_0 = m^2 + \delta m^2_0 + o(g^4)$.  $\delta m^2_0 \sim
g^2$ can be found from a renormalization condition for the physical
fermion mass.

A natural condition for fitting $m^2_0$ is that the effective
hamiltonian at the scale $\lambda_0$ has fermionic eigenstates with
eigenvalues of the form $p^- = (p^{\perp 2} + {\tilde m}^2) /p^+$, where
$\tilde m$ denotes the physical fermion mass.  In analogy to Eqs.
(3.23) and (3.24) one obtains

$$ {\tilde m}^2 = m^2_\lambda - \int[x\kappa] gf(z^2_\lambda)
{\kappa^2+(1+x)^2 m^2 \over x ({\cal M}^2 - m^2 ) } g f(z^2_\lambda)
+ o(g^4). \eqno (3.35) $$

\noindent So,

$$ m^2_0 = {\tilde m}^2 + g^2 \int[x\kappa] f^2(z^2_{\lambda_0})
 {\kappa^2+(1+x)^2 m^2 \over x({\cal M}^2 - m^2)} + o(g^4).
\eqno (3.36) $$

\noindent The initial $m^2_\epsilon$ can be calculated in terms
of $m^2$, $\tilde m^2$, $g$ and $\epsilon$ from Eq. (3.32). The
effective fermion mass parameter in the interacting hamiltonian
of width $\lambda$ is

$$ m^2_\lambda = {\tilde m}^2 + g^2 \int[x\kappa] f^2(z^2_{\lambda})
{\kappa^2+(1+x)^2 m^2 \over x ({\cal M}^2 - m^2) }  + o(g^4).
\eqno (3.37) $$

Analogous equation in the case of nucleons coupled to pions is the same
as Eq.  (3.37) except for $(1-x)^2$ instead of $(1+x)^2$ in the
numerator and the isospin factor 3 in front of the integral.  In the
limit of a $\theta$-function for the similarity factor $f$ one obtains
the result $m^2 = {\tilde m}^2 + o(g^2)$, ${\tilde m} = m_N$, and

$$ m^2_\lambda = m^2_N + 3 g^2 \int[x\kappa] \theta(\lambda^2 + 3m^2
-{\cal M}^2) {\kappa^2+(1-x)^2 m^2 \over x ({\cal M}^2 - m^2) } +
o(g^4) . \eqno (3.38) $$

\noindent For $\lambda^2 = (m + n_\pi \, \mu_\pi)^2 - 3m^2$, where
$n_\pi$ is a small integer one obtains ($\alpha = g^2/ 4\pi$)

$$  m^2_\lambda = m^2_N + m_N^2 {3 \alpha \over 4 \pi} c \,\, .
  \eqno(3.39) $$

\noindent $c= 4/3 (n_\pi \, \mu_\pi / m_N)^3 +o(\mu_\pi^4)$.  The
expansion formula for $c$ shows the correction is small for small meson
masses.  Note that $\lambda^2$ must be negative for small $n_\pi$, as
explained below Eq.  (2.20).  The exact result for $ n_\pi = 3 $ gives
$c \sim 0.03 $ and $ n_\pi = 4 $ gives $c \sim 0.12$.  Even for quite
large couplings the effective mass parameter in the hamiltonian deviates
only a little from the physical nucleon mass if the hamiltonian width
allows momentum changes of the order of a few meson masses only.
Moreover, the physically relevant off-shell effects are not given
directly by the above numbers but by the difference between these and
the effects of the interactions present in the effective hamiltonian.
The combined effect is zero for the nucleon mass itself to order $g^2$.
Eq.  (3.39) suggests that the self-interaction effects can be calculable
in perturbation theory.  This is encouraging for the program outlined in
Ref.  \cite{BGP}.

If we used Eq.  (2.20) in the $\theta$-function limit with an
infinitesimal $u_0$ and $\lambda^2 = \tilde \lambda ^2 /\sqrt{u_0}$ for
$n=1$ the $\theta$-function under the integral in Eq.  (3.38)
would be replaced by $\theta(\tilde \lambda^2 + m^2 - {\cal M}^2)$.

\vskip.3in
\centerline {\bf Fermion-fermion interaction}
\vskip.1in

Our next example is the second order calculation of the effective
hamiltonian term which contains products of two creation and two
annihilation operators for fermions.  The differential equation we need
to consider results from Eq.  (2.29) for the two-fermion terms;

$$ {d \over d\lambda} {\cal G}_{22\lambda} = \left[ f_\lambda {\cal
G}_{12\lambda} \left\{ {d \over d\lambda} (1-f_\lambda) {\cal
G}_{21\lambda} \right\}_{{\cal G}_{1\lambda}} - \left\{ {d \over
d\lambda} (1-f_\lambda) {\cal G}_{12\lambda} \right\}_{{\cal
G}_{1\lambda}} f_\lambda {\cal G}_{21\lambda} \right]_{22} . \eqno
(3.40) $$

\noindent The subscript 22 denotes a term with two creation and two
annihilation operators for fermions. 21 denotes a term with one
annihilation operator and one creation operator for fermions and one
creation operator for bosons. 12 denotes a term which annihilates a
fermion and a boson and creates a fermion.  For a hermitean hamiltonian,
we have ${\cal G}_{12} = {\cal G}_{21}^\dagger$.

The right-hand side of Eq.  (3.40) does not contain disconnected
interactions (it never does, cf.  Section 3.a) and one can isolate the
terms with two creation and two annihilation operators for fermions by
contracting one creation operator and one annihilation operator for
bosons.  The only term which contributes is the ordered and regularized
third term on the right-hand side of Eq.  (3.3).  Thus, in Eq.  (3.40),
we have

$$ {\cal G}_{21\lambda} = \sum_{\sigma \sigma_f} \int [k_1 k k_2]
16\pi^3 \delta(k_1 + k - k_2)\, g{\bar u}_{m k_1 \sigma_f}u_{m k_2
\sigma}\, r(\epsilon e_f/\Lambda^2)\, r(\epsilon e_b/\Lambda^2)
b^\dagger_{k_1 \sigma_f} a^\dagger_{k} b_{k_2 \sigma} $$ $$ = \int [P]
{1 \over P^+} \sum_{\sigma \sigma_f}\int [x \kappa] g{\bar u}_{m
xP+\kappa \sigma_f} u_{m P \sigma} \, r(\epsilon e_f/\Lambda^2)\,
r(\epsilon e_b/\Lambda^2) \, b^\dagger_{xP+\kappa
\sigma_f}a^\dagger_{(1-x)P-\kappa} b_{P \sigma} \, , \eqno(3.41)$$

\noindent where $ e_f = ( \kappa^2 + m^2 ) / x $ and $ e_b = ( \kappa^2
+ \mu^2 ) / (1-x) $. This representation illustrates appearance of the
parent and daughter momentum variables in the interaction term.  The
factor $g\, {\bar u}_{m k_1 \sigma_f}u_{m k_2 \sigma}\, r(\epsilon
e_f/\Lambda^2)\, r(\epsilon e_b/\Lambda^2) b^\dagger_{k_1 \sigma_f}$
will be denoted by $g_{21}$.  The analogous factor in ${\cal
G}_{12\lambda}$ will be denoted by $g_{12}$.  Similarly, in the case of
the four-fermion interaction term, we have

$$ {\cal G}_{22\lambda}
= \int [P] {1 \over P^+} \sum_{\sigma_1 \sigma_2 \sigma_3 \sigma_4}
\int [x\kappa][y\rho] \, g_{22\lambda} \,
b^\dagger_{xP+\kappa \sigma_1}b^\dagger_{(1-x)P-\kappa \sigma_3}
b_{yP+\rho \sigma_2}b_{(1-y)P-\rho \sigma_4}
, \eqno(3.42)$$

\noindent where $g_{22\lambda}$ is a function of the daughter momentum
variables $x$, $\kappa^\perp$, $y$, $\rho^\perp$ and the fermion spin
projections on the z-axis:  $\sigma_1$, $\sigma_2$, $\sigma_3$ and $
\sigma_4$.  Details of the notation will become clear shortly.  To order
$g^2$, only $f_\lambda$ depends on $\lambda$ on the right-hand side of
Eq.  (3.40) and it can be written in terms of the coefficient functions
as

$$ {d \over d\lambda} g_{22} = \left[ f \{ -f'\} - \{
- f'\} f \right] \left[ g_{12} g_{21} \right]_{22} .
\eqno (3.43) $$

\noindent The subscript $\lambda$ and arguments of the functions are not
indicated, to simplify the formula.  Expression in the first bracket is
called {\it the inner similarity factor} for ${\cal G}_{22}$.  The word
``inner'' is used to distinguish this factor from the form factor
introduced by the operation $F_\lambda$ when this operation is applied
to $G_{22}$.  The latter form factor can be called the {\it outer}
similarity factor because it depends on the incoming and outgoing
invariant masses only, independently of the internal structure of the
interaction.

Momentum variables in Eq.  (3.43) can be expressed by the daughter
momentum variables from Eq.  (3.42).  One needs to express the parent
and daughter variables of ${\cal G}_{12}$ and ${\cal G}_{21}$, and the
energy denominators, in terms of $x$, $\kappa^\perp$, $y$ and
$\rho^\perp$ from Eq.  (3.42).  This is done as follows.

The momentum labels of the fermion annihilation operators in Eq.  (3.42)
are denoted by $k_2$ and $k_4$ and the momentum labels of the fermion
creation operators in Eq.  (3.42) are denoted by $k_1$ and $k_3$ . The
numbers assigned to the fermion momenta correspond to the numbers
labeling their spin projections on the z-axis in Eq.  (3.42).  In terms
of the origin of the annihilation and creation operators in Eq.  (3.42),
${\cal G}_{21}$ provides the fermion creation operator with momentum
$k_1$ and the fermion annihilation operator with momentum $k_2$.  ${\cal
G}_{12}$ provides the fermion creation operator with momentum $k_3$ and
the fermion annihilation operator with momentum $k_4$.  There is a
change of sign due to the reordering of the fermion operators.  The
boson operators from ${\cal G}_{12}$ and ${\cal G}_{21}$ are contracted
and provide factors similar to the factors associated with an
intermediate particle in the old-fashioned hamiltonian calculations of
the $S$-matrix.

It is useful to think about the effective hamiltonian in terms of a
scattering amplitude with two vertices but the reader should remember
that the formula we are describing is not for an $S$-matrix matrix
element.  Therefore, the ``scattering'' language has a limited meaning.
The fine point is that, after evaluation of ${\cal G}_\lambda$, one has
to replace the bare creation and annihilation operators by the effective
ones in order to obtain $G_\lambda$.  The term in ${\cal G}_\lambda$ is
not directly related to any scattering process before the replacement is
made.  The replacement prevents confusion between the hamiltonian
calculus which uses the bare operators, and the $S$-matrix calculus
which uses the effective hamiltonian and the corresponding incoming,
outgoing and intermediate states of effective particles.  The scattering
language becomes particularly confusing in theories with gauge symmetry,
spontaneous symmetry breaking and confinement.  None of these features
appear here in the calculation to order $g^2$.  Therefore, the
scattering language is useful in the current example.

The intermediate boson momentum is defined for $+$ and $\perp$
components as $k_5 = k_3 - k_4$ and $k^-_{5\mu} = (k^{\perp 2}_5 +
\mu^2)/k^+_5$.  These four components form the four-momentum of the
exchanged boson. It is denoted by $k_{5\mu}$ to indicate the mass which
determines the minus component.  The same result for $k_{5\mu}$ is
obtained by subtracting $k_1$ from $k_2$ instead of $k_4$ from $k_3$.
It is so because the translational invariance of the hamiltonian on the
light-front implies momentum conservation for the $+$ and $\perp$
components.

Thus, the inner similarity factor in Eq. (3.43) is

$$ \left[ f \{ - f'\} - \{ - f'\} f \right]
 =  f(z^2_{12 \lambda}) {[-f(z^2_{21 \lambda})]'\over \Delta E_{21}}
 - {[-f(z^2_{12 \lambda})]' \over \Delta E_{12}} f(z^2_{21 \lambda})
 . \eqno (3.44) $$

\noindent The prime denotes differentiation with respect to $\lambda$.

In the case of ${\cal G}_1$ in Eq.  (3.4), the whole inner similarity
factor of the analogous structure was equal to the derivative of
$f^2(z^2_\lambda)$ divided by a single denominator.  For both functions
$f$ in Eq.  (3.4) had the same argument $z^2_\lambda$ and the two
corresponding energy denominators were the same.  In Eq.  (3.44) we have
two different arguments of the similarity functions $f$ and two
different energy changes.  Namely, $z_{12 \lambda}$ and $\Delta E_{12}$
for the vertex of ${\cal G}_{12}$ with momenta $k_{3m}$, $k_{4m}$ and
$k_{5\mu}$, and $z_{21 \lambda}$ and $\Delta E_{21}$ for the vertex of
${\cal G}_{21}$ with momenta $k_{1m}$, $k_{2m}$ and $k_{5\mu}$.

The parent momentum for the vertex of ${\cal G}_{12}$ is $P_{12} =
(k_{5\mu} + k_{3m} + k_{4m})/2$ so that for the $+$ and $\perp$
components we have $P_{12} = k_3$.  Similarly, the parent momentum for
the vertex of ${\cal G}_{21}$ is $P_{21} = (k_{5\mu} + k_{1m} +
k_{2m})/2$ so that for the $+$ and $\perp$ components we have $P_{21} =
k_2$.

Now, the rules provided by Eqs.  (2.12) to (2.20) imply the following
formulae for the arguments of the similarity functions $f$.

$$ \Delta {\cal M}^2_{12} =
(k_{5\mu} + k_{4m})^2 - k^2_{3m} =
2 (k_{5\mu} + k_{4m} - k_{3m}) P_{12}
 . \eqno(3.45) $$

$$ \Sigma {\cal M}^2_{12} = {\cal M}^2_{12} + 2m^2 . \eqno (3.46) $$

$$ z_{12 \lambda} = { \Delta {\cal M}^2_{12} \over
            \Sigma {\cal M}^2_{12} + \lambda^2 } . \eqno(3.47)$$

$$ \Delta {\cal M}^2_{21} =
 k^2_{2m} - (k_{5\mu} + k_{1m})^2 =
 - 2 (k_{5\mu} + k_{1m} - k_{2m}) P_{21}
 . \eqno(3.48) $$

$$ \Sigma {\cal M}^2_{21} = - {\cal M}^2_{21} + 2m^2 . \eqno (3.49) $$

$$ z_{21 \lambda} = { \Delta {\cal M}^2_{21} \over
                     \Sigma {\cal M}^2_{21} + \lambda^2 } .
\eqno(3.50)$$

Equations (2.12) to (2.16) imply

$$  \Delta E_{12} = {\Delta {\cal M}^2_{12} \over P^+_{12}} \eqno(3.51)$$

\noindent and

$$  \Delta E_{21} = {\Delta {\cal M}^2_{21} \over P^+_{21}}.
\eqno(3.52)$$

In terms of the familiar parameters $x$, $\kappa^\perp$, $y$ and
$\rho^\perp$ from Eq.  (3.42), i.e.

$$ P = k_1 + k_3 = k_2 +k_4,  \eqno (3.53) $$

$$ x = k^+_1/P^+, \eqno (3.54) $$

$$ \kappa^\perp = k_1^\perp - x P^\perp, \eqno (3.55) $$

\noindent and

$$ y = k^+_2/P^+, \eqno (3.56) $$

$$ \rho^\perp = k_2^\perp - y P^\perp \, , \eqno (3.57) $$

\noindent the mass differences which determine the arguments of the
similarity functions read as follows.

$$ \Delta {\cal M}^2_{12} = (1-x) \left[
{(\kappa^\perp-\rho^\perp)^2 + \mu^2 \over y-x} +
{\rho^{\perp 2} + m^2 \over 1-y} -
{\kappa^{\perp 2} + m^2 \over 1-x} \right] . \eqno(3.58)
$$

$$ \Delta {\cal M}^2_{21} = y \left[
{\rho^{\perp 2} + m^2 \over y}
- {\kappa^{\perp 2} + m^2 \over x} -
{(\kappa^\perp-\rho^\perp)^2 + \mu^2 \over y-x} \right] .\eqno(3.59) $$

\noindent The corresponding energy denominators are

$$ \Delta E_{12} = \left[ {(\kappa^\perp-\rho^\perp)^2 + \mu^2 \over
y-x} + {\rho^{\perp 2} + m^2 \over 1-y} - {\kappa^{\perp 2} + m^2 \over
1-x} \right]/P^+ \eqno(3.60)$$

\noindent and

$$ \Delta E_{21} = \left[ {\rho^{\perp 2} + m^2 \over y} -
{\kappa^{\perp 2} + m^2 \over x} - {(\kappa^\perp-\rho^\perp)^2 + \mu^2
\over y-x} \right]/P^+ . \eqno(3.61)$$

In Eqs.  (3.40) to (3.61) always $y > x$.  Evaluating matrix elements of
the effective interaction between states of indistinguishable fermions
one obtains results in which the momentum and spin variables are
properly symmetrized (antisymmetrized) as dictated by the statistics.

Evaluation of the second bracket in Eq.  (3.43) gives

$$ [g_{12} g_{21}]_{22} \, = \, - \, g \, {\bar u}_{m (1-x)P-\kappa
\sigma_3} u_{m (1-y)P-\rho \sigma_4} \, r(\epsilon e_4/\Lambda^2)
r(\epsilon e_{12}/\Lambda^2)$$
$${1 \over (y-x)P^+} \, g \, {\bar u}_{m xP+\kappa \sigma_1} u_{m
yP+\rho \sigma_2} \, r(\epsilon e_1/\Lambda^2) r(\epsilon
e_{21}/\Lambda^2) \, . \eqno(3.62)$$

\noindent The arguments of the regularization factors appear in the mass
differences.  Namely,

$$ \Delta {\cal M}^2_{12}= e_4 + e_{12} - m^2 =
   {\kappa^{\perp 2}_{12} + m^2 \over x_{12} } +
   {\kappa^{\perp 2}_{12} + \mu^2 \over 1 - x_{12} }
   -m^2 ,  \eqno(3.63.a) $$

\noindent where

$$  x_{12} = {1-y \over 1-x} , \eqno(3.63.b) $$

$$  \kappa^\perp_{12} = -\rho^\perp + x_{12} \kappa^\perp,
\eqno(3.63.c)$$

\noindent and

$$ \Delta {\cal M}^2_{21}= m^2 - e_1 + e_{21} =
   m^2 - {\kappa^{\perp 2}_{21} + m^2 \over x_{21} } -
   {\kappa^{\perp 2}_{21} + \mu^2 \over 1 - x_{21} }
  ,  \eqno(3.64.a) $$

\noindent where

$$  x_{21} = {x \over y} , \eqno(3.64.b) $$

$$  \kappa^\perp_{21} = \kappa^\perp - x_{21} \rho^\perp .
\eqno(3.64.c)$$

\noindent Eqs.  (3.62) to (3.64) contain daughter energies in the
notation introduced already in Eqs.  (2.48) and (2.49).  Similar
subscript notation will be used to label the regularization factors.
Equation (3.43) reads, in the abbreviated notation, as follows.

$$ {d \over d\lambda} g_{22} =
   \left[
    {y \over y-x}\, {f_{12} f'_{21} \over \Delta {\cal M}^2_{21}}
- {1-x \over y-x}\, {f'_{12} f_{21} \over \Delta {\cal M}^2_{12}}
  \right]\theta(y-x) \,
g{\bar u}_3 u_4 \, r_4 r_{12} \, g{\bar u}_1 u_2 \, r_1 r_{21}
 . \eqno (3.65) $$

\noindent In the familiar limit of Eq.  (3.8) where the similarity
function $f$ approaches the $\theta$-function, i.e.  $f_{12} =
\theta[\lambda^2 + 2m^2 - \Delta{\cal M}^2_{12}]$ and $f_{21} =
\theta[\lambda^2 + 2m^2 + \Delta{\cal M}^2_{21}]$, the derivatives of
the similarity functions become $\delta$-functions and one can integrate
Eq.  (3.65) using the relation $\int_a^\infty ds \theta(s-b) \delta(s-c)
= \theta(c-b) \theta(c-a)$.  The result is

$$ g_{22\lambda} = g_{22\epsilon} + \left[
{ y \theta_{21-12} \theta_{21-\lambda} \over
(y-x) |\Delta {\cal M}^2_{21}|} +
{(1-x) \theta_{12-\lambda} \theta_{12-21} \over
(y-x) |\Delta {\cal M}^2_{12}|} \right] \theta(y-x)
\, g{\bar u}_3 u_4 \, r_4 r_{12} \, g{\bar u}_1 u_2 \,
r_1 r_{21} . \eqno (3.66.a) $$

\noindent The symbols $\theta$ with various subscripts denote the
following functions:

$$ \theta_{12-21} = 1 - \theta_{21-12} = \theta(|\Delta {\cal M}^2_{12}|
- |\Delta {\cal M}^2_{21}|) , \eqno(3.66.b)$$

$$\theta_{12-\lambda} = \theta(
|\Delta {\cal M}^2_{12}| - \lambda^2 - 2m^2), \eqno(3.66.c)$$

\noindent and

$$ \theta_{21-\lambda} =
\theta( |\Delta {\cal M}^2_{21}| - \lambda^2 - 2m^2). \eqno(3.66.d)$$

\noindent The initial value term of $g_{22\epsilon}$ at $\lambda =
\infty$ is absent in the canonical hamiltonian.  It is equal zero if
matrix elements of $H_{22\lambda} = F_\lambda \left[ G_{22\lambda}
\right]$ between finite free energy states have a limit when $\epsilon
\rightarrow 0$.  If the limit does not exist due to a diverging
$\epsilon$-dependence a counterterm containing nonzero $g_{22\epsilon}$
is required to remove the divergence.

If we used Eq.  (2.20) in the $\theta$-function limit with an
infinitesimal $u_0$ and $\lambda^2 = \tilde \lambda ^2 /\sqrt{u_0}$ for
$n=1$ then, $\lambda^2 + 2 m^2$ in the $\theta$-function arguments above
would be replaced by $\tilde \lambda^2$.  This feature will be used
later in the case of nonrelativistic bound states.

The easiest momentum configuration to analyze $g_{22\epsilon}$ is the
one where the sum of free energies for the momenta of creation operators
equals the sum of free energies for the momenta of annihilation
operators:  $(k_{1m} + k_{3m})^2 = (k_{2m} + k_{4m})^2 = {\cal M}^2$.
We will refer to this configuration as {\it the energy-diagonal part of
the interaction}.  In the energy-diagonal part of the interaction, we
have $\Delta E_{12} = - \Delta E_{21}$, and

$$ {\rho^{\perp 2} + m^2 \over 1-y} - {\kappa^{\perp 2} + m^2 \over 1-x}
 = {\kappa^{\perp 2} + m^2 \over x} - {\rho^{\perp 2} + m^2 \over y} .
\eqno (3.67) $$

\noindent Thus,

$$ {|\Delta{\cal M}^2_{12}|\over 1-x} =
   {|\Delta{\cal M}^2_{21}|\over y}   =
   {\mu^2 + {\vec q}^{\, 2} \over y-x} , \eqno(3.68) $$

\noindent where ${\vec q}^{\, 2} = (\kappa^\perp - \rho^\perp)^2 +
(y-x)^2 {\cal M}^2$.  These relations imply $\theta_{12-21} = 1 -
\theta_{21-12} = \theta(1-x-y)$, $\theta_{12-\lambda}= \theta[\mu^2 +
{\vec q}^{\, 2} - (y-x)(\lambda^2 + 2m^2)/(1-x)]$ and
$\theta_{21-\lambda}= \theta[\mu^2 + {\vec q}^{\, 2} - (y-x)(\lambda^2 +
2m^2)/y]$.  Therefore, $\theta_{12-\lambda} = 1$ in the same momentum
range where $\theta_{21-\lambda} = 1$.  Thus, the energy-diagonal part
of the fermion-fermion effective interaction order $g^2$ is

$$ g_{22\lambda} = g_{22\epsilon} + { g{\bar u}_3 u_4 \, r_4 r_{12}
\, g{\bar u}_1 u_2 \, r_1 r_{21} \over \mu^2 + {\vec q}^{\, 2}}
\theta(y-x) \theta \left[ \mu^2 + {\vec q}^{\, 2} - {y-x\over
max(y,1-x)} (\lambda^2 + 2m^2)  \right] . \eqno (3.69) $$

Equation (3.69) is helpful because it provides insight into the more
complicated interaction from Eq.  (3.66).  When the momentum transfer is
sufficiently large and $(y-x)(\lambda^2 + 2m^2)/max(y,1-x)$ is
negligible so that the $\theta$-functions and the regularization
functions $r_4$, $r_{12}$, $r_1$ and $r_{21}$ in Eq.  (3.69) equal 1,
the second term on the right-hand side of Eq.  (3.69) is equal to
Feynman's expression for the one boson-exchange scattering amplitude for
two fermions.  Namely, the numerator factors are standard for the Yukawa
interaction and the denominator equals $ \mu^2 - (k_{3m} - k_{4m})^2 =
\mu^2 - (k_{2m} - k_{1m})^2$ (the necessary antisymmetrization for
identical fermions requires evaluation of matrix elements of the
hamiltonian term under consideration).  However, there is a difference
between the energy diagonal part of the effective hamiltonian matrix
elements and the on-shell Feynman scattering amplitude due to the
$\theta$-functions and the regularization factors in Eq.  (3.69) where
they differ from 1. In other words, our theory is regularized ab initio
and the resulting amplitudes contain the regularizing factors.  The
width dependent factor of the effective hamiltonian does not belong in a
physical scattering amplitude and we will explain how it goes away when
one evaluates $S$-matrix elements later in this Section.

The $\theta$-functions force the momentum transfer carried by the
intermediate boson, $|{\vec q}|$, to be larger than
$\left[(y-x)(\lambda^2 + 2m^2)/y - \mu^2\right]^{1/2}$.  The size of
this number depends on how large $\lambda^2$ and the ratios of the
longitudinal momenta are.  If $\lambda^2$ is negative and compensates
$2m^2$, the lower bound on the momentum transfer is absent.  For larger
$\lambda^2$, the ratio of $y-x$ to the parent fermion $y$ has to be
smaller than $\mu^2(\lambda^2 + 2m^2)^{-1}$ for the lower bound on the
momentum transfer to be absent.  Otherwise, the momentum transfer is
limited from below.  This means that the effective interaction term does
not include the long distance part of the Yukawa potential.

The regularization factors in the limit $\epsilon \rightarrow 0$
converge pointwise to 1. No diverging cutoff dependence is obtained when
evaluating matrix elements of $G_{22\lambda}$ between states of finite
invariant masses ${\cal M}^2$ so the matrix elements of $H_{22\lambda}$
are free from divergences.  Therefore, $g_{22\epsilon} = 0$.  We can
replace the regularization factors in the limit $\epsilon \rightarrow 0$
by 1.

We proceed to the analysis of Eq.  (3.66).  No divergences appear in the
finite matrix elements of the effective hamiltonian of width $\lambda$
when $\epsilon \rightarrow 0$.  One can see that this is the case using
Eqs.  (3.63.a) and (3.64.a).  Namely, the arguments of the
regularization factors are finite for finite $\Delta {\cal M}^2_{12}$
and $\Delta {\cal M}^2_{21}$ and they approach 0 when $\epsilon
\rightarrow 0$.  One demands that the free invariant masses of the
states of fermions used to calculate the matrix elements are finite.
The only possibility for $\Delta {\cal M}^2_{12}$ or $\Delta {\cal
M}^2_{21}$ to diverge emerges when $x$ approaches $y$, i.e. when the
longitudinal momentum transfer between the fermions approaches zero.  In
such case, $e_{12}$ and $e_{21}$ approach infinity even for a vanishing
transverse momentum transfer because the meson mass squared is greater
than zero.

Now, the remaining factors of spinors and energy denominators, the
latter multiplied by the boson phase-space factor of $y-x$, are all
finite in the limit $x \rightarrow y$.  The regularization factors
$r_{12}$ and $r_{21}$ deviate from 1 only in the small region in the
momentum space where $|x-y| < \epsilon\Lambda^2/\mu^2$ (or in a still
smaller region for a nonzero meson transverse momentum).  All other
factors in the interaction are finite in this region.  Therefore, for
finite wave packets or bound state wave functions used in the evaluation
of the matrix element, this small region produces a contribution which
is proportional to $\epsilon$.  Thus, it vanishes in the limit $\epsilon
\rightarrow 0$.  Consequently, the matrix elements of $g_{22\epsilon}$
are equal 0 and the regularization factors can be replaced by 1.

The full result for the effective fermion-fermion interaction in the
limit $\epsilon \rightarrow 0$ is

$$ H_{22\lambda} = F_\lambda\left[G_{22\lambda}\right] $$
$$ = \int [P] {1 \over
P^+} \sum_{\sigma_1 \sigma_2 \sigma_3 \sigma_4} \int [x\kappa][y\rho]\,
g_{22\lambda}\, f(z^2_{22\lambda})\, b^\dagger_{\lambda xP+\kappa
\sigma_1}b^\dagger_{\lambda (1-x)P-\kappa \sigma_3} b_{\lambda
yP+\rho \sigma_2}b_{\lambda (1-y)P-\rho \sigma_4} \, ,
\eqno(3.70.a)$$

\noindent where

$$ g_{22\lambda} = \left[ { \theta_{12-21} \theta_{21-\lambda} \over
\mu^2 - q^2_{21} } + { \theta_{12-\lambda} \theta_{21-12}\over \mu^2 -
q^2_{12} } \right] \theta(y-x) g{\bar u}_3 u_4 g{\bar u}_1 u_2 \, , \eqno
(3.70.b)  $$

$$ q_{12} = k_{3m} - k_{4m} , \eqno(3.70.c)$$

$$ q_{21} = k_{2m} - k_{1m} , \eqno(3.70.d)$$

$$ \theta_{12-21} = 1 - \theta_{21-12} =
   \theta\left[(1-x)(\mu^2-q^2_{12})  - y(\mu^2-q^2_{21})\right]
   , \eqno(3.70.e) $$

$$ \theta_{12-\lambda} = \theta\left[(1-x)(\mu^2-q^2_{12}) -
(y-x)(\lambda^2 + 2m^2)\right] , \eqno(3.70.f) $$

$$ \theta_{21-\lambda} = \theta\left[y(\mu^2-q^2_{21}) -
(y-x)(\lambda^2 + 2m^2)\right] . \eqno(3.70.g) $$

\noindent The argument of the outer similarity factor $
f(z^2_{22\lambda}) $ that limits the width of the effective interaction
in momentum space, is

$$ z_{22\lambda} = {\Delta{\cal M}^2_{22}\over
\Sigma{\cal M}^2_{22} + \lambda^2 }. \eqno(3.70.h) $$

\noindent The mass difference, $\Delta{\cal M}^2_{22} = {\cal M}^2_{24}
- {\cal M}^2_{13}$, and the mass sum, $\Sigma{\cal M}^2_{22} = {\cal
M}^2_{24} + {\cal M}^2_{13}$, are expressed by the fermion momenta
through relations ${\cal M}^2_{24} = (k_{2m} + k_{4m})^2$ and ${\cal
M}^2_{13} = (k_{1m} + k_{3m})^2$.

Equations (3.70.a) to (3.70.h) explain the structure of the
fermion-fermion effective interaction order $g^2$ in terms of the two
four-momentum transfers, $q_{12}$ and $q_{21}$.  The transfer $q_{12}$
appears in the vertex where the intermediate boson is annihilated and
the transfer $q_{21}$ appears in the vertex where the boson is created.
The $\theta$-functions exclusively select which momentum transfer
appears in the denominator.  The lower bounds on the momentum transfers
depend on the ratio of $|x-y|$ to $y$ and $1-x$ and on the masses and
$\lambda^2$.

We can now evaluate matrix elements of the $T$-matrix between effective
two-fermion states using the hamiltonian of width $\lambda$ to second
order in $g$;

$$  T(E) = H_{I\lambda} + H_{I\lambda} {1\over E -
H_{0\lambda} + i\varepsilon} H_{I\lambda} \, . \eqno(3.71) $$

\noindent We have $H_{0\lambda} = G_{1\lambda}$ and $H_{I\lambda} =
F_\lambda\left[ G_{22\lambda} + G_{12\lambda} + G_{21\lambda}\right]$.
The first term on the right-hand side of Eq.  (3.71) contributes solely
through $H_{22\lambda}$.  In the second term, only $H_{12\lambda} +
H_{21\lambda}$ contributes in $H_{I\lambda}$.

The first term in Eq.  (3.71) has its matrix element given by the
antisymmetrization of the right-hand side of Eq.  (3.70.b).  The
multiplication by $f(z^2_{22\lambda})$ does not matter because
$f(z^2_{22\lambda}) = 1$ in the energy-diagonal matrix elements and only
the energy-diagonal part contributes to the cross section.  The
energy-diagonal part of $g_{22\lambda}$ is given by

$$ g_{22\lambda} = { \, g{\bar u}_3 u_4 \, g{\bar u}_1 u_2 \, \over
\mu^2 - q^2} \, \theta(y-x) \, \theta_\lambda \, , \eqno (3.72) $$

\noindent where $ \theta_\lambda = \theta \left[ max(y,1-x)(\mu^2 - q^2)
- (y-x)(\lambda^2 + 2m^2) \right]$ and $q = q_{12} = q_{21}$.  The
antisymmetrization of the right-hand side of Eq.  (3.72) produces the
contribution of the first term in Eq.  (3.71) to the scattering
amplitude.

The second term in Eq.  (3.71) provides the one-boson exchange amplitude
with form factors in the fermion-boson vertices.  The form factors are
the similarity functions $f_\lambda$.  The resulting amplitude is given
by the antisymmetrization of Eq.  (3.72) with $\theta_\lambda$ replaced
by the product of the form factors.  In the $\theta$-function limit, the
vertex form factors equal $1 - \theta_{21-\lambda}$ and $1 -
\theta_{12-\lambda}$.  Their product equals $1- \theta_\lambda$.  Thus,
the second term provides the same contribution as the first term but the
factor $\theta_\lambda$ is replaced by $1 - \theta_\lambda$.

The sum of both terms in Eq.  (3.71) produces the matrix element of the
scattering matrix on-energy-shell which is independent of $\lambda$.
Our complete on-shell result in the effective theory is equal to the
well known Feynman result for the one boson exchange scattering
amplitude.

There is an important property of the second order calculation above
which is worth a separate note.  When the hamiltonian width in the mass
difference becomes small the effective meson emission can no longer
occur.  Thus, the effective theory describes fermions interacting by
potential forces.  The potentials are given by factors
$f(z^2_{22\lambda})g_{22\lambda}$ in $H_{22\lambda}$.  The form factors
$f(z^2_{22\lambda})$ are known off-energy shell.  $g_{22\lambda}$
contains also the inner similarity factors which force the intermediate
boson to form a sufficiently high invariant mass state but if the width
is small enough these factors are equal 1. $f(z^2_{22\lambda})
g_{22\lambda}$ is the generalized relativistic potential term that
equals Yukawa potential in the nonrelativistic limit.  Thus, we have
accomplished a derivation of the boost invariant potential term order
$g^2$ in the effective Yukawa theory.

The nonrelativistic Yukawa theory is obtained when the width $\lambda$
is such that the allowed energy transfers are much smaller than the
effective fermion masses.  This condition limits only the relative
motion of the effective fermions.  It does not limit their total
momentum which can still be chosen arbitrarily by taking advantage of
the boost invariance.  The reduction of the fermion dynamics in Yukawa
theory with small $\lambda$ to the Schr\"odinger equation in second
order perturbation theory is further discussed in Ref.
\cite{Wieckowski}.  Initial studies of 4th order similarity in a Yukawa
model are given in Ref.  \cite{4thorder}.

Note that in the energy diagonal part of the effective potential as well
as in the on-energy shell scattering amplitude the outer similarity
factor equals 1 independently of the size of the momentum transfer.  In
other words, one cannot see the outer similarity factor in the physical
scattering amplitude order $g^2$ and the only signs of the effective
nature of the potential are the form factors in the interaction
vertices.

If we use the interaction $g \bar \psi i\gamma_5 \vec \tau \psi \vec
\phi$ instead of $g \bar \psi \psi \phi$ in writing the initial
hamiltonian of Eq.  (3.3), the resulting effective potential corresponds
to the one-pion exchange between nucleons.  Since the formalism is not
limited to the nonrelativistic domain of the fermion momenta or to the
lowest order perturbation theory, one can investigate this type of
potentials in a wide range of applications in meson-baryon and
quark-pion physics.

There exists a possibility that the similarity flow of hamiltonians may
lead to growth of coupling constants for small width.  The outer
similarity factor reduces the strength of the effective interactions
when $\lambda$ decreases.  Effective hamiltonians with small width may
have the same bound state eigenvalues as hamiltonians of similar
structure with large widths and small couplings if the effective
coupling constants become large for small widths.  The range of coupling
constants requires investigation in order to establish if the size of
coupling constants required in the meson-nucleon phenomenology can be
explained this way.

\vskip.3in
\centerline {\bf Fermion-anti-fermion interaction}
\vskip.1in

The fermion-anti-fermion interaction order $g^2$ satisfies a
differential equation which is analogous to Eq.  (3.40) but more terms
appear.  The operator subscripts must distinguish fermions and
anti-fermions and one has to include terms which result from the
annihilation channel.  The fermion-anti-fermion effective interaction
term is

$$ {\cal G}_{1\bar 1 \bar 1 1 \lambda}
= \int [P] {1 \over P^+} \sum_{\sigma_1 \sigma_2 \sigma_3 \sigma_4}
\int [x\kappa][y\rho] \, g_{1\bar 1 \bar 1 1 \lambda} \,
b^\dagger_{xP+\kappa \sigma_1} \, d^\dagger_{(1-x)P-\kappa \sigma_3}
\, d_{(1-y)P-\rho \sigma_4} \, b_{yP+\rho \sigma_2}
. \eqno(3.73)$$

\noindent Note the change of order of the spin numbering and momentum
assignments in comparison to Eq.  (3.42) for fermions.  The new order
results from the operator ordering including anti-fermions as defined in
Section 2.a.  Momenta $k_1$ and $k_2$ are used for fermion and $k_3$ and
$k_4$ for anti-fermion operators with even subscripts for annihilation
operators and odd subscripts for creation operators.

There are three terms contributing to the derivative of $g_{1\bar 1 \bar
1 1 \lambda}$ with respect to $\lambda$:  one due to the annihilation
channel and two due to the exchange of a boson.  One of the latter two
contributions results from the emission of the boson by the fermion and
absorption by the anti-fermion and the other one from the emission by
the anti-fermion and absorption by the fermion.  In each of the terms
there are two similarity functions with different arguments.  We have

$$ { d g_{1\bar 1 \bar 1 1 \lambda} \over d \lambda} =
S_1 g\bar u_1 v_3 \, g \bar v_4 u_2 \, r_{11} r_{13} r_{14} r_{12}
{1 \over P^+} $$
$$ -\left\{
 S_2 r_{21} r_{2512} r_{24} r_{2534} {\theta(y-x) \over (y-x)P^+}
+S_3 r_{32} r_{3512} r_{33} r_{3534} {\theta(x-y) \over (x-y)P^+}
    \right\} g\bar u_1 u_2 \, \, g \bar v_4 v_3 \, .
\eqno(3.74) $$

\noindent The inner similarity factors are

$$ S_i = f(z^2_{i2}) {[-f(z^2_{i2})]'\over \Delta E_{i2}} -
{[-f(z^2_{i1})]' \over \Delta E_{i1}} f(z^2_{i2}) . \eqno (3.75) $$

\noindent Equation (3.75) is similar to Eq.  (3.44) (the subscript
$\lambda$ is skipped for clarity).  The second subscript of the
arguments of the similarity function $f$ denotes the vertex, i.e. 1
stands for the vertex where the boson was annihilated and 2 stands for
the vertex where the boson was created.  In Eq.  (3.74), the fermion
regularization factors first subscript is the same as the corresponding
inner similarity factor subscript (i.e. the subscript of $S$) and the
second subscript is the same as the corresponding fermion momentum
subscript.  The boson regularization factors are distinguished by the
subscript $5$ following the convention from Eqs.  (2.50).  Their first
subscript is also the same as the corresponding inner similarity factor
subscript.  Last two subscripts of the boson regularization factors
equal subscripts of the fermion momenta from the vertex where the boson
regularization factor originated.  Arguments of the regularization
factors have the same subscripts as the regularization factors
themselves, i.e.  $r_i = r(\epsilon e_i /\Lambda^2)$.  The daughter
energies in the arguments are calculated according to the rules given in
Eqs.  (2.47) to (2.50).  We give the results below for completeness.
The same arguments will appear in all theories of physical interest.

$$e_{11} = {\kappa^{\perp \, 2} + m^2 \over x} \, . \eqno(3.76.a) $$

$$e_{13} = {\kappa^{\perp \, 2} + m^2 \over 1-x} \, . \eqno(3.76.b) $$

$$e_{14} = {\rho^{\perp \, 2} + m^2 \over y} \, . \eqno(3.76.c) $$

$$e_{12} = {\rho^{\perp \, 2} + m^2 \over 1-y} \, . \eqno(3.76.d) $$

$$ e_{21} = {\kappa^{\perp \, 2}_{212} + m^2 \over x_{212}} \, .
\eqno(3.77.a)$$

$$ e_{2512} = {\kappa^{\perp \, 2}_{212} + \mu^2 \over 1-x_{212}} \, .
\eqno(3.77.b)$$

$$ x_{212} = {x \over y}        \, . \eqno(3.77.c)$$

$$ \kappa^{\perp}_{212} = \kappa^\perp - x_{212} \rho^\perp \, .
\eqno(3.77.d)$$

$$ e_{24} = {\kappa^{\perp \, 2}_{234} + m^2 \over x_{234}} \, .
\eqno(3.77.e)$$

$$ e_{2534} = {\kappa^{\perp \, 2}_{234} + \mu^2 \over 1-x_{234}} \, .
\eqno(3.77.f)$$

$$ x_{234} = {1-y \over 1-x}        \, . \eqno(3.77.g)$$

$$ \kappa^{\perp}_{234} = -\rho^\perp + x_{234} \kappa^\perp \, .
\eqno(3.77.h)$$

$$ e_{32} = {\kappa^{\perp \, 2}_{312} + m^2 \over x_{312}} \, .
\eqno(3.78.a)$$

$$ e_{3512} = {\kappa^{\perp \, 2}_{312} + \mu^2 \over 1-x_{312}} \, .
\eqno(3.78.b)$$

$$ x_{312} = {y \over x}        \, . \eqno(3.78.c)$$

$$ \kappa^{\perp}_{312} = \rho^\perp - x_{312} \kappa^\perp \, .
\eqno(3.78.d)$$

$$ e_{33} = {\kappa^{\perp \, 2}_{334} + m^2 \over x_{334}} \, .
\eqno(3.78.e)$$

$$ e_{3534} = {\kappa^{\perp \, 2}_{334} + \mu^2 \over 1-x_{334}} \, .
\eqno(3.78.f)$$

$$ x_{334} = {1-x \over 1-y}        \, . \eqno(3.78.g)$$

$$ \kappa^{\perp}_{334} = -\kappa^\perp + x_{334} \rho^\perp \, .
\eqno(3.78.h)$$

\noindent Arguments of the similarity functions and energy denominators
which appear in Eq.  (3.75) are calculated according to the rules given
in Eqs.  (2.12) to (2.19) and (2.24) to (2.26).  The results are
universal for all one-particle-exchange two-particle interactions and
are given below for completeness.

$$ \Delta {\cal M}^2_{11} = (k_1 + k_3)^2_\mu - (k_{1m} + k_{3m})^2$$
$$   = \mu^2 - e_{11} - e_{13} \, . \eqno(3.79.a) $$

$$ \Sigma {\cal M}^2_{11} = - \Delta {\cal M}^2_{11} + 2\mu^2
\, . \eqno(3.79.b) $$

$$ \Delta E_{11} = \Delta {\cal M}^2_{11}/P^+
\, . \eqno(3.79.c) $$

$$ \Delta {\cal M}^2_{12} = (k_{2m} + k_{4m})^2 - (k_2 + k_4)^2_\mu$$
$$   = e_{14} + e_{12} - \mu^2  \, . \eqno(3.79.d) $$

$$ \Sigma {\cal M}^2_{12} = \Delta {\cal M}^2_{12} + 2\mu^2
\, . \eqno(3.79.e) $$

$$ \Delta E_{12} = \Delta {\cal M}^2_{12}/P^+
\, . \eqno(3.79.f) $$

$$ \Delta {\cal M}^2_{21} = (k_{2534\mu}+ k_{4m})^2 - k_{3m}^2 $$
$$   =  e_{2534} + e_{24} - m^2  \, . \eqno(3.80.a) $$

$$ \Sigma {\cal M}^2_{21} = \Delta {\cal M}^2_{21} + 2m^2
\, . \eqno(3.80.b) $$

$$ \Delta E_{21} = \Delta {\cal M}^2_{21}/(1-x)P^+
\, . \eqno(3.80.c) $$

$$ \Delta {\cal M}^2_{22} = k_{2m}^2 - (k_{2512\mu} + k_{1m})^2 $$
$$   = m^2 - e_{2512} - e_{21} \, . \eqno(3.80.d) $$

$$ \Sigma {\cal M}^2_{22} = - \Delta {\cal M}^2_{22} + 2\mu^2
\, . \eqno(3.80.e) $$

$$ \Delta E_{22} = \Delta {\cal M}^2_{22}/yP^+
\, . \eqno(3.80.f) $$

$$ \Delta {\cal M}^2_{31} = (k_{3512\mu}+ k_{2m})^2 - k_{1m}^2 $$
$$   =  e_{3512} + e_{32} - m^2 \, . \eqno(3.81.a) $$

$$ \Sigma {\cal M}^2_{31} = \Delta {\cal M}^2_{31} + 2m^2
\, . \eqno(3.81.b) $$

$$ \Delta E_{31} = \Delta {\cal M}^2_{31}/xP^+
\, . \eqno(3.81.c) $$

$$ \Delta {\cal M}^2_{32} = k_{4m}^2 - (k_{3534\mu} + k_{3m})^2 $$
$$   = m^2 - e_{3534} - e_{33} \, . \eqno(3.81.d) $$

$$ \Sigma {\cal M}^2_{32} = - \Delta {\cal M}^2_{32} + 2\mu^2
\, . \eqno(3.81.e) $$

$$ \Delta E_{32} = \Delta {\cal M}^2_{32}/(1-y)P^+
\, . \eqno(3.81.f) $$

\noindent In all cases, the arguments of the similarity functions are
given by Eq.  (2.18), i.e.  $ z_i = {\Delta {\cal M}^2_i /( \Sigma {\cal
M}^2_i + \lambda^2) }$ for all subscripts appearing in Eq.  (3.75).

The same reasoning is used to integrate Eq.  (3.74) as in the case of
Eq.  (3.65) for the fermion-fermion interaction.  For the similarity
function $f$ approaching the $\theta$-function with $u_0 = 1/4$ in Eq.
(2.20), we have $f(z^2_i) = \theta(\lambda^2 + 2m_i^2 - |\Delta {\cal
M}^2_i|)$ with $m_i^2 = \mu^2$ in the first, and $m_i^2 = m^2$ in the
second and third inner similarity factors in Eq.  (3.74).

Integration of Eq. (3.74) gives

$$ g_{1\bar 1 \bar 1 1 \lambda} = g_{1\bar 1 \bar 1 1 \epsilon}
+ c_1 \, g\bar u_1 v_3 \, g \bar v_4 u_2 \, r_{11} r_{13} r_{14}
r_{12}$$
$$ +\left[
 c_2 \, r_{21} r_{2512} r_{24} r_{2534} \, \theta(y-x)
+c_3 \, r_{32} r_{3512} r_{33} r_{3534} \, \theta(x-y)
   \right]\, g\bar u_1 u_2 \, \, g \bar v_4 v_3 \, , \eqno(3.82.a) $$

\noindent where the coefficients are,

$$ c_1 =
  {\theta_{12-11} \theta_{12-\lambda} \over |\Delta {\cal M}^2_{12}|}
+ {\theta_{11-\lambda} \theta_{11-12} \over |\Delta {\cal M}^2_{11}|}
\, , \eqno(3.82.b) $$

$$ c_2 =
  {y \theta_{22-21} \theta_{22-\lambda} \over (y-x)|\Delta {\cal
M}^2_{22}|}
+ {(1-x)\theta_{21-\lambda} \theta_{21-22} \over (y-x)|\Delta {\cal
M}^2_{21}|}
\, , \eqno(3.82.c) $$

$$ c_3 =
  {(1-y) \theta_{32-31} \theta_{32-\lambda} \over (x-y)|\Delta {\cal
M}^2_{32}|}
+ {x \theta_{31-\lambda} \theta_{31-32} \over (x-y)|\Delta {\cal
M}^2_{31}|}
\, . \eqno(3.82.d) $$

\noindent The symbols for $\theta$-functions have the following meaning.
$\theta_{i-j} = \theta (|\Delta {\cal M}^2_i| - |\Delta {\cal M}^2_j|)$
and $\theta_{i-\lambda} = \theta (|\Delta {\cal M}^2_i| - 2m^2_i -
\lambda^2)$ with $m^2_i$ equal $\mu^2$ in $c_1$ and $m^2$ in $c_2$ and
$c_3$.

The next step is the construction of the interaction $F_\lambda\left[
G_{1\bar 1 \bar 1 1 \lambda}\right]$ from ${\cal G}_{1\bar 1 \bar 1 1
\lambda}$ of Eq.  (3.73) using Eqs.  (2.8) and (2.9).

Then, one has to find out if matrix elements of $F_\lambda\left[
G_{1\bar 1 \bar 1 1 \lambda}\right]$ between finite free invariant mass
states have a limit when $\epsilon \rightarrow 0$.  Stated differently,
one checks if the existence of the limit requires the initial value of
$g_{1\bar 1 \bar 1 1 \epsilon}$ to differ from zero to cancel potential
divergences in the limit.  Following the same steps as in the case of
Eqs.  (3.66) and (3.70), one can check that no divergences arise.
Therefore, $g_{1\bar 1 \bar 1 1 \epsilon} = 0$.

The final answer for the effective fermion-anti-fermion interaction is

$$ H_{1\bar 1 \bar 1 1 \lambda} = F_\lambda\left[ G_{1\bar 1 \bar 1 1
\lambda}\right] $$
$$ = \int [P] {1 \over P^+} \sum_{\sigma_1 \sigma_2 \sigma_3 \sigma_4}
\int [x\kappa][y\rho] \, g_{1\bar 1 \bar 1 1 \lambda}
f(z^2_{1\bar 1 \bar 1 1 \lambda}) \,
b^\dagger_{\lambda xP+\kappa \sigma_1} \,
d^\dagger_{\lambda(1-x)P-\kappa \sigma_3}
\, d_{\lambda (1-y)P-\rho \sigma_4} \, b_{\lambda yP+\rho \sigma_2}
, \eqno(3.83.a)$$

\noindent where

$$ g_{1\bar 1 \bar 1 1 \lambda} = c_1\, g\bar u_1 v_3 \, g \bar v_4 u_2
\,
   + \left[ c_2 \, \theta(y-x) + c_3 \, \theta(x-y) \right] \,
   g\bar u_1 u_2 \, \, g \bar v_4 v_3 \, . \eqno(3.83.b) $$

\noindent In terms of the fermion momenta,

$$ c_1 = {\theta ( s - 3\mu^2 - \lambda^2) \over s - \mu^2 } \, ,
\eqno(3.83.c) $$

\noindent with $ s = max( {\cal M}^2_{13}, {\cal M}^2_{24} )$, and

$$ c_2 \theta(y-x) + c_3 \theta(x-y) =
{\theta_{a-b} \theta_{a-\lambda} \over a} +
{\theta_{b-a} \theta_{b-\lambda} \over b} \, , \eqno(3.83.d)$$

\noindent with $ a = \mu^2 - q_{12}^2 $, $ b = \mu^2 - q_{34}^2 $,
$\theta_{a-b} = 1- \theta_{b-a} = \theta (m_{xy} a - m_{1-x1-y} b) $,
$\theta_{a-\lambda} = \theta \left[ m_{xy} a - |x-y|(2m^2 + \lambda^2)
\right] $, $\theta_{b-\lambda} = \theta \left[ m_{1-x1-y} b - |x-y|(2m^2
+ \lambda^2)\right] $, $m_{xy} = max(x,y)$ and $m_{1-x1-y} =
max(1-x,1-y)$.  The argument of the outer similarity factor in Eq.
(3.83.a), i.e.  $ z_{1\bar 1 \bar 1 1 \lambda}$, is equal to
$z_{22\lambda}$ from Eq.  (3.70.h).  Note that Eqs.  (3.83.a-d) provide
the generalization of Eqs.  (3.70.a-h) to the case of effective
interactions of distinguishable fermions.

When $\lambda^2$ is reduced below $4m^2 - 3\mu^2$, the internal
similarity factor in the annihilation term stays equal 1 independently
of the value of $\lambda$.  The effective interaction term provides the
full contribution of the annihilation channel to the
fermion-anti-fermion scattering amplitude of order $g^2$.  The
fermion-anti-fermion-boson term in the effective hamiltonian which could
contribute acting twice in the scattering is zero for such low values of
$\lambda^2$ because the mass gap between the boson and the fermion pair
is larger than $\lambda$ allows.

The internal similarity factor in the exchange term becomes equal 1
independently of $\lambda$ only when $\lambda^2$ becomes smaller than
$-2m^2 + 2 m\mu + \mu^2$.  The lower bound on $\lambda^2$ is $ - m^2 - (
m + \mu )^2$ (see the discussion of Eqs.  (2.18) and (2.20)).  In the
lower bound region, the effective boson emission and absorption vanish
and the exchange interaction term provides the full scattering amplitude
due to the one-boson exchange.  The amplitude is equal to the standard
result on-shell where the outer similarity factor equals 1.

If the boson mass is much smaller than the fermion mass the fermion
energies for small momenta are quadratic functions of momentum while the
boson energy is a linear function of momentum.  Therefore, for
sufficiently small momenta, the boson energy is large in comparison to
the fermion kinetic energies and their changes.  Thus, the
one-boson-exchange interaction is mediated by a relatively high energy
intermediate state.  Consequently, it is contained in a potential term
in the effective hamiltonian.

For small $\lambda$, the effective hamiltonian contains potentials which
are equal to standard scattering amplitudes in the Born approximation.
The potentials differ from the Born amplitudes off-shell in a unique way
which is dictated by principles of the hamiltonian quantum mechanics and
the similarity renormalization group:  the outer similarity factor
reduces the strength of the interaction off-energy-shell.  In the
light-front dynamics the off-shellness is measured in terms of the free
invariant mass.

Outside the lower bound region for $\lambda$ the scattering amplitudes
obtain also contributions from the effective interactions which change
the number of bosons by one in the transition through the intermediate
states.  Analysis of Eq.  (3.71) in application to the
fermion-anti-fermion scattering follows the same steps as for the
fermion-fermion scattering in the previous Section.  The resulting
on-shell scattering amplitude is independent of $\lambda$.  The
amplitude is equal to the well known perturbative result in Yukawa
theory to order $g^2$.

\vskip.3in
{\bf 3.c QED}
\vskip.1in
\nopagebreak

This Section describes calculations of the effective mass squared term
for photons, the effective mass squared term for electrons and the
effective interaction between electrons and positrons in QED.  The
calculated terms are order $e^2$.

The initial expression which we use to calculate the renormalized
hamiltonian of QED is obtained from the lagrangian ${\cal L} = -{1 \over
4} F^{\mu \nu} F_{\mu \nu} + \bar \psi (i\not\!\!D - m)\psi$ by the
procedure of evaluating the energy-momentum tensor $T^{\mu \nu}$ and
integrating $T^{+-}$ over the light-front.  \cite{Y} We have

$$ H_{QED} = \int dx^- d^2 x^\perp \left[ \bar \psi_m \gamma^+
{-\partial^{\perp 2} + m^2 \over 2 i\partial^+} \psi_m -
{1\over 2} A^\nu_0 \partial^{\perp 2} A_{0\nu} \right.$$
$$ \left.  + e \bar \psi_m \not\!\!A_0 \psi_m + e^2 \bar \psi_m \not\!\!A_0
{\gamma^+ \over 2i\partial^+} \not\!\!A_0 \psi_m + {e^2 \over 2} \bar
\psi_m \gamma^+ \psi_m { 1 \over (i\partial^+)^2} \bar \psi_m \gamma^+
\psi_m \right]_{x^+=0} \, ,  \eqno(3.84) $$

\noindent where $\psi_m$ is a free fermion field with mass $m$ and
$A^\nu_0$ is a free massless photon field with $A^+_0 = 0$.

We replace fields $\psi_m(x)$ and $A^\nu_0(x)$ for $x^+=0$ by the
Fourier superpositions of creation and annihilation operators, we order
the operators in all terms and we drop terms containing divergent
integrals which result from contractions.  This is done in the same way
as in the Yukawa theory but more terms need to be considered.  Then, we
introduce the regularization factors.

The ultraviolet regularization factors are already familiar and the same
as in the Yukawa theory.  The additional regularization is required due
to the infrared singularities.  Photons have diverging polarization
vectors when their $+$-momentum approaches 0. The corresponding seagull
term, i.e. the 5th term in Eq.  (3.84), is diverging too.  We introduce
the infrared regularization factors $(1 + \delta/x)^{-1}$ as described
in Section 2.b.

We also introduce a photon mass term $\mu^2_\epsilon = \mu^2_\delta$ by
adding it to $ - \partial^{\perp \, 2}$ in the second term in Eq.
(3.84).  A fixed value of $\mu_\delta$ leads to the conclusion in
perturbation theory that the photon eigenstates have masses equal to
$\mu_\delta$ when the charge $e$ approaches $0$.  Therefore, we will be
forced to consider the limit $\mu_\delta \rightarrow 0$ in order to
discuss physical photons to order $e^2$.  Also, the nonzero mass squared
term for photons leads to additional divergences when $\delta
\rightarrow 0$ and the limit of $\mu_\delta \rightarrow 0$ removes
those.

The infrared finiteness of QED suggests that physical results in our
approach should be independent of $\mu_\delta$ when it is sufficiently
small.  We introduce the photon mass $\mu_\delta$ and investigate the
limit $\mu_\delta \rightarrow 0$.  The second order calculations in this
paper lead to results which are independent of $\mu^2_\delta$ when it
tends to zero.

\vskip.3in
\centerline {\bf Photon mass squared}
\vskip.1in

The same procedure from Section 2 which led to Eqs.  (3.4) and (3.5) in
Yukawa theory leads in QED to

$$ {\cal G}_{1\, photon \,\lambda} = \sum_\sigma \int[k] { k^{\perp 2} +
\mu^2_\lambda \over k^+ } a^\dagger_{k\sigma} a_{k\sigma} \, . \eqno
(3.85)
$$

\noindent No correction arises to the term $k^{\perp 2} /k^+$ because
our regularization preserves the kinematical symmetries of light-front
dynamics.

A new feature in comparison to the Yukawa theory is the polarization of
photons.  With the kinematical symmetries explicitly preserved, only
terms diagonal in the photon polarization emerge.  For example, terms
proportional to $k^i \varepsilon^i_{\sigma_1} \, k^j
\varepsilon^j_{\sigma_2}$ with $\sigma_1 \neq \sigma_2$ cannot appear
because the regularization and similarity factors do not introduce
dependence on the photon momentum.  Note that such terms are allowed by
the power counting.  \cite{W3}

The net result of the photon self-interaction is an effective photon
mass squared term which is independent of the photon momentum but varies
with the effective hamiltonian width $\lambda$.  One obtains more
complicated results for the effective photon free energy if the
regularization or similarity factors violate kinematical symmetries of
light-front dynamics.  \cite{W3}

The dependence of $\mu^2_\lambda$ on $\lambda$ is determined to order
$e^2$ by the equation

$$ {d \mu^2_\lambda \over d\lambda} \delta^{\sigma_1 \sigma_2} = e^2
\int[x\kappa] {d f^2(z^2_\lambda) \over d\lambda} { Tr
\not\!\varepsilon^*_{k \sigma_1} (\not\!k_{1m} + m) \not\!\varepsilon
_{k \sigma_1} (\not\!k_{2m} - m) \over {\cal M}^2 - \mu^2_\delta}
r_{\epsilon }(x,\kappa) \,  \eqno (3.86) $$

\noindent and the initial condition at $\lambda = \infty$.  However, it
is also sufficient to know the effective mass squared at any single
value of $\lambda$ to determine its value at other values of $\lambda$
using Eq.  (3.86).  The initial condition at $\lambda = \infty$ is
distinguished only because it provides connection with standard
approaches based on the local lagrangian for electrodynamics.

In Eq.  (3.86), ${\cal M}^2 = (\kappa^{\perp \, 2} + m^2)/x(1-x)$,
$\Delta {\cal M}^2 = {\cal M}^2 - \mu^2_\delta$ and $\Sigma {\cal M}^2 =
{\cal M}^2 + \mu^2_\delta$ so that $z_\lambda = ({\cal M}^2 -
\mu^2_\delta) /({\cal M}^2 + \mu^2_\delta + \lambda^2)$.  In the limit
of Eq.  (3.8), we have $f^2(z^2_\lambda) = \theta(\lambda^2 +
3\mu^2_\delta - {\cal M}^2)$.  In fact, Eq.  (3.86) is free from
infrared singularities and we could skip the introduction of
$\mu^2_\delta$ by letting it go to zero at this point.  However, the
systematic approach defines the hamiltonian of QED including the
infrared regulator mass for photons and we can keep it here for
illustration.  The regularization factor $r_\epsilon (x,\kappa)$ in Eq.
(3.86) is the same as in Eq.  (3.9) in Yukawa theory because only
fermion regularization factors enter Eq.  (3.86), according to Eqs.
(2.47) to (2.49), and these factors are the same in both theories.

Evaluation of the spin factor gives

$$ {d \mu^2_\lambda \over d\lambda} = e^2 \int[x\kappa] \, {d
f^2(z^2_\lambda) \over d\lambda} \, { 2 {\cal M}^2 - 4 \kappa^{\perp \,
2} \over {\cal M}^2 - \mu^2_\delta} \, r_{\epsilon }(x,\kappa) , \eqno
(3.87) $$

\noindent which is the QED analog of Eq.  (3.6) from Yukawa theory.

Integration of Eq.  (3.87) is carried out through the same steps as in
the Yukawa theory.  We can use Eq.  (3.22) to calculate the effective
photon mass squared $\mu^2_\lambda$ knowing its value $\mu^2_0$ at some
value of $\lambda = \lambda_0$.

The value of $\mu^2_0$ is found by requesting that the effective
hamiltonian eigenvalues for photon states contain the physical photon
mass $\tilde \mu$, expected to be 0. However, solving the eigenvalue
equation to second order in the coupling constant $e$ through the same
steps as in the case of mesons in Yukawa theory in Eqs.  (3.23) to
(3.25), leads to the physical photon mass ${\tilde \mu} = \mu_\delta$.
$\mu_\delta$ is small but finite and it is considered, in terms of
powers of $e$, to be of order $e^0 = 1$ when $e \rightarrow 0$.

In the $\theta$-function limit for the similarity function $f$ with $u_0
= 1/4$ one obtains

$$ \mu^2_0 = \mu^2_\delta + {\alpha \over 4\pi} \int_0^1dx \int_0^\infty
d\kappa^2 { 2 {\cal M}^2 - 4 \kappa^{\perp \, 2} \over x(1-x){\cal M}^2
- \mu^2_\delta} \theta(\lambda^2_0 + 3\mu^2_\delta -{\cal M}^2) + o(e^4)
\, , \eqno(3.88) $$

\noindent Thus, at other values of $\lambda$, the effective photon mass
squared is

$$ \mu^2_\lambda = \mu^2_\delta + {\alpha \over 4\pi} \int_0^1dx
\int_0^\infty d\kappa^2 { 2 {\cal M}^2 - 4 \kappa^{\perp \, 2} \over
x(1-x){\cal M}^2 - \mu^2_\delta} \theta(\lambda^2 + 3\mu^2_\delta -{\cal
M}^2) + o(e^4) \, . \eqno(3.89) $$

\noindent This result naturally depends on the infrared regularization
parameter $\mu^2_\delta$ but no singularity appears when this parameter
is set equal to zero.  For $ \lambda^2 + 3\mu^2_\delta \leq 4m^2$, where
$4m^2$ is the lowest possible free invariant mass squared for the two
intermediate fermions, the photon mass is independent of the hamiltonian
width $\lambda$ and it equals $\mu^2_\delta$.  For larger values of
$\lambda^2$, the effective photon mass grows with the width $\lambda$ so
that its larger value compensates effects of the interactions which
become active for the larger width.  The net result is that the photon
eigenstates have eigenvalues with masses squared equal $\mu^2_\delta$
independently of $\lambda$.  Finally, the result favored by experimental
data is obtained in the limit $\mu^2_\delta \rightarrow 0$ at the end of
the calculation.

\vskip.3in
\centerline {\bf Electron mass squared}
\vskip.1in

Electron and positron self-interactions through emission and
reabsorbtion of transverse photons result in the fermion free energy
terms of the form exactly the same in QED as in Eq.  (3.27) in Yukawa
theory.  However, the effective mass of electrons and positrons depends
on the width differently than in the case of Yukawa theory.  Instead of
Eq.  (3.28), one obtains now

$$ {d m^2_\lambda \over d\lambda} = e^2 \sum_{\tilde \sigma}
\int[x\kappa] {df^2(z^2_\lambda) \over d\lambda}
{ {\bar u}_{m\sigma k} \not\!\varepsilon_{\tilde k \tilde \sigma}
(\not\!p_m + m)
\not\!\varepsilon^*_{\tilde k \tilde \sigma} u_{m\sigma k} \over
{\cal M}^2 - m^2} \, r_{\epsilon \delta}(x,\kappa) \, ,\eqno (3.90) $$

\noindent where ${\tilde k} = (k - p)_0$, $p^+ = xk^+$, $p^\perp =
xk^\perp + \kappa^\perp$, ${\cal M}^2 = (m^2 + \kappa^2)/ x  +
(\mu_\delta^2 + \kappa^2)/(1-x)$, $\Delta {\cal M}^2 = {\cal M}^2 -
m^2$, $\Sigma {\cal M}^2 = {\cal M}^2 + m^2$ and $z_\lambda = \Delta
{\cal M}^2 /(\Sigma {\cal M}^2 + \lambda^2)$.  The regularization factor
of Eq.  (2.49) for the intermediate particles and the infrared regulator
for the intermediate photon, as given at the end of Section 2.b, imply

$$ r_{\epsilon \delta}(x,\kappa) = \left[ 1+{\epsilon\over\Lambda^2}
{\cal M}^2 + \left({\epsilon\over\Lambda^2}\right)^2{\kappa^2 +m^2 \over
x}{\kappa^2 +\mu^2_\delta \over 1-x}\right]^{-2} \left(1 + {\delta \over
1-x} \right)^{-2} \, .\eqno(3.91) $$

\noindent The sum over photon polarizations in Eq.  (3.90) produces the
well known expression

$$ \sum_{\tilde \sigma} \varepsilon^\alpha_{\tilde k \tilde \sigma}
\varepsilon^{* \beta}_{\tilde k \tilde \sigma} = -g^{\alpha \beta} +
{{\tilde k}^\alpha g^{+\beta} + g^{+\alpha} {\tilde
k}^\beta \over {\tilde k}^+} \,\, , \eqno(3.92) $$

\noindent and the spin factor in Eq. (3.90) is

$$ {\bar u}_{m\sigma k} \gamma_\alpha (\not\!p_m + m) \gamma_\beta
u_{m\sigma k} \left[-g^{\alpha \beta} + {{\tilde k}^\alpha g^{+\beta} +
g^{+\alpha} {\tilde k}^\beta \over {\tilde k}^+} \right] = {2\over
x}\left[ (1-x)^2 m^2 + \kappa^2 {1 + x^2 \over (1-x)^2}\right] \, .
\eqno(3.93)$$

\noindent The new feature of this expression, in comparison to the
Yukawa theory, is the divergence for $x \rightarrow 1$, i.e. where the
photon longitudinal momentum approaches 0.

The rate of change of the electron mass term versus the effective
hamiltonian width in the $\theta$-function limit for the similarity
function with $u_0 = 1/4$ is

$$ {d m^2_\lambda \over d\lambda^2} = {\alpha \over 4\pi} \int_0^1 dx
\int_0^\infty du \,\, \delta(3m^2 + \lambda^2 - {\cal M}^2) \, {m^2
{2(1-x)^2 \over x} + u {2(1 + x^2) \over 1-x} \over {\cal M}^2 - m^2} \,
r_{\epsilon \delta}(x,\kappa) \, ,\eqno (3.94.a) $$

\noindent where $u = \kappa^2/x(1-x)$,

$$ r_{\epsilon \delta}(x,\kappa) = \left[ 1+{\epsilon\over\Lambda^2}
{\cal M}^2 + \left({\epsilon\over\Lambda^2}\right)^2 \left[(1-x)u + {m^2
\over x}\right]\left[x u + {\mu^2_\delta \over 1-x}\right]\right]^{-2}
\left(1 + {\delta \over 1-x} \right)^{-2} \, ,\eqno(3.94.b) $$

\noindent and

$$ {\cal M}^2 = u + {m^2 \over x} + {\mu_\delta^2 \over 1-x}
\, .\eqno(3.94.c) $$

The divergence structure of the effective electron mass in Eq.  (3.94.a)
is obscured by the fact that the whole effective mass term is merely a
number dependent on $\lambda$ while three cutoff and regularization
parameters appear in the integral:  $\epsilon$, $ \delta$ and
$\mu_\delta$.  The only available condition is that the effective
electron mass should have a limit when $\epsilon \rightarrow 0$.
However, this condition has to be satisfied without generating
divergences in the physical electron mass (i.e. in the electron
eigenvalue energy) when we remove the infrared regularization.  Since
other contributions to the physical electron mass may diverge as
$\delta$ or $\mu_\delta$ tend to 0, and only the sum is finite in the
limit, one needs to keep track of the infrared structure in defining the
$\epsilon$-independent (i.e. ultraviolet finite) part of the
counterterm.

The divergences due to $\epsilon \rightarrow 0$, $\delta \rightarrow 0$
and $\mu_\delta \rightarrow 0$, are not resolved in the single mass
constant.  Many elements of a complete analysis overlap in producing the
final answer and many simplifications are possible.  We will proceed in
this Section with a simplified analysis.  A more extended analysis will
be required for other hamiltonian terms where the outcome of the
procedure is not reduced to finding only one number in the effective
interaction.  For example, in the electron-positron interaction term
the external momenta of fermions introduce a whole range of additional
parameters. That case will be illustrated in the next Section.

The $\delta$-function under the integral on the right-hand side of Eq.
(3.94.a) forces ${\cal M}^2$ to be equal $\lambda^2 + 3m^2$.  The
smallest possible value of $\cal M$ is $m + \mu_\delta$ and the negative
value of $\lambda^2 = (m + \mu_\delta)^2 - 3m^2 = - m^2 - ( m +
\mu_\delta )^2 + 2[ (m + \mu_\delta)^2 - m^2]$ is required to reach this
lower bound (see the comments about Eq.  (2.20)).  Below this bound the
right-hand side of Eq.  (3.94.a) vanishes, no emission or absorption of
photons by electrons is possible and the effective electron mass stays
constant.  The smallest possible value of $\lambda^2$ allowed by Eq.
(2.18) is $ - m^2 - ( m + \mu_\delta)^2$.  The difference between these
bounds vanishes when the photon mass goes to zero.

In the next Section about electron-positron interaction we will also
consider the case of the infinitesimal $u_0$ (see the discussion below
Eq.  (2.20)), which leads to $\delta(\tilde \lambda^2 + m^2 - {\cal
M}^2)$ in Eq.  (3.94.a), instead of $\delta(3m^2 + \lambda^2 - {\cal
M}^2)$.

For $\lambda^2 > ( m + \mu_\delta)^2 - 3m^2 $, the right-hand side of
Eq.  (3.94.a) is positive.  Therefore, the effective electron mass
squared term grows together with the width of the hamiltonian.  This
growth combines with the growing negative contributions of the
corresponding effective transverse photon emission and reabsorption so
that the physical electron mass is independent of the hamiltonian width.

For $\lambda^2$ close to the lower bound, $u$ must be close to 0 and $ x
\sim 1 - \mu_\delta/(m + \mu_\delta)$.  But ${\cal M}^2$ is limited and
determined by the value of $\lambda$.  Quite generally, as long as $\cal
M$ remains limited, $u$ and $\mu^2_\delta/(1-x)$ are limited.  The
invariant mass denominator in Eq.  (3.94.a) equals $\lambda^2 + 2m^2$.
The denominator is small only when $\lambda^2$ is close to the lower
bound.  Then, the integration range is small too.  The denominator, when
expressed in terms of $\lambda$, can be pulled out and put in front of
the integral.  The integration over $u$ sets $u = \tilde u(x) =
\lambda^2 + 3m^2 - m^2/x -\mu^2_\delta/(1-x)$ and forces the condition
$\tilde u(x) > 0$.  This condition implies the following limits on the
integration over $x$, provided $ \lambda^2 + 3m^2 > (m + \mu_\delta)^2 $
since otherwise the integral is 0.

$$ x_0 - \Delta x < x < x_0 + \Delta x \, , \eqno(3.95.a) $$

$$ x_0 = {1\over 2}\left( 1 + {m^2 - \mu_\delta^2 \over \lambda^2 +
3m^2}\right) \, , \eqno(3.95.b) $$

$$ \Delta x = \sqrt{ x_0^2 - {m^2\over \lambda^2 + 3m^2}} \, .
\eqno(3.95.c) $$

\noindent Within these limits, $\tilde u(x)$ varies from the minimal
value of 0 at the lower bound $x_0 - \Delta x$ through a maximum of
$\lambda^2 + 3m^2 - (m + \mu_\delta)^2$ at $x = m/(m + \mu_\delta)$ to
the minimal value of 0 again at the upper bound $x_0 + \Delta x$.  In
the case of infinitesimal $u_0$, one replaces $\lambda^2 + 2 m^2$ in
these formulae by $\tilde \lambda^2$.  Eq.  (3.94.a) reads

$$ {d m^2_\lambda \over d\lambda^2} = {\alpha \theta [3m^2 + \lambda^2 -
(m + \mu_\delta)^2] \over 4\pi(\lambda^2 + 2m^2)} \int_{x_0 - \Delta
x}^{x_0 + \Delta x} dx \left[m^2 {2(1-x)^2\over x} + {\tilde u}(x) {2(1
-+ x^2) \over 1-x}\right] \, r_{\epsilon \delta}(x,\tilde u(x)) \,
, \eqno (3.96.a) $$

\noindent where

$$ r_{\epsilon \delta}(x, \tilde u(x) ) = \left[
1+{\epsilon\over\Lambda^2} (3m^2 + \lambda^2) +
\left({\epsilon\over\Lambda^2}\right)^2 \left[(1-x) \tilde u(x) + {m^2 \over
x}\right] \left[x \tilde u(x) + {\mu^2_\delta \over 1-x}\right] \right]^{-2}$$
$$ \times \left(1 + {\delta \over 1-x} \right)^{-2} \, .\eqno(3.96.b) $$

The upper limit of integration over $x$ for $\mu^2_\delta$ much smaller
than $m^2$ and $ 2 m^2 + \lambda^2$ , equals $ 1 - \mu^2_\delta/(2m^2 +
\lambda^2)$ and approaches 1 when $\mu_\delta \rightarrow 0$.  For $x$
close to 1, the factor $(1-x)^{-1}$ in the square bracket of the
integrand is large and leads to a logarithmic dependence of the integral
on the upper integration limit.  The logarithm would become infinite for
$\mu_\delta \rightarrow 0$ if $\delta$ were equal 0. Therefore, the
limit $\mu_\delta \rightarrow 0$ is sensitive to the presence of the
regularization factor with $\delta \neq 0$.  For the finite
$\mu_\delta$, the region of $ x \rightarrow 1$ is regulated by
$\epsilon$ and can be considered an ultraviolet limit.  A counterterm to
the diverging $\epsilon$ dependence could remove the divergence due to
$x \rightarrow 1$.  Then, a separate cutoff parameter $\delta$ would not
be needed but the resulting terms would diverge for $\mu^2_\delta
\rightarrow 0$.  For finite $\mu^2_\delta$, the derivative of the
electron mass with respect to $\lambda$ is finite.  For $\mu^2_\delta =
0$, the integrand in the region $x \sim 1$ is regulated solely by the
infrared regularization factor $[1 +\delta/(1-x)]^{-2}$ since the upper
limit of integration over $x$ is equal 1.

For finite $\lambda^2$ all three terms in ${\cal M}^2$, i.e.  $\tilde u
(x)$, $m^2/x$ and $\mu^2_\delta/(1-x)$ are limited.  Therefore, for
finite $\mu_\delta$, one can take the limit $\epsilon \rightarrow 0$ in
the integrand.  The factor $[1 - \delta/(1-x)]^{-2}$ remains and
additionally cuts off the integration region at $x \sim 1 - \delta$.
The ratio of $\delta$ to $\mu^2_\delta/(2m^2 +\lambda^2)$ determines the
size of contributions obtained from the upper range of integration over
$x$.  For finite fixed values of $\mu_\delta$, one can take the limit
$\delta \rightarrow 0$ and $\log{[(2m^2 +\lambda^2)/\mu^2_\delta]}$
appears in the answer.

The right-hand side of Eq.  (3.96.a) contains terms which behave for
large $\lambda$ as a constant, as $\lambda^{-2}$ with factors of
logarithms of $\lambda$ and as functions vanishing faster than
$\lambda^{-2}$.  One integrates Eq.  (3.96.a) over $\lambda$ from
$\lambda_0$ to infinity in order to express the effective electron mass
squared term at $\lambda_0$, denoted $m^2_0$, in terms of the initial
$m^2_\epsilon$.  Clearly, the integration over $\lambda$ would diverge
without the regularization factor which depends on $\epsilon$.  The
integration produces terms behaving as $\epsilon^{-1}$, $\log \epsilon$
and terms convergent in the limit $\epsilon \rightarrow 0$.
$m^2_\epsilon$ in the initial hamiltonian must be supplied with a
counterterm to subtract the diverging $\epsilon$-dependent terms in the
effective hamiltonians.

In summary, the infrared divergence due to $\mu_\delta \rightarrow 0$
and $\delta \rightarrow 0$ appears in the derivative of $m^2_\lambda$
with respect to $\lambda$.  Therefore, even if one requests that the
electron mass term is finite at some value of $\lambda$, the effective
masses of electrons in the neighboring hamiltonians with even slightly
different widths will diverge when $\mu_\delta$ and $\delta$ approach 0.
We have to abandon the requirement that the effective electron mass term
at any value of $\lambda$ remains finite when the infrared
regularization is removed.  The effective masses diverge in the limit
$\mu_\delta \rightarrow 0$ and $\delta \rightarrow 0$.  The only
condition we can fulfill through the ultraviolet renormalization is that
the effective electron masses for finite $\lambda$ are independent of
$\epsilon$.

Mathematical details of the effective electron mass term calculation are
more complicated than in the Yukawa theory because the infrared
regularization parameters are present.  Otherwise, the calculation is
essentially the same and we skip the description here.  We only stress
that the counterterm and the effective masses of electrons and positrons
depend on the infrared cutoffs and they diverge when the cutoffs are
being removed.

Thus, the effective electron mass squared term in the limit $\epsilon
\rightarrow 0$ is

$$ m^2_\lambda = m^2_0 + e^2 \int [x\kappa] \left[f^2(z^2_\lambda) -
f^2(z^2_{\lambda_0})\right] { m^2 {2(1-x)^2\over x} + {\kappa^2 \over
x(1-x)} {2(1 + x^2) \over 1-x} \over {\cal M}^2 - m^2} \left(1+{\delta
\over 1-x}\right)^{-2}  + o(e^4) \, .\eqno (3.97) $$

\noindent The finite term $m^2_0$ has a limit when $\epsilon \rightarrow
0$.  Its dependence on the infrared regularization is not displayed.
$m^2_0$ is found from a suitable renormalization condition.

The natural condition to be satisfied by $m^2_0$ is that the effective
hamiltonian of some width $\lambda$ has the electron eigenstates with
eigenvalues equal $(p^{\perp \, 2} + {\tilde m}^2)/p^+$, where $p$
denotes the electron momentum and $\tilde m$ is the physical electron
mass.  The eigenvalue equation for electrons can be solved in
perturbation theory in the same way as for bosons in Yukawa theory in
Eqs.  (3.23) - (3.26) and fermions in Eqs.  (3.35) - (3.39), or in QED
for photons in Eqs.  (3.88) - (3.89).  One obtains the condition

$$ {\tilde m}^2 = m^2_0 - e^2 \int [x\kappa]
f^2(z^2_{\lambda_0}) { m^2 {2(1-x)^2\over x} + {\kappa^2 \over
x(1-x)} {2(1 + x^2) \over 1-x} \over {\cal M}^2 - m^2} \left(1+{\delta
\over 1-x}\right)^{-2}  + o(e^4) \, ,\eqno (3.98) $$

\noindent and one can calculate $m_0$ from this condition.
Consequently,

$$ m^2_\lambda = {\tilde m}^2 + e^2 \int [x\kappa] f^2(z^2_\lambda) {
m^2 {2(1-x)^2\over x} + {\kappa^2 \over x(1-x)} {2(1 + x^2) \over 1-x}
\over {\cal M}^2 - m^2} \left(1+{\delta \over 1-x}\right)^{-2} + o(e^4)
\, ,\eqno (3.99) $$

\noindent and $m^2 = {\tilde m}^2 + o(e^2)$.

The physical electron mass is independent of the infrared regularization
because the regularization dependent $m^2_\lambda$ and the effective
emission and absorption of photons combine to the regularization
independent result.

\vskip.3in
\centerline {\bf Electron-positron interaction}
\vskip.1in

Calculation of the effective electron-positron interaction to order
$e^2$ is of interest as a way to derive the Coulomb force in quantum
electrodynamics - this interaction is responsible for the formation of
positronium.  Also, effective interactions between quarks and
anti-quarks in QCD have a number of similar features and the QED
calculation provides an introduction to the QCD case.

Generally speaking, the QED calculation of the effective
electron-positron interaction proceeds in the same way as in the case of
fermion-anti-fermion interaction in Yukawa theory except for three new
elements.

The first is that photons have polarization vectors which enter in the
vertex factors and introduce additional dependence on the exchanged
photon momentum.  This dependence leads to infrared divergences for
small longitudinal momenta of exchanged photons.

The second feature is that the infrared divergences require additional
regularization factors.  We use the parameter $\delta$ and the photon
mass squared $\mu_\delta \neq 0$.  The limits $\delta \rightarrow 0$ and
$\mu_\delta \rightarrow 0$ are generally understood as to be taken at
the end of a calculation of observables and not in the effective
hamiltonian itself.  However, it may also be possible to take the limits
in matrix elements of the hamiltonian between states which do not induce
infrared divergences, i.e. do not involve small $x$ photons.

The third feature is that the one-photon exchange interaction needs to
be combined with the 5th term from Eq.  (3.84) to obtain the standard
results for the electron-positron scattering in the Born approximation.
The 5th term from Eq.  (3.84) provides the initial condition for the
renormalization group flow of the effective hamiltonians.  In order
$e^2$ this term is only supplied with the outer similarity factor by the
operation $F_\lambda$.  It does not change in the flow beyond this
factor because it is order $e^2$ itself.  The initial condition provides
a contribution which is needed to obtain the Coulomb potential.  This is
a different situation than in Yukawa theory where no four-fermion
seagull interactions appeared and the one-meson exchange interaction was
sufficient to produce the Yukawa potential in the effective hamiltonians
of small widths.

The effective electron-positron interaction term has the analogous
structure as the fermion-anti-fermion interaction in Eq.  (3.73), i.e.

$$ {\cal G}_{1\bar 1 \bar 1 1 \lambda}
= \int [P] {1 \over P^+} \sum_{\sigma_1 \sigma_2 \sigma_3 \sigma_4}
\int [x\kappa][y\rho] \, g_{1\bar 1 \bar 1 1 \lambda} \,
b^\dagger_{xP+\kappa \sigma_1} \, d^\dagger_{(1-x)P-\kappa \sigma_3}
\, d_{(1-y)P-\rho \sigma_4} \, b_{yP+\rho \sigma_2}
\, . \eqno(3.100)$$

\noindent The coefficient function $g_{1 \bar 1 \bar 1 1 \lambda}$
of order $e^2$ satisfies the differential equation

$$ { d g_{1\bar 1 \bar 1 1 \lambda} \over d \lambda} = S_1
\sum_{\sigma_5} e\bar u_1 \not\!\varepsilon_{k_5 \sigma_5} v_3 \, e \bar
v_4 \not\!\varepsilon^*_{k_5 \sigma_5} u_2 \, r_{11} r_{13} r_{14} r_{12}
{1 \over P^+} $$
$$ - S_2 r_{21} r_{2512} r_{24} r_{2534} r_{3/5} r_{2/5}
 {\theta(y-x) \over (y-x)P^+}
\sum_{\sigma_5} e \bar v_4 \not\!\varepsilon_{k_5 \sigma_5} v_3
e \bar u_1 \not\!\varepsilon^*_{k_5 \sigma_5} u_2 \, $$
$$ - S_3 r_{32} r_{3512} r_{33} r_{3534} r_{1/5} r_{4/5}
{\theta(x-y) \over (x-y)P^+}
\sum_{\sigma_5} e \bar u_1 \not\!\varepsilon_{k_5 \sigma_5} u_2 \,
e \bar v_4 \not\!\varepsilon^*_{k_5 \sigma_5} v_3
\, , \eqno(3.101) $$

\noindent which is the QED analog of Eq.  (3.74) from Yukawa theory.
Notation is the same as in Eqs.  (3.74) to (3.81) with the exception
that $\mu^2$ is replaced by $\mu^2_\delta$.  The new elements are the
infrared regularization factors of Section 2.b, i.e.  $r_{3/5} = r(
k_3^+ \delta /k_5^+)$, $r_{2/5} = r( k_2^+ \delta /k_5^+)$, $r_{1/5} =
r( k_1^+ \delta /k_5^+)$, $r_{4/5} = r( k_4^+ \delta /k_5^+)$, and the
photon polarization vectors.  The sum over photon polarizations gives

$$ \sum_{\sigma_5} \varepsilon_{k_5 \sigma_5}^\alpha
\varepsilon^{* \beta}_{k_5 \sigma_5} =
- g^{\alpha \beta} + {k^\alpha_{50} g^{\beta +} + g^{\alpha +}
k^\beta_{50} \over k^+_\beta}
\, . \eqno(3.102)$$

\noindent The terms proportional to the four-vector $k_{50}$ can be
rewritten using the Dirac equation for free fermions of mass $m$.  For
example, in the second term on the right-hand side of Eq.  (3.101) we
have

$$ \bar u_1 \not\!k_{50} u_2 \, = \bar u_1 \left[ \not\!k_{2m}
- \not\!k_{1m} + {1\over 2} \gamma^+ [ (k_2 - k_1)^-_0 - k_{2m}^- +
k_{1m}^-] \right] u_2 \, = \bar u_1 \gamma^+ u_2 \, { -(k_{2m}-k_{1m})^2 \over
2(k_2^+ - k_1^+)} \, . \eqno(3.103) $$

\noindent Using similar relations for all vertex factors involved one
obtains

$$ { d g_{1\bar 1 \bar 1 1 \lambda} \over d \lambda} = S_1 \left[
-g^{\mu \nu} - g^{\mu +} g^{\nu +} {s_{13} + s_{24} \over 2 P^{+\,2} }
\right] e\bar u_1 \gamma_\mu v_3 \, e \bar v_4 \gamma_\nu u_2 \, r_{11}
r_{13} r_{14} r_{12} {1 \over P^+} $$
$$ - \left\{
     S_2 r_{21} r_{2512} r_{24} r_{2534}r_{3/5}
     r_{2/5}{\theta(y-x) \over (y-x)P^+}
  + S_3 r_{32} r_{3512} r_{33} r_{3534}r_{1/5}
     r_{4/5} {\theta(x-y) \over (x-y)P^+}
     \right\} $$
$$  \times \left[ -g^{\mu \nu} - g^{\mu +} g^{\nu +} {q^2_{12} +
q^2_{34}
     \over 2 (x-y)^2 P^{+\,2} } \right]
     e \bar u_1 \gamma_\mu u_2 \,  e \bar v_4 \gamma_\nu v_3 \,
\, , \eqno(3.104) $$

\noindent We use the notation $ s_{ij} = (k_i + k_j)^2 $ and
$ q^2_{ij} = (k_i - k_j)^2 $.

Integration of Eq.  (3.104) proceeds in the same way as in the case of
Eq.  (3.74).  The initial condition at $\lambda = \infty$ includes the
seagull term.

$$ H^{seagull}
= \int [P] {1 \over P^+} \sum_{\sigma_1 \sigma_2 \sigma_3 \sigma_4}
\int [x\kappa][y\rho] \, g^{seagull}_{1\bar 1 \bar 1 1} \,
b^\dagger_{xP+\kappa \sigma_1} \, d^\dagger_{(1-x)P-\kappa \sigma_3}
\, d_{(1-y)P-\rho \sigma_4} \, b_{yP+\rho \sigma_2}
\, , \eqno(3.105.a)$$

\noindent where

$$ g^{seagull}_{1\bar 1 \bar 1 1} =
    - \left[
           r_{21} r_{2512} r_{24} r_{2534} r_{3/5}
  r_{2/5} { \theta(y-x) \over (y - x)^2}  \right. $$
$$ \left. + r_{32} r_{3512} r_{33} r_{3534} r_{1/5}
           r_{4/5} { \theta(x-y) \over (x-y)^2} \right]
  e \bar u_1 \gamma_\mu u_2 \,  e \bar v_4 \gamma_\nu v_3 \,
  { g^{\mu+} g^{\nu +} \over P^{+\,2}}  $$
$$    + e\bar u_1 \gamma_\mu v_3 \,  \, e \bar v_4 \gamma_\nu u_2 \,
     r_{11} r_{13} r_{14} r_{12}
     { g^{\mu+} g^{\nu +} \over P^{+\,2}}
\, . \eqno(3.105.b)$$

The result of the integration of Eq.  (104) is

$$ g_{1\bar 1 \bar 1 1 \lambda} =
   g^{counterterm}_{1\bar 1 \bar 1 1 \epsilon}
   - c_1 e\bar u_1 \gamma^\mu v_3 \,  \, e \bar v_4 \gamma_\mu u_2 \,
         r_{11} r_{13} r_{14} r_{12} $$
$$ +\left[ -{1\over 2} c_1 (s_{13} + s_{24}) + 1 \right]
   {e\bar u_1 \gamma^+ v_3 \,  \, e \bar v_4 \gamma^+ u_2 \, \over P^{+ \, 2}}
         r_{11} r_{13} r_{14} r_{12} $$
$$ - \left[ c_2 r_{21} r_{2512} r_{24} r_{2534}
r_{3/5} r_{2/5} \theta(y-x)
 + c_3 r_{32} r_{3512} r_{33} r_{3534}
r_{1/5} r_{4/5}  \theta(x-y) \right]
     e \bar u_1 \gamma^\mu u_2 \,  e \bar v_4 \gamma_\mu v_3 \,  $$
$$ - \left\{
 \left[ {1 \over 2} c_2 (q_{12}^2 + q_{34}^2) + 1\right]
  r_{21} r_{2512} r_{24} r_{2534}
  r_{3/5}  r_{2/5} \theta(y-x) \right.$$
$$\left.
+  \left[ {1 \over 2} c_3 (q_{12}^2 + q_{34}^2) + 1\right]
r_{32} r_{3512} r_{33} r_{3534}
r_{1/5} r_{4/5}  \theta(x-y) \right\}
{e \bar u_1 \gamma^+ u_2 \,  e \bar v_4 \gamma^+ v_3 \,  \over (x-y)^2
 P^{+\, 2} } \, . \eqno(3.106)$$

\noindent The coefficients $c_1$, $c_2$ and $c_3$ are given by universal
Eqs.  (3.82.b) to (3.82.d) which were derived already in Yukawa theory,
with the replacement of the meson mass by the photon mass, $\mu^2 =
\mu^2_\delta$.

This result will be now analysed term by term for illustration in the
electron-positron scattering in second order perturbation theory.  The
first term in Eq.  (3.71) for the $T$-matrix calculated to order $e^2$
has matrix elements equal to the matrix elements of the effective
interaction from Eq.  (3.106).  When evaluating the $S$-matrix elements,
one considers configurations where the free energy of incoming fermions
equals the free energy of the outgoing fermions.  This configuration
selects the energy-diagonal part of the effective interaction:  $s_{13}
= s_{24} = s$ and $q^2_{12} = q^2_{34} = q^2 $. Also, the external
similarity factor which appears in the effective hamiltonian as an
additional factor to $g_{1\bar 1 \bar 1 1 \lambda}$ equals 1. Thus,
$g_{1\bar 1 \bar 1 1 \lambda}$ in the energy diagonal part can be viewed
as the scattering amplitude.  It simplifies in the energy diagonal part
to

$$
   g_{1\bar 1 \bar 1 1 \lambda} =
   g^{counterterm}_{1\bar 1 \bar 1 1 \epsilon}
   - c_1 e\bar u_1 \gamma^\mu v_3 \,  \, e \bar v_4 \gamma_\mu u_2 \,
$$
$$
   +\left[ - c_1 s + 1 \right]
   {e\bar u_1 \gamma^+ v_3 \,  \, e \bar v_4 \gamma^+ u_2 \, \over
   P^{+ \, 2}}
$$
$$
- \left[ c_2 r_{2512} r_{2534} r_{3/5} r_{2/5} \theta(y-x)
       + c_3 r_{3512} r_{3534} r_{1/5} r_{4/5} \theta(x-y) \right]
     e \bar u_1 \gamma^\mu u_2 \,  e \bar v_4 \gamma_\mu v_3 \,
$$
$$
- \left\{
  \left[ c_2 q^2 + 1\right] r_{2512} r_{2534} r_{3/5} r_{2/5}
  \theta(y-x)
  \right.
$$
$$
  \left.
+ \left[ c_3 q^2 + 1\right] r_{3512} r_{3534} r_{1/5} r_{4/5}
  \theta(x-y)
  \right\}
{e \bar u_1 \gamma^+ u_2 \,  e \bar v_4 \gamma^+ v_3 \,  \over (x-y)^2
 P^{+\, 2} } \, . \eqno(3.107)$$

\noindent We have removed the regularization factors which equal 1 for
fermions with finite free energy.

In the limit $\mu_\delta \rightarrow 0$ and for the cutoff $\lambda$
close to $-2m^2$, a number of simplifications occur.  We display the
result for the case where the intermediate photon momentum fraction
$|x-y| >> (x,y,1-x,1-y) \delta$, i.e. when the photon momentum is not
negligible in comparison to the fermion momenta.  In this case, the
infrared regularization factors for the photon which are still kept in
Eq.  (3.107) equal 1 and

$$
   g_{1\bar 1 \bar 1 1 \lambda} =
   g^{counterterm}_{1\bar 1 \bar 1 1 \epsilon}
   - {u_1 \gamma^\mu v_3 \,  \, e \bar v_4 \gamma_\mu u_2 \over s}
$$
$$ + \theta\left[ |q^2| - |x-y|(2m^2
+\lambda^2)/max(x,y,1-x,1-y)\right]
   {e \bar u_1 \gamma^\mu u_2 \,  e \bar v_4 \gamma_\mu v_3 \over q^2}
$$
$$ - \theta\left[ |x-y|(2m^2
+\lambda^2)/max(x,y,1-x,1-y) - |q^2| \right]
{e \bar u_1 \gamma^+ u_2 \,  e \bar v_4 \gamma^+ v_3 \,  \over (x-y)^2
 P^{+\, 2} } \, . \eqno(3.108)$$

The first term is the potentially necessary counterterm which we have
not yet determined.  Since the remaining terms are not sensitive to
$\epsilon$, the counterterm matrix element is equal zero.

The second term is the well known expression for the electron-positron
annihilation channel scattering amplitude in the Born approximation.  No
limits on the fermion momenta appear because the invariant mass squared
of two fermions is larger than $4m^2$ which is the minimal invariant
mass difference in the transition between the two fermions and a one
massless photon state.  Since we assume the width $\lambda^2$ to be
small the effective hamiltonian contains the full amplitude for
transition through the intermediate photon state.

The third term is equal to the standard second order expression for the
electron-positron scattering amplitude via one-photon exchange, except
for the $\theta$-function factor which forces the momentum transfer to
be sufficiently large.

The meaning of this restriction is visible in the $T$-matrix.  The third
term contributes to the $e^+e^-$-scattering amplitude through the first
term in Eq.  (3.71).  The same contribution would originate from the
second term in Eq.  (3.71) if we were using the initial hamiltonian to
calculate the $T$-matrix.  In contrast, the effective hamiltonian with
the small width $\lambda$ limits the effective photon emissions and
absorptions to small momentum transfers and, therefore, it is not able
to provide this contribution through the second term in Eq.  (3.71).
This contribution is then contained in the effective hamiltonian above
and comes in the scattering matrix through the first term in Eq.
(3.71).

The fourth term is unusual in the sense that it should not appear in the
electron-positron scattering at all.  The fourth term distinguishes the
$z$-axis in its structure and diverges when $x \rightarrow y$.  The
$\theta$-function factor in the fourth term is equal 1 where the
$\theta$-function factor of the third term is equal 0. And vice versa,
the fourth term $\theta$-function equals 0 where the third term
$\theta$-function equals 1.

The need for the fourth term becomes clear when one recalls that the
second term in Eq.  (3.71) also contributes to the electron-positron
scattering amplitude.  The relevant contribution comes through the
effective one-photon exchange which results from the double action of
$H_{I\lambda}$.  $H_{I\lambda}$ is given by the operation $F_\lambda$
applied to the third term of the QED hamiltonian from Eq.  (3.84), (see
Eq.  (2.11)).  The operation $F_\lambda$ multiplies the photon emission
and absorption vertices by the factor $f_\lambda$.  This factor was set
equal to a $\theta$-function in the current example.  Each interaction
provides one factor of the $\theta$-function.  The resulting factor in
the second term of Eq.  (3.71) is the same in QED as in Yukawa theory in
Eq.  (3.72), except for the antisymmetrization effect which leads to the
$\theta$-function which stands in front of the fourth term in Eq.
(3.108).

Now, the second term in Eq.  (3.71) contains spin factor which is the
same as in the last term in Eq.  (3.104).  The $g^{\mu\nu}$ part
complements the third term in Eq.  (3.108) and produces the full well
known one-photon exchange scattering amplitude which is free from the
$\theta$-function factor.  The remaining part provides the term which
cancels the odd fourth term in Eq.  (3.108).  Thus, the effective
hamiltonian calculated to second order in powers of the charge $e$
contains an odd term and $\theta$-functions which are required to
compensate for the odd contributions to the scattering amplitude from
the effective hamiltonian order $e$ acting twice.  The above analysis of
the energy diagonal part of the second order effective hamiltonian
explains the role of different terms in Eqs.  (3.106) to (3.108).

The analysis also suggests that apparently infrared diverging terms in
the effective hamiltonian may mutually compensate their diverging
contributions in the scattering amplitude on energy-shell.  In the
current example, we see the interplay between the second-order seagull
term and the double action of the first order emission and absorption of
photons.  The first order hamiltonian matrix elements diverge when the
photon longitudinal momentum approaches zero.  The second order seagull
term compensates this divergence in the on-energy-shell $T$-matrix
elements.

The remaining point to make here is that the result of Eq.  (3.106) with
the counterterm equal 0 leads to the effective light-front hamiltonian
version of the Coulomb force in the limit of small $\lambda^2 + 2m^2 \ll
\alpha m^2$.  The key elements in deriving this conclusion are the outer
similarity factors and the smallness of $\alpha$.  The outline of the
derivation is following (cf.  Refs.  \cite{JPG} and \cite{Wieckowski}).

If only the small energy transfers are allowed by the outer similarity
factor, i.e. transfers much smaller than the electron mass, then the
wave functions of the lowest mass eigenstates of the effective
hamiltonian are strongly peaked at small relative electron momenta and
they fall off very rapidly as functions of the relative momentum.  This
is not true without the outer similarity factor because the function
$g_{1\bar 1 \bar 1 1 \lambda}$ alone is too large at the large energy
transfers and it would produce singular contributions in the large
relative momentum region making the eigenvalue problem sensitive to the
ultraviolet regularization cutoffs.

Below the width scale the wave functions fall off as dictated by the
eigenvalue equation with small $\alpha$.  Above the width scale the fall
off is very fast due to the similarity factor which justifies
restriction to momenta much smaller than $m$ and the nonrelativistic
approximation for all factors in Eq.  (3.106) becomes accurate.

In the nonrelativistic approximation, $q_{12}^2 = q_{34}^2 = q^2$ and
Eq.  (3.107) applies.  Further, the $\theta$-functions in Eq.  (3.108)
become effectively equal 1 and 0, respectively.  The last term is not
leading to important contribution despite its divergent longitudinal
structure because it is canceled by the effective massless photon
exchange as described earlier in this Section.

The dominant contributions are provided by the second and the third
terms from Eq.  (3.108) which are well known to have the right
nonrelativistic structure for predicting positronium properties in the
Schr\"odinger equation.  Now, the outer similarity factor becomes
irrelevant to the spectrum in the leading approximation because the
coupling constant is very small (cf.  \cite{JPG} and \cite{Wieckowski}).
In the dominant region the electron velocity is order $\alpha$, the
nonrelativistic approximation to the full dynamics produces wave
functions with relative momenta order $\alpha m$ and the outer
similarity factor in the effective interaction can be replaced by 1.

We can use the infinitesimal $u_0$ in Eq.  (2.20) and replace $\lambda^2
+ 2m^2$ by $\tilde \lambda^2$ (see the discussion below Eq.  (2.20)).
When $\tilde \lambda$ is order $\alpha m$ and $x-y$ is order $\alpha$
the momentum transfer $\vec q ^{\, 2}$ is typically order $\alpha^2 m^2$
which is much larger than $2 (x-y)\tilde \lambda^2$ in Eq.  (3.108).
Thus, the $\theta$-function is equal 1 and the second term in Eq.
(3.108) becomes equal to the standard Coulomb interaction with the well
known Breit-Fermi structure of the spin factors.  This step completes
the derivation of the Coulomb potential.  The derivation explains the
effective nature of the Schr\"odinger equation with the Coulomb
potential in the light-front hamiltonian formulation of QED.

\newpage
{\bf 3.d QCD}
\vskip.1in

The main conceptual complication in hamiltonian calculations in QCD is
confinement which is not yet fully understood.  To order $g^2$ the
manifestation of the confinement problem is the lack of a well defined
initial condition for the renormalization group flow of the effective
hamiltonians.  It will require an extended research effort to find the
class of acceptable initial conditions.  For example, the on-mass-shell
renormalization conditions for the quark and gluon mass terms are
questionable.

We stress the urgent need for the higher order calculations by
describing some details of the second order calculation of the $q \bar
q$ effective interaction.  The calculation is similar to the one in QED
above, except for the option for a different treatment of the last term
in Eq.  (3.108).  Perry \cite{Perry} suggested that the long distance
part of this term may remain uncanceled in the effective QCD dynamics
because of the gluon non-abelian gauge interactions.  If the last term
in Eq.  (3.108) would remained uncanceled it could be claimed to
generate confinement in the light-front hamiltonian approach to QCD.
\cite{Perry} We describe the structure of this term in the present
approach since it is different than in Ref.  \cite{Perry}.  The
differences result from the different definitions of the similarity
transformation, boost invariance and not invoking coupling coherence.
The coupling coherence arguments are replaced by a plain perturbative
renormalization condition for quark mass terms.

\vskip.3in
\centerline {\bf Quark and gluon mass terms}
\vskip.1in

Results one obtains  from Eq.  (2.38) in QCD in second order in ${\cal
G}_{2\lambda}$ can be illustrated by two equations for the effective
masses of quarks and gluons. The range of widths and coupling constants
for which these equations can pertain to physics are not known yet.

We do not write regularization factors in detail.  One can write
them easily using results of Section 2 and previous examples in Section
3. In the case of effective masses a number of simplifications occur as
explained in the previous Sections.  Let us consider an infinitesimal
$u_0$ in Eq.  (2.20) and simplify notation by replacing $\tilde \lambda$
by $\lambda$ itself.  Then, we can write

$$ {d \over d\lambda} {\cal G}_{1\lambda} =
\left[  {\cal G}_{12\lambda}
{d f^2(z_\lambda / \lambda^2) / d\lambda \over  {\cal G}_{1\lambda}
- E_{1\lambda} } {\cal G}_{21\lambda} \right]_{11} \, .\eqno(3.109)$$

\noindent $E_{1\lambda}$ is the eigenvalue of ${\cal G}_{1\lambda}$
corresponding to the subscript 11.  A set of arguments $z_\lambda$ is
needed.  Namely,

$$  z_1 = {\kappa^2 + \mu^2_\lambda \over x(1-x) } - \mu^2_\lambda \, ,
\eqno(3.110) $$

$$ z_2 = {\kappa^2 + m^2_\lambda \over x(1-x) } - \mu^2_\lambda \, ,
\eqno(3.111) $$

$$ z_3 = {\kappa^2 + m^2_\lambda \over x} +
{\kappa^2 + \mu^2_\lambda \over 1-x } - m^2_\lambda \, . \eqno(3.112) $$

\noindent $m_\lambda$ and $\mu_\lambda$ are the effective quark and
gluon masses, respectively. Then,

$$ {d m^2_\lambda \over d\lambda} = \int [x\kappa] g^2_{q\lambda}
 z_3^{-1} {d f^2 (z_3/\lambda^2)\over d\lambda }
\left[\kappa^2[2/x +4/(1-x)^2] + 2 m^2_\lambda (1-x)^2/x \right]
r_{qg\epsilon}(x,\kappa) \eqno(3.113) $$

\noindent and

$$ {d \mu^2_\lambda \over d\lambda} = 3 \int [x\kappa]g^2_{g\lambda}
z_1^{-1} {d f^2 (z_1/\lambda^2)\over d\lambda }
\kappa^2[4/x^2 +2]  r_{gg\epsilon}(x,\kappa)  $$
$$ + \int [x\kappa]g^2_{q\lambda}
z_2^{-1} {d f^2 (z_2/\lambda^2)\over d\lambda }
\left[ {\kappa^2 + m^2_\lambda \over x(1-x)} - 2 \kappa^2 \right] \,
r_{qq\epsilon}(x,\kappa) \, .\eqno(3.114) $$

\noindent The gluon couples to the quark-anti-quark pairs and pairs of
gluons while the quark couples only to the quark-gluon pairs.  The
number of colors above is equal to $3$ and the number of flavors to $1$.

These equations are not further studied here for two major reasons.  The
first one is that we do not know the initial conditions to use for such
study.  The second is that the equations involve two running couplings
which are not known yet.  The third and fourth order calculations are
required to find them.

The importance of effective mass issue for quarks and gluons is
illustrated below by the calculation of the small energy transfer
effective forces between quarks and anti-quarks.

\vskip.3in
\centerline {\bf Quark-anti-quark effective interaction}
\vskip.1in

The second order results are similar to QED.  For heavy quarkonia one
can directly look at Eq.  (3.108).  The only change required is the
color $SU(3)$ matrices sandwiched between color vectors of quarks and
summed over colors of the exchanged gluons.  The counterterm is 0. The
second term gives the annihilation channel potential but the color
matrix is traceless and this excludes the single gluon annihilation
channel from the dynamics of color singlet $Q \bar Q$ states.  The third
term leads to the color Coulomb potential with the well known
Breit-Fermi spin factors.

About the last term in Eq.  (3.108) it was suggested by Perry
\cite{Perry} that a term of this kind may become a seed for confining
interactions if it is not canceled.  A cancelation occurs in
perturbation theory when one evaluates the model interaction in the
$Q\bar Q$ sector in the explicit expansion in powers of $g$ to second
order using the operation $R$ and Eq.  (1.3) in the nonrelativistic
approximation for the quark relative momentum, exactly the same way as
for electrons in QED with massless photons.

However, there are reasons for effective gluons to acquire a gap through
the non-abelian terms which are absent in effective QED.  One can assume
that the gluon energies in the $Q\bar Q g$ sector may be lifted up so
that the gluons cannot contribute to the model $Q\bar Q$ interaction in
the way the abelian massless photons can in the model electron-positron
interaction in QED.  As a result of this assumption one obtains the last
term in Eq.  (3.108) acting in the effective $Q\bar Q$ sector.

In the nonrelativistic approximation, the Coulomb term and the term in
question are (see Eq. (3.108))

$$ -\, \theta\left[ {\vec q}^{\,2} - |x-y|\,2\tilde \lambda^2 \right] \,\,
{g^2 4 m^2 \over \vec q^{\,2}}\quad - \quad \theta\left[ |x-y|\,2\tilde
\lambda^2 - {\vec q}^{\,2} \right] \,\, {g^2 \over (x-y)^2 } \,\, .
\eqno(3.115)$$

\noindent The regularization factors are the same in both terms and they
are not displayed.  The only effect of their presence which matters here
is that $|x-y|$ is limited from below by about $\delta/2$.  The initial
infrared regularization parameter $\delta$ comes in through the initial
condition in the renormalization group flow of the seagull term.  The
flow is limited in the second order calculation to the dependence of the
outer similarity factor on $\tilde \lambda$.  The factor $1/2$ results
from $|x-y|/x_{quark}$ being limited from below by $\delta$ and the
quarks having $x_{quark} \sim 1/2$.  In fact, the lower bound on $|x-y|$
is given by $ \delta \, max(x,1-x,y,1-y)$.  This is different from Ref.
\cite{Perry} where instead of the ratios of the $+$-momentum fractions a
separate frame dependent scale for $+$-momentum is introduced.

The third component of the exchanged gluon momentum is $q_3 = (x-y) 2m$.
Thus, we see that the uncanceled singular term is represented by the
potential which is analogous to the Coulomb potential except for that
the factor $ - 1 /{\vec q}^{\,2}$ is replaced by $ - 1 / q_3^2$ and both
terms have mutually excluding and complementary supports in the momentum
transfer space.

The seagull $\theta$-function can be rewritten as $ \theta[ \omega^2 -
(|q_3| - \omega)^2 - q^{\perp\,2}]$, where $\omega = \tilde \lambda^2 /2
m$.  The support of this function is two spheres of radius $\omega$
centered at $q^\perp = 0$ and $q_3 = \pm \omega$.  The spheres touch
each other at the point $q^\perp = q_3 = 0$.  In this point, $q_3^2$ in
the denominator produces a singularity.

Let us initially consider both terms in Eq.  (3.115) as the actual
interaction in the model $Q\bar Q$ sector, i.e. as if they were not
affected by the operation $R$ in Eq.  (1.3).  The Coulomb term works
outside the two spheres in the $\vec q$-space and the singular seagull
term works inside.

In the region of the singularity, both $q^\perp $ and $q_3$ are small in
comparison to $\omega$.  In this rough analysis one can neglect the
outer similarity factor $\theta(\tilde \lambda^2 - |k^2 -k'^2|)$ since
it is equal 1 when $\vec q = \vec k - \vec k'$ approaches 0. $\vec k$ is
the relative momentum of the created $Q\bar Q$ pair and $\vec k'$ is the
relative momentum of the annihilated $Q\bar Q$ pair.  $\omega = (\tilde
\lambda / m) \tilde \lambda /2 \ll \tilde \lambda/2$ and the spheres
have the radius about $\tilde \lambda / m$ times smaller than the outer
similarity factor width in the quark relative momenta,
i.e. the relative size of the spheres in comparison to the outer
similarity factor support approaches 0 when $\tilde \lambda/m
\rightarrow 0$.

Since $q^{\perp \, 2}$ is order $q_3$ in the singular region the
divergence when $q_3 \rightarrow 0$ is logarithmic.  The lower limit of
integration over $|q_3|$ for a given $x$ is given by $ 2 m\, \delta \,
max(x,1-x)$. However, we assume $x = 1/2 + o(g^2)$ and we neglect terms
of higher order than $g^2$ in the model $Q \bar Q$ hamiltonian.

The potential resulting from the uncanceled seagull term is
given by the following expression (cf. \cite{Perry}),

$$ V(\vec r) \quad \sim \quad - \,\, \int {d^3 q \over (2\pi)^3}\,\,
\exp{(i\,\, \vec q \, \vec r)} \,\,\, { \theta(2\omega |q_3| - {\vec q}^{\,2})
\,\,\theta (|q_3| - 2 m \delta) \over q_3^2 } \, .\eqno(3.116) $$

\noindent The sign $\sim$ means that the diverging dependence on
$\delta$ is subtracted and the same coefficient stands in front of the
integral as in the Coulomb potential term.  The argument for the
infrared subtraction goes as follows.

If the gluons cannot cancel the last term in Eq.  (3.108) they
presumably cannot contribute to the model quark self-energies either for
the same reason.  Because the size of the quark mass in the effective
hamiltonian is unknown one may propose that its value is chosen in the
second order calculation in the same way as for nucleons in the Yukawa
theory in Eq.  (3.37) or electrons in QED in Eq.  (3.99).  A would-be
quark eigenstate has a finite constituent quark mass when the gluons are
allowed to contribute in the whole range of momenta from zero up in the
eigenvalue equation.  This setting is equivalent to the solution Perry
proposed for his coupling coherence condition for the quark
self-energies.  \cite{Perry} The argument also illustrates the urgency
of questions concerning the initial conditions and higher order analysis
in the similarity renormalization group flow.

There is nothing wrong with the mass adjustment despite the infrared
divergence.  We have noticed in the previous Section that the arbitrary
finite parts of the ultraviolet counterterms can be infrared divergent.
This time, however, the positive and infrared logarithmically divergent
part of the effective quark mass term in the model eigenvalue equation
for heavy quarkonia remains uncanceled when the transverse gluons with
$2\omega|q_3| - \vec q^{\,2} > 0$ are declared to be absent from the
model dynamics.  The uncanceled part of the effective quark mass term
stands in the eigenvalue problem.  The point is it can now cancel the
diverging $\delta$ dependence in the seagull term which is not canceled
because the gluon exchange below $\tilde \lambda$ is missing.  The new
cancelation between the incomplete masses and the seagull occurs in the
colorless states.  It is analysed here in the nonrelativistic limit.

We describe the cancelation mechanism in the case of equal masses of
quarks.  The mechanism is similar but not identical to that in Ref.
\cite{Perry}.  The infrared divergent mass squared term comes into the
quarkonium eigenvalue equation divided by $x(1-x)$.  But in the second
order analysis the mass divergence appears only as a logarithmically
divergent constant and the $x$-dependence is of higher order.  The same
diverging constant with the opposite sign is generated by the seagull
term.

The infrared divergent terms and their cancelation are not directly
related to the ultraviolet renormalization procedure.  They appear in
the ultraviolet-finite effective small width hamiltonian dynamics.  Note
also that the introduction of the gluon mass $\mu_\delta$ in the
regularization could matter for the lower bound on $|x-y|$ and it could
even eliminate the whole contribution when the upper bound of $\omega$
meets the lower bound of $\mu_\delta$.  We assume here $\mu_\delta = 0$.

The divergent part in Eq.  (3.116) is independent of $\vec r$ and it is
easily removed by subtracting 1 from $\exp{i \vec q \vec r}$.
Evaluation of the integral leads to the answer that for large $r$
the seagull term produces a logarithmic potential of the form

$$ V(\vec r) \quad  \sim  \quad
{2 \, \omega \, a(\hat e_r)  \over \pi}\,\, \log{r} \,\, , \eqno
(3.117) $$

\noindent where $a$ is equal 1 for the radial versor $\hat e_r$ along
the $z$-axis and it equals 2 when $\vec r$ is purely transverse.  This
potential is confining.  It is also boost invariant.  But the rotational
asymmetry of the potential raises doubts.  It suggests that an important
piece of physics is missing in the reasoning used to derive it.  The
obvious sources of questions are the mechanism of blocking the effective
gluon emissions and absorptions, role of the operation $R$, the role of
the nonrelativistic approximation, the size of the quark and gluon
masses and the strong dependence of the term on the width $\tilde
\lambda$.  The most urgent question is what happens in higher order
calculations.

\vskip.5in
{\bf 4. CONCLUSION}
\vskip.1in

We have defined and illustrated on a few perturbative examples a general
method of calculating light-front hamiltonians which can be used for the
relativistic description of interacting particles.  The starting point
in the calculation is a field theoretic expression for the bare
hamiltonian density.  This expression is multiply divergent in the
physically interesting cases.  Therefore, the hamiltonian theory
requires renormalization.

In the renormalization process, one calculates a whole family of
effective hamiltonians as functions of the width parameter $\lambda$
which determines the range of the effective interactions on the energy
scale.

An effective hamiltonian of a small width $\lambda$ is much different
from the initial bare hamiltonian.  It couples only those states whose
masses differ by less than a prescribed amount.  Thus, the effective
theory contains only near-neighbor interactions on the energy scale.  No
scale is removed in the calculation but the correlations between
dynamics at significantly different energy scales are integrated out.
Therefore, in principle, the effective eigenvalue problem can be solved
scale by scale using standard techniques for finite matrices which
describe dynamics at a single scale.

Our formalism is based on the earlier work on renormalization of
hamiltonians from Refs.  \cite{GW1}
and \cite{GW2} where the hamiltonians are defined by their matrix elements
in a given set of basis states. Wegner has developed similar
equations for hamiltonian matrix elements in solid state physics. \cite{WEG}
The present approach to renormalization of hamiltonians
introduces the following features.

Our similarity transformation is defined in terms of creation and
annihilation operators.  Consequently, calculations of counterterms in
perturbation theory can be performed without knowing details of the
specific Fock states which are needed to evaluate the matrix elements.
This is useful because a large number of Fock states needs to be
considered.  The renormalization scheme is free from practical
restrictions on the Fock space sectors.

Expressing the effective hamiltonians in terms of the creation and
annihilation operators of effective particles and showing that the
effective interactions are connected is a prerequisite to obtain the
cluster decomposition property.  \cite{WE1} The effective interactions
in our approach do not contain disconnected terms.  The number of
creation and annihilation operators in a single term is limited in
perturbation theory by $2+n(V-2)$ where $n$ is the order of a
perturbation theory and $V$ is the number of operators in the perturbing
term.

The physically motivated assumptions about the model space of effective
states included in solving a particular problem are introduced after the
effective hamiltonian is calculated.  The interaction terms in the
effective hamiltonian contain the similarity factors which diminish the
dynamical significance of the Fock sectors with numbers of effective
particles considerably different from the number of effective particles
in the dominant sectors.

The present operator formulation does not introduce spectator dependent
interactions, even in the case where we include the sums of the
invariant masses for incoming and outgoing particles in the similarity
factors.  The sums are useful for estimates of cutoff dependence in
perturbation theory.

The formalism explicitly preserves kinematical symmetries of the
light-front frame.  The structure of counterterms is constrained by
these symmetries, including boost invariance.  Hence, the number of
possible terms is greatly limited.  Preserving boost invariance is
particularly important because it is expected to help in understanding
the parton model and constituent quark model in QCD, simultaneously.

It is essential to include the running of the coupling constants in the
calculation of the small width dynamics.  The examples of second order
calculations we described in this article do not include the running
coupling constant effects.  Inclusion of these effects requires higher
order calculations.

Wegner's equation can be adapted to building an operator approach
similar to what we described in the present article.  The initial
equation which replaces our Eq.  (2.29) when one uses the Wegner
generator of the similarity transformation is

$$ {d {\cal H}_\lambda \over d\lambda^2} \quad = \quad {-1\over
\lambda^4}\,\, \left[ \, [{\cal H}_{1 \lambda}, {\cal H}_{2 \lambda}],
\, {\cal H}_\lambda \right]\,\,.\eqno(4.1)$$

\noindent However, there is little flexibility left in the equation so
that the widening of the hamiltonian band is not readily available.

There exists a class of generalized equations for the flow of hamiltonian 
matrix elements described in Ref. \cite{GW4} and already studied in a 
simple numerical model.  These equations allow widening of the effective 
hamiltonian matrix at large energies.  The generalized equations can also 
be adapted for the construction of the creation and annihilation operator 
calculus.  Namely,

$$ {d {\cal H}_\lambda \over d\lambda} \quad  = \quad
\left[ F\{{\cal H}_{2 \lambda}\}, \, {\cal H}_{\lambda} \right]
\,\, . \eqno(4.2)$$

\noindent These equations require detailed definitions of the similarity
factors generated by the operation $F$. \cite{GW4}

In summary, the present formalism for renormalization of hamiltonians in
the light-front Fock space provides a tool for working on a host of
theoretical issues in particle dynamics. Second order applications produce
boost invariant Yukawa potential, Schr\"odinger equation for internal
bound state dynamics and logarithmically confining quark-anti-quark
interaction. However, it remains to be verified if the formalism can 
lead to quantitative improvements in our description of particles.  
Rotational symmetry and infrared singularities in gauge theories require 
further studies.  Most urgent are the calculations of effective hamiltonians 
in the third and fourth order perturbation theory.  

\vskip.5in
{\bf Acknowledgment}
\vskip.1in

The author thanks Ken Wilson, Bob Perry, Billy Jones, Tomek
Mas{\l}owski, Marek Wi{\c e}ckowski, Martina Brisudov\'a and 
Brent Allen for numerous discussions.

\end{document}